\documentclass[report]{revtex4-1}
\usepackage{url}
\usepackage{graphicx}
\usepackage{dcolumn}
\usepackage{bm}
\usepackage{latexsym}
\usepackage{ulem}
\usepackage{amsmath}
\usepackage{amssymb}
\usepackage{xcolor}
\usepackage{array}
\usepackage{tabularx}
\usepackage{longtable}

\begin{document} 

\title{\bf Length control of long cell protrusions: rulers, timers and transport}

\author{Swayamshree Patra }
\email{Corresponding author; Present address: Department of Molecular, Cellular and Developmental Biology, Yale University,\\ New Haven, CT 06511, USA; Electronic address: swayamshree.patra @yale.edu}
\author{Debashish Chowdhury}
\email{Electronic address: debch@iitk.ac.in}
\affiliation{Department of Physics, Indian institute of Technology Kanpur, Kanpur 208 016, India}

\author{Frank J\"ulicher}
\email{Electronic address: julicher@pks.mpg.de}
\affiliation{Max-Planck Institute for the Physics of Complex Systems, N\"othnitzer Strasse 38, 01187 Dresden, Germany}
\date{\today}

\begin{abstract}
{\bf Abstract:} A living cell uses long tubular appendages for locomotion and sensory purposes. Hence, assembling and maintaining a protrusion of correct length is crucial for its survival and overall performance. Usually the protrusions lack the machinery for the synthesis of building blocks and imports them from the cell body. What are the unique features of the transport logistics which facilitate the exchange of these building blocks between the cell and the protrusion? What kind of ‘rulers’ and ‘timers’ does the cell use for constructing its appendages of correct length on time? How do the multiple appendages coordinate and communicate among themselves during different stages of their existence? How frequently do the fluctuations drive the length of these dynamic protrusions out of the acceptable bounds?  These questions are addressed from a broad perspective in this review which is organized in three parts. In part-I the list of all known cell protrusions is followed by a comprehensive list of the mechanisms of length control of cell protrusions reported in the literature. We review not only the dynamics of the genesis of the protrusions, but also their resorption and regrowth as well as regeneration after amputation. As a case study in part-II, the specific cell protrusion that has been discussed in detail is eukaryotic flagellum (also known as cilium); this choice was dictated by the fact that flagellar length control mechanisms have been studied most extensively over more than half a century in cells with two or more flagella. Although limited in scope, brief discussions on a few non-flagellar cell protrusions in part-III of this review is intended to provide a glimpse of the uncharted territories and challenging frontiers of research on subcellular length control phenomena that awaits vigorous investigations.

\end{abstract}

\maketitle

\tableofcontents{}

\section{Introduction}

The question of the size of living systems has fascinated humans for centuries; the stories of ``Gulliver's Travels'' and ``Alice in Wonderland'' are the prime examples.  In a classic essay, titled ``on being the right size'', J.B.S. Haldane \cite{haldane25} first analyzed the physical reasons that would explain why ``for every type of animal there is a convenient size''. Haldane focused his analysis on the size of whole organisms \cite{bonnerbook}. The mechanisms that ensure the ``convenient'' size of a cell \cite{marshall16,rafelski08} and sub-cellular structures \cite{marshall02,marshall15} have become a very active field of research in recent years. In this article, instead of a multi-cellular organism or a single cell, we consider the size of subcellular structures. Long protrusions and membrane-bound organelles, that appear as cell appendages, are prominent among such sub-cellular structures. The advantage of choosing these long protrusions for studying the mechanisms of subcellular size control is that these are effectively one-dimensional systems for which the length characterizes the size \cite{marshall04}. Thus, our review is restricted to the study of length control of cell protrusions.

Phenomena related to size of living systems have been studied with at least four different motivations:
\begin{itemize}
\item (I) One of the motivations is to find out the relations between the size and various other parameters that characterize the structure or functions of the organism; the results have been presented usually in the form of scaling relations that are referred to as {\it allometric relations} \cite{westbook,bonnerbook}.
\item (II) Another related motivation, particularly from the perspective of physical sciences, is to derive the  allometric relations from the laws of physics and chemistry \cite{westbook,bonnerbook}. 
\item (III) As Dobzhansky’s famous statement summarizes, ``nothing in biology makes sense except in the light of evolution’’  \cite{dobzhansky}; the structural design of the cell protrusions are also no exception. Their structures observed in contemporary experiments have naturally resulted over millions of years by `tinkering’, following the Darwinian principles of evolution and selection. Naturally, from the perspective of biology, a major motivation for studying size is to explore not only the `proximate' and `ultimate' {\it causes} \cite{mayr61} of selection of the specific convenient size, but also the functional {\it consequences} of any deviation from that size that may be caused, possibly, by mutation.
\item (IV) From the perspective of biological physics, a strong motivation is to understand the (a) dynamics of growth up to a given convenient size, (b) mechanisms for stopping further growth, and maintaining that size subsequently, and (c) dynamics of length changes, if required either as part of the cell cycle or in its response to environmental stress, etc. The curiosities on the identity of the molecular rulers used by the cell to measure protrusion length and the logistics of intra-protrusion transport for the delivery of the structure-building materials are also of current interest in biological physics.\\
It is the motivation of type (IV) that captures the spirit of this review.  
\end{itemize}

Although length may appear to be a mere geometric characteristic of a protrusion, there are many interesting questions on its dynamics and function that are intimately related to its size. The uniformity in the size of a particular type of protrusions in a given species of cells suggests that individual cells sense the size of these protrusions in relation to their respective target size. A fundamental question that arises here is: what type of molecular rulers does a cell use to measure the length of a protrusion and how is the information on the measured length sensed by the cell? How does a cell regulate the rate of growth or shrinkage of a protrusion?  Given a particular protrusion, does it elongate by adding subunits at its base or at its distal tip? In the latter case, how does the cell transport structure building components to the distal tip? In case the cell deploys molecular motorized transport vehicles for transporting protrusion building materials between the base and the tip of the protrusion, how do its rates of growth and shrinkage depend on this transport? Is the rate of growth or shrinkage, or both, dependent on the length of the protrusion and, if so, what is the functional form of their length-dependence? A key question on such a regulatory mechanism is: how does the cell stop further growth when it attains its target length?  In case of cells with multiple copies of the same appendage, for example a multiflagellated cell, there are even more intriguing questions: how does a multiflagellated cell simultaneously regulate the growth and/shrinkage of more than one flagellum? Are the dynamics of growth and shrinkage of different flagella of a cell correlated to each other and if so, how do different flagella communicate with one another and how do they share the structure building materials from  common pools in the cell? We critically review the literature that have addressed such questions on the mechanisms that control and regulate the lengths of long cell protrusions.

Components of a cell can undergo wear and tear during the life time of a cell and, therefore, almost continuous turnover of building blocks of most cell appendages take place to maintain their structural integrity. The long protrusions of a cell can also suffer damage or injury caused by external agents. If an appendage is severed, the materials that constituted its structure are lost and, unless the wound heals soon, further loss of more materials  from the punctured cell can prove fatal. Even when the wound heals by the cell's self-repair mechanism the cell may not regain its full functional ability if the appendage cannot be regenerated to its original form. We would like to emphasize that wound healing and regeneration are two distinct aspects of self-repair; healing simply stops further loss of intracellular material whereas regeneration is the process of rebuilding of an appendage to replace the lost one. Not all self-healing cells can regenerate lost structures. Beginning with the ancient mythologies, {\it regeneration} has captured the imagination for millenia. Scientific studies over the last century has established that regeneration can occur at different levels of biological organization, namely, at the levels of subcellular protrusions, cells in tissues, organs and even whole organisms \cite{tang17,slack17}. 
In this article, in the context of regeneration, we focus exclusively at the subcellular level and  review how some types of long protrusions are regenerated by a cell.

In the context of our discussion on regeneration of cell protrusions, the relevant questions are: 
How does a cell get the information that a protruding appendage has been removed? How does that signal trigger the protein synthesis programs? In principle, a cell could know the loss of a structure by sensing (i) mere loss of function of the structure, (ii) severe cellular stresses caused by the absence of the lost structure, or (iii) the disappearance of a signal that the intact structure produces. Once a cell gets confirmed information of the loss 
can it regenerate a retracted or amputated appendage and what conditions must be fulfilled so that the regenerating appendage grows to its original full length? What is the mechanism of regeneration and, for a specific protrusion, is it  completely different from the mechanism of its original genesis? In the case of, say, a multi-flagellated cell, how does the ongoing regeneration of a flagellum affect the lengths of other flagella of the same cell?

Throughout this article the main experimental observations on length control of cell protrusions are summarized to motivate the formulation of the corresponding theoretical models. But, we do not discuss the details of the methods and protocols followed in those experiments. In contrast, we delve somewhat deeper into the mathematical and computational methods used in analyzing the corresponding theoretical models because of our quest for the models that quantitatively account for as many of the observed phenomena as possible.  
This review is divided into three parts. In part-I all known cell protrusions are listed along with a summary of their major components and the key features of their architectural design, dynamics and function. This overview is followed by a comprehensive list of the length control mechanisms and their pedagogical explanation. Most of the questions posed in the preceding paragraphs are addressed in this part of the article. The physical mechanisms that govern dynamics of genesis, loss and regeneration of the protrusion are critically reviewed to establish not only their strengths but also to expose possible weaknesses. Part-II deals exclusively with eukaryotic flagellum which has served as the most popular system in the experimental studies of length control. It serves as a testing ground for the abstract models and theoretical predictions on length control.  Finally, limited purpose of part-III is to present a few fascinating examples of non-flagellar protrusions whose length control mechanisms need wider attention of investigators cutting across disciplinary boundaries.

\begin{widetext}
\part{General features of long cell protrusions}
\end{widetext}

In this part we present a list of cell protrusions along with their important properties that are relevant for highlighting `unity among the diversity'.  We summarize various mechanisms of length control, modes of protrusion loss and regeneration, the concept of length fluctuations of a single protrusion and cooperativity among the multiple copies of the protrusions present in a  cell.

\begin{figure*}
\begin{center}
\includegraphics[width=0.7\textwidth]{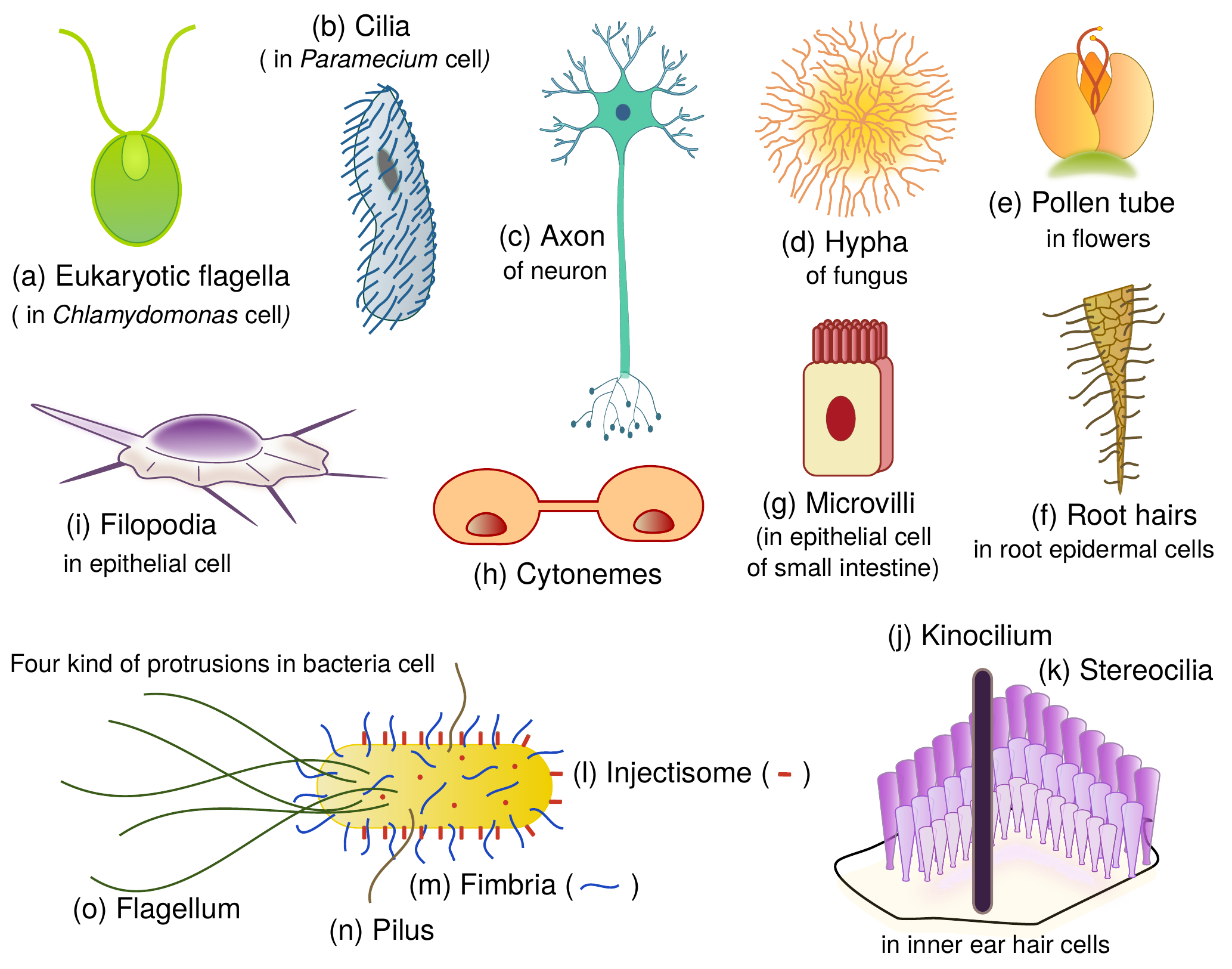}
\end{center}
\caption{ \textbf{Long protrusions of living cells and unicellular microorganisms:} Protrusions of (a-k) eukaryotic and (l-o) prokaryotic cells.}
\label{fig-protrusion_types}
\end{figure*}

\section{Unity in the diversity of cell appendages}
\subsection{Diversity}
We begin by reviewing the diversity in various characteristics of long cell protrusions. \\
\subsubsection{Geometric diversity}
The three characteristics which give the complete geometric description of a cell protrusion are the following :\\
(i) {\bf Length} in the steady state as well as its time dependence while away from the steady state; \\
(ii) {\bf Number of protrusions} at different stages of the life of a cell. Even at a given instant, not all the copies of a particular type of protrusion of a cell may be equally long \cite{marshall01,aizawa98,mcinally19,dotti88};\\
(iii) {\bf Positions on the cell surface} is important for proer biological function of the protrusions. For cells with multiple copies of a type of protrusion, different copies may be deployed at different positions on the cell surface. In the steady state they may be placed either symmetrically  about some axis \cite{mcinally19} or  asymmetrically at  random locations on the cell surface \cite{schuhmacher15}. 

\subsubsection{Compositional and structural diversity}

The eukaryotic protrusions can be broadly classified into two groups based on the dominant cytoskeletal filament type that is enclosed in a protrusion. These filaments provide structural strength and also act as track for the  motor mediated active transport. The two groups of protrusions are: \\

(i) {\bf Microtubule-based protrusions}: Microtubules are tubular stiff filaments which are polymers of subunit proteins called tubulin.  Eukaryotic flagellum (also called cilium) \cite{pedersen12}  is one of the most common examples of microtubule based long cell protrusions.\\
(ii) {\bf Actin-based protrusions:} Filamentous actin are polymers of monomeric subunit proteins called actin. These filaments are often present in branched form in many locations in a cell including some protrusions. But in some long cell protrusions, bundles of linear actin filaments are aligned with the axis of the tubular protrusion. Filopodia \cite{gallop20}, microvilli \cite{bretscher83} and stereocilia \cite{mcgrath17,schwander10,velez-ortega19} are example of such actin-based long cell protrusions. 

In contrast to the eukaryotic cells, bacteria lack microtubules and actin filaments.  Based on the type of major constituent proteins which polymerize to form the bacterial protrusions, we can classify them into the following groups:\\
(i) {\bf Flagellin-based protrusions:} Thousands of flagellin monomers polymerize sequentially to form the bacterial flagellum \cite{chen17,renault17} which has a hollow cylindrical structure. \\ 
(ii) {\bf Pilin-based protrusions:} Pili and fimbriae are formed by the polymerization of the pilin monomers \cite{ramirez20}.

\subsubsection{Dynamical diversity}

\begin{figure}
\begin{center}
\includegraphics[width=0.8\textwidth]{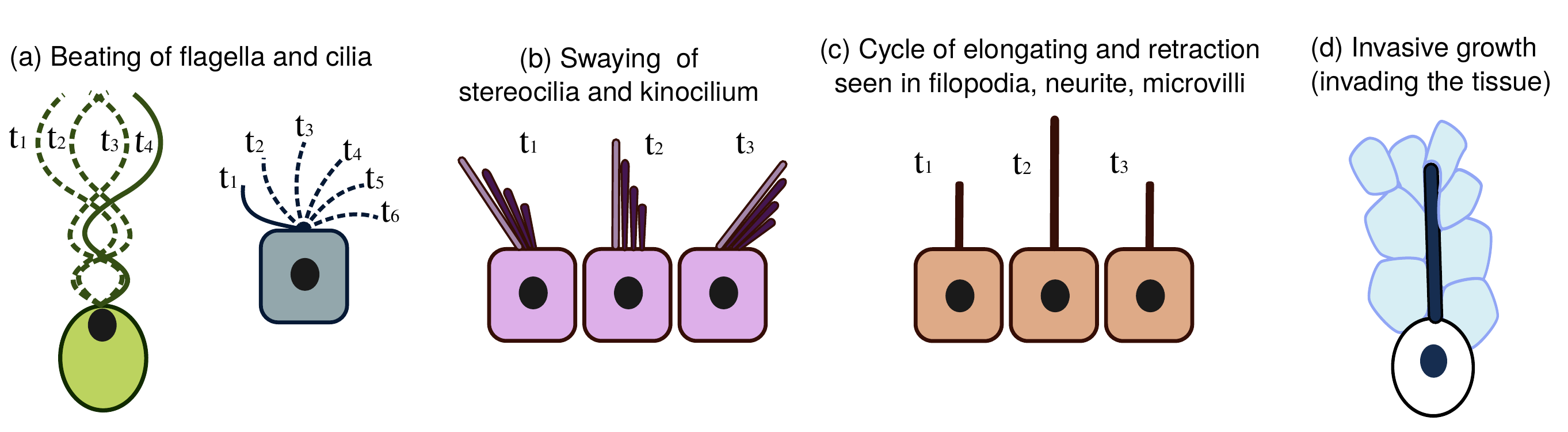}
\end{center}
\caption{ \textbf{Dynamics of protrusions :} Snapshots of dynamic protrusions depicted schematically at different time $t_1<t_2<t_3<t_4$. (a) Beating of  flagella and   cilia. (b) Swaying of  kinocilium and the stereocila. (c) Elongation and retraction dynamics as seen in filopodia, neurites and microvilli. (d) Invasive growth as seen in root hairs, pollen tubes and filamentous fungi invading through the surrounding tissue. } 
\label{fig-protrusions_dynamics}
\end{figure}

Four types of dynamic movements of the protrusions have been identified (See Fig.\ref{fig-protrusions_dynamics}(a)-(d)); a given protrusion, however, may exhibit more than one type of movement on different time scales of observation, depending on the functional necessity.\\
(i)  {\bf Immotile} protrusion remains static on sufficiently long duration of observation.  \\
(ii) {\bf Beating} is a whip-like wavy motion without change of length whereas {\it swaying}, which usually involves a cluster or aggregate of protrusions, is a rhythmic movement from one side to another (See Figs.\ref{fig-protrusions_dynamics}(a) and (b)).\\
 (iii) {\bf Elongation and retraction}  leads to growth and shrinkage of the protrusion (See Fig.\ref{fig-protrusions_dynamics}(c)).\\ 
 (iv) {\bf Invasion} \cite{nezhad13} manifests as a persistent push through the surrounding medium assisted by mechanical stress generated by an elongating tip that often {\it splits} and {\it branches} out (See Fig.\ref{fig-protrusions_dynamics}(d)).

\subsubsection{Lifetime diversity}

Based on their lifetime, the protrusion can be divided into the following two classes :\\
(i) {\bf Transient protrusions} retract completely and may re-emerge multiple times during the entire lifetime of a cell.  For example, filopodia, appear and disappear multiple times before the cell completes one full cycle.  \\
(ii) {\bf Permanent protrusions} emerge at a certain stage and remain intact throughout the lifetime of the cell unless some extraordinary situation arises (like accident or attack from a predator). Stereocilia and axon are two protrusions which exist for the entire lifetime of the cell. 

\subsubsection{Functional diversity}

Cells design and build their own protrusions for different functions that are highly diverse and depend on the cell type. We list only a few of the typical functions performed by cell protrusions \cite{nezhad13} to emphasize their diversity:\\
(i) {\bf Cell motility and migration}: Protrusions are needed for crawling (with lamellopodium) \cite{krause14}, swimming (with flagellum)  \cite{goldstein92},  etc. \\
(ii) {\bf Circulating surrounding fluid}: Many types of cells use cilia for circulating fluids, like mucus, over their surfaces  \cite{shinohara17,gilpin20}. \\
(iii) {\bf Cargo delivery}:  long protrusions, like pollen tubes of plants \cite{zhang17} and membrane nanotubes in animals \cite{davis08,abounit12} serve to deliver molecular cargo to distant locations. \\ 
(iv) {\bf Connecting distant locations within a multicellular organism}: Long protrusions of a cell can enable it to connect distant locations of the same multicellular organism. One of the most important cells of this category is the neuron \cite{nezhad13}; the long axons enable different parts of an animal body to be linked to the brain.\\
(v) {\bf Intercellular communication}: Long nanotubes known as cytonemes \cite{kornberg14,yamashita18} are known to serve as conduits for communication between two cells through exchange of matter and information. \\
(vi) {\bf Collecting cues and signals}: Many protrusive structures, like filopodia and stereocilia, are used for exploring the environment in search of different biochemical cues and mechanical signals \cite{winans16}.\\
(vii) {\bf Nutritional needs}: Some type of cells explore their surroundings, comprising of biotic and abiotic media, for organic and inorganic materials needed for its nutrition. Microvilli in the small intestine and fungal hyphae are the most familiar examples of this type of cell protrusions \cite{riquelme18, steinberg17}.\\
(viii) {\bf Mating}: Sex pili on bacterial surface are exclusively built for conjugation as they serve as conduits for transferring genetic material \cite{ramirez20}. Eukaryotic flagella are also used for mating \cite{wan21}.\\

In Table.\ref{tab-protrusion-diversity}, we list the properties of several long cell protrusions. In spite of the diversities summarized above, there are some common features in various  aspects of their structure and dynamics which we discuss next.  \\


\begin{longtable*} {|>{\centering\arraybackslash}m{1.8cm}|>{\centering\arraybackslash}m{2cm}|>{\centering\arraybackslash}m{1.5cm}|>{\centering\arraybackslash}m{1.3cm}|>{\centering\arraybackslash}m{2cm}|>{\centering\arraybackslash}m{2cm}|>{\centering\arraybackslash}m{3.5cm}|>{\centering\arraybackslash}m{2cm}|}
\hline 
\multicolumn{8}{c}{{ \bf Diversity of long cell protrusions (Length, Lifetime, Structure, Dynamics and Function)} }\\
\hline
{\bf Protrusion} & {\bf  Host cell} & {\bf Length} & {\bf Copies} & {\bf  Lifetime} & {\bf  Internal structure } & {\bf Function} & {\bf Dynamics} \\
\hline
{\bf Eukaryotic flagellum}  & { Green algae, Giardia, Sperm, Trypanosome \cite{vincensini11} } & 10-20 $\mu$m \cite{marshall01,mcinally19} & 1-16 & Less than one cell cycle to multiple cell cycle & Mictrotubule based  \cite{prevo17} & Swimming, Chemosensing, Mating, Food assimilation & Beating and gliding \cite{goldstein15} \\
\hline
{\bf Eukaryotic cilium}  & { Paramecium, mammalian cells} & 2-10 $\mu m$  & 1-6000 & Multiple cell cycle & Mictrotubule based &  Chemo / Osmo / Photo / Odour sensing,  Food assimilation, Circulating and directing the fluid, Swimming &  Immotile, Elongation and retraction \\
\hline
{\bf Neurites} & Neuron & 10-20 $\mu$m \cite{dotti88} & 4-6 \cite{dotti88} &  Transforms into dendrites \& axon after few hours \cite{dotti88} & Microtubule based \cite{winans16}  & Searching targets using guiding cues \cite{winans16}  & Elongation and retraction \cite{winans16}\\
\hline
{\bf Axon} & Neuron & Few $\mu$m (insects) - few meters (girrafe) & 1 &  Entire lifetime of the cell & Microtubule and neurofilament based \cite{yuan17,leterrier17} & Connecting distant locations with brain \& spinal cord & Elongation \& retraction \\
\hline
{\bf Filopodia } & Migrating cells, Fibroblasts, Macrophages, Growth cones of neurons & 10-30 $\mu$m & 10-70 & $<$ 10 mins & Actin based & Probing and sensing the surrounding, tethering  and grabbing & Elongation and retraction\\
\hline
{\bf Stereocilia } & Auditory hair cells & 1-100 $\mu$m \cite{manor08} & 30-300 \cite{rzadzinska04} &  Entire lifetime of the cell \cite{narayanan15} & Actin based \cite{lin05} & Mechanoelectrical transduction \cite{hudspeth00} & Swaying \cite{lin05}\\
\hline
{\bf Kinocilium } & Auditory hair cells \cite{hudspeth00} & 2-3 times longer than stereocilium & 1 & Entire lifetime of the cell & Microtubule based & Supporting the stereocilium and mechnosensing \cite{hudspeth00} & Swaying\cite{lin05}\\
\hline
{\bf Microvilli } & epithelial cells \cite{sauvanet15}, enterocytes \cite{crawley14}, trophoblasts,  oocytes \cite{courjaret16}, lymphocytes \cite{orbach20}. & 0.3-2 $\mu$m & 200  \cite{young} - 3000  \cite{fisher67}  & Few minutes \cite{gorelik03} or entire lifetime of the cell & Actin based \cite{sauvanet15} & Increasing cell surface area for absorption and adhesions, Mechanosensors \cite{sauvanet15} & Immotile, Elongation and retraction \cite{gorelik03}\\
\hline
{\bf Pollen tube} & Pollen grain \cite{obermeyerbook} & Few mm to few feet & 1 & Till it reaches female gamete & Actin based \cite{cai14,chebli13} & Conduit for transferring sperm \cite{obermeyerbook} & Invasive growth \cite{nezhad13} \\
\hline
{\bf Root hairs} & Trichoblast cells \cite{salazar16}  & 80-1500 $\mu$m & 1 & 2-3 weeks & Actin based & Absorption of nutrients and water & Invasive growth \cite{nezhad13} \\
\hline
{\bf Hyphae} & Filamentous fungi \cite{steinberg17,riquelme18} & & & Transforms to septae \cite{steinberg17,riquelme18} & Microtubule based \cite{steinberg17,riquelme18} & Colonization by invasive growth \cite{steinberg17,riquelme18} & Invasive growth \cite{nezhad13} \\
\hline
{\bf Bacterial flagellum} & Bacteria \cite{schuhmacher15} & 5-10 $\mu$m \cite{schuhmacher15} & 1-25 \cite{schuhmacher15} & Multiple cell cycles cite{aizawa98} & Flagellin based \cite{zhuang20b} & Swimming & Beating \\
\hline
{\bf Flagellar hook} & Bacteria \cite{cornelis06}  & 55 nm \cite{cornelis06} & 1 per flagellum &  & Flagellin based & Connecting the cell and the flagellum & \\
\hline
{\bf Fimbrae} & Bacteria \cite{proft18}  & 3 $\mu$m & 1000 & Multiple cell cycles & Pilin based & Adhesion , Aggregation, Resistance to external factors cite{proft18} & Elongation and retraction \cite{burrows05} \\
\hline
{\bf Pili} & Bacteria \cite{proft18}  & 0.5-20 \cite{ramirez20} $\mu$m & 1-4 & Multiple cell cycles & Pilin based & Conjugation & Elongation and retraction \cite{burrows05} \\
\hline
{\bf Injectisome} & Bacteria \cite{cornelis06} & 40-60 nm \cite{cornelis06} & 10-30 \cite{schuhmacher15} & Multiple cell cycles & & Adhesion and smooth passage for transferring materials across the membranes \cite{cornelis06} & Immotile\\
\hline
\caption{{\bf Diversity in long cell protrusions:} }
\label{tab-protrusion-diversity}
\end{longtable*}

\begin{table*}
\begin{tabularx}{\textwidth}{|>{\centering \setlength\hsize{1\hsize}\setlength\linewidth{\hsize}}X|>{\centering \setlength\hsize{1\hsize}\setlength\linewidth{\hsize}}X|>{  \setlength\hsize{1\hsize}\setlength\linewidth{\hsize}}X|}
\hline
{\bf Base} & {\bf Shaft} & {~~~~~~~~~~~~~~~~~~~~~~~~\bf Tip} \\
\hline
\multicolumn{3}{|c|}{\bf Eukaryotic flagellum and cilium}\\
\hline
Basal body and transition zone
\begin{itemize}
\item Assembling the IFT trains
\item A barrier which prevents molecules to diffuse into and out of the protrusion.
\item A gatekeeper which checks and modifies the cargo of the trains before dispatching them into the flagellum.
\end{itemize} &
Flagellum
\begin{itemize}
\item Axoneme, which is an arrangement of microtubule doublets,  forms the core structure of flagellum. 
\item Intraflagellar transport (IFT)  consists of molecular motors which pull the IFT trains that carry building blocks and signalling proteins for assembling, maintaining and disassembling the flagellum.
\end{itemize} &
Flagellar tip
\begin{itemize}
\item Precursors are incorporated at the tip during the assembly.
\item Due to the constant turnover of precursors at the tip, they are returned back to the cell.
\item Houses specialised machineries for sensory purposes.
\end{itemize}\\
\hline
\hline
\multicolumn{3}{|c|}{\bf Axon of a neuron}\\
\hline
Axon initial segment \cite{huang18}
\begin{itemize}
\item Barrier between somatodendritic and axonal compartment. 
\item Sorting the cargoes for anterograde and retrograde transport.
\item Shapes the axon potential.
\end{itemize} &
Axon \cite{encalada14}
\begin{itemize}
\item Axonal transport consists of molecular motors which walk on the parallelly laid microtubules and carry vesicles and signalling proteins and establish connection between the cell body and the synapse. 
\item Parallel networks of neurofilaments provide strength. 
\end{itemize} &
~~~~~~~~~Synapse and axon terminal
\begin{itemize}
\item Identifies the correct target while establishing connection. 
\item Connects the neuron to other neurons of the circuit and to other sensory tissues.
\end{itemize}\\
\hline
\multicolumn{3}{|c|}{\bf Actin based protrusions: Microvilli, Filopodia and Stereocilia }\\
\hline
Actin rootlet
\begin{itemize}
\item Generates protrusive force for the elongation of the protrusion \cite{orly14}. 
\item Depolymerization of the actin at the base \cite{orly14}.
\end{itemize} &
Shaft
\begin{itemize}
\item Actin bundle forms the core structure \cite{svitkina18}.
\item Actin undergoes retrograde flow.
\item Motors walk on the actin filaments towards the tip carrying  precursors and signalling proteins \cite{nambiar10}. 
\end{itemize} &
~~~~~~~~~~~~~~~~~~~~~~~~~~~Tip
\begin{itemize}
\item Actin regulating proteins control  turnover and polymerization \cite{orly14}. 
\item Linkers at the tip for interprotrusion adhesion \cite{lin05, crawley14}.
\item Myosin motor complex for adhesion and force generation \cite{sousa05} 
\end{itemize}\\
\hline
\multicolumn{3}{|c|}{\bf Protrusion with invasive lifestyle : Filamentous fungi, root hairs and pollen tube}\\
\hline
\begin{itemize}
\item No concept of base. 
\item Houses a vacuole which generates turgor pressure.
\end{itemize} &
Sub-apex 
\begin{itemize}
\item Motors walking on the microtubule and actin filaments carry vesicles from the Golgi bodies to the tip \cite{cai14,chebli13,steinberg17}.\\	 
\item Vesicles are recycled by endocytosis at the sub apex region \cite{commer21}. \\
\end{itemize} &
~~~~~~~~~~~~~~~~~~~Tip at the apex
\begin{itemize}
\item Extension of protrusion by exocytosis at the tip \cite{rounds13} by using vesicles from a pool maintained at the tip. (Spitzenk\"orper in fungi  \cite{steinberg17,riquelme18}, clear zone and vesicle supply center \cite{obermeyerbook, pei12}) 
\item Force generation for invasion \cite{nezhad13}. 
\end{itemize}\\
\hline
\hline

\multicolumn{3}{|c|}{\bf Bacterial protrusions I : Flagellum, Flagellar hook and Injectisome}\\

\hline
Secretion system \cite{zhuang20b}
\begin{itemize}
\item Secretion apparatus for unfolding and secreting the precursors and various other virulence factors into the protrusion using ion motive force. 
\end{itemize} &
Flagella / Hook / Needle complex  \cite{zhuang20b}
\begin{itemize}
\item In these conduit like protrusions, the secreted precursors and move towards the tip by pure diffusion.\\	 
\end{itemize} &
~~~~~~~~~Tip  \cite{zhuang20b}
\begin{itemize}
\item Precursors are incorporated at the tip for elongating the protrusion.
\item Special apparatus at the tip for sensing and adhering to the host cell.
\end{itemize}\\
\hline
\hline
\multicolumn{3}{|c|}{\bf Bacterial protrusions II : Fimbrae and Pili}\\

\hline
~~~~~~~~~Base 
\begin{itemize}
\item Secretion apparatus for unfolding and secreting the precursors and other virulence factors into the protrusion using ion motive force \cite{burrows05}. 
\item Elongation and retraction by polymerization and depolymerization at the base \cite{ramirez20}.
\end{itemize} &
Shaft
&
~~~~~~~~~Tip at the apex
\begin{itemize}
\item Special apparatus at the tip for sensing and adhering to the host cell.elongation \cite{proft18}. 
\end{itemize}\\
\hline
\end{tabularx}
\caption{\bf{Universal features of protrusion architecture.} }
\label{tab-unity}
\end{table*}

\subsection{Unity}
Now we point out the features which are common in all the cell protrusions discussed above.
\subsubsection{Universal compartments in the architectural design of cell protrusions}
\begin{figure}
\begin{center}
\includegraphics[width=0.5\textwidth]{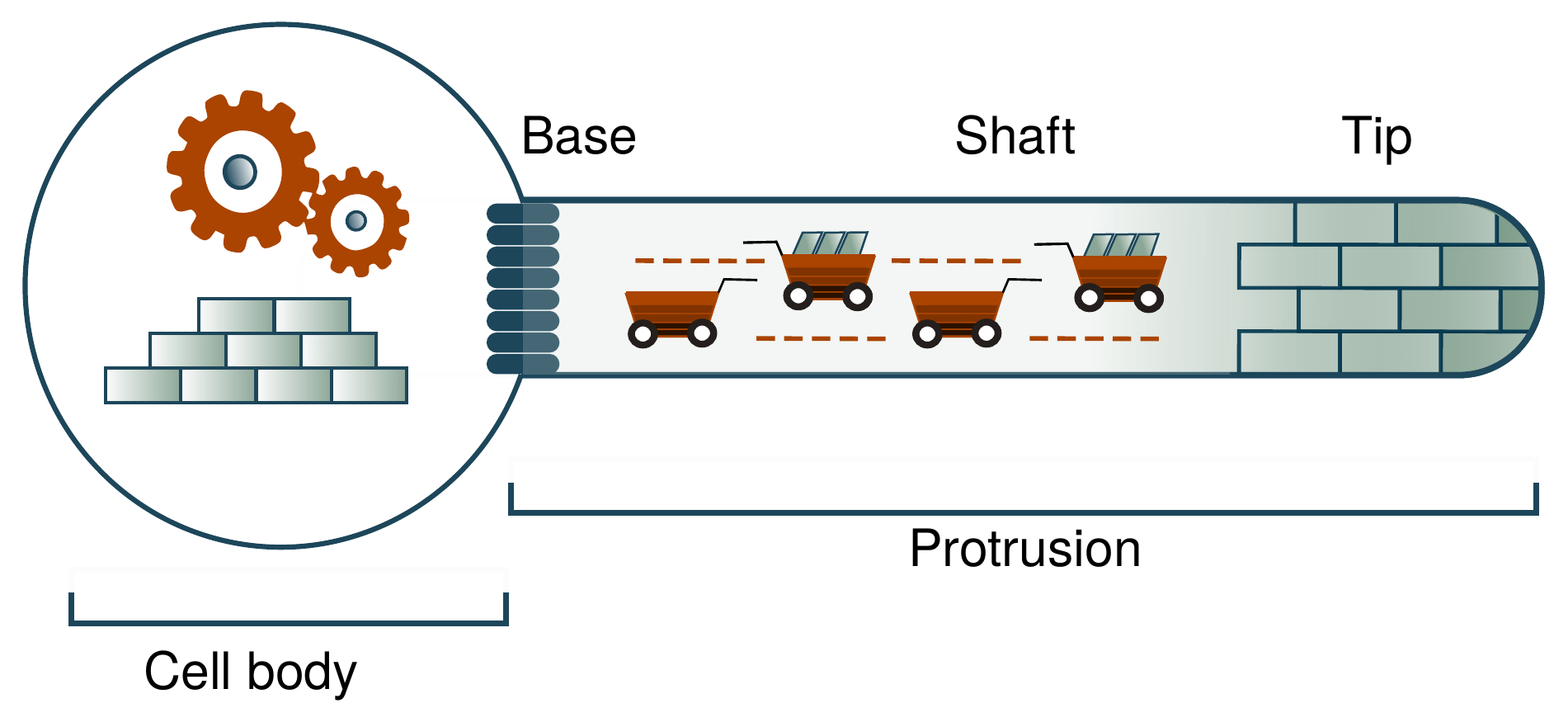}
\end{center}
\caption{ \textbf{Architectural design of a cell protrusion:} Three major compartments of the protrusion are base, shaft and tip. Transport logistics facilitate exchange of materials between the cell body and the protrusion.  } 
\label{fig-protrusions_architecture}
\end{figure}

Cell protrusions share common compartments in their architectural design. The three major compartments which serve as the functional modules of a long  protrusion are its (i) base, (ii) shaft and (iii) tip, as shown in Fig.\ref{fig-protrusions_architecture}. Here we summarize briefly the role of each of these compartments:  \\
(i) {\bf Base:} The base acts as a gatekeeper between the protrusion and the cell body. It ensures that only the proteins specific to the protrusion enter its tubular interior. In some cases, the base modifies these proteins  either chemically or structurally for the benefit of the protrusion. It also acts as a barrier against accidental undesirable crossing of specific proteins across this gate by diffusion.\\
(ii) {\bf Shaft:} The extension between the base and the tip is the shaft. It acts as a conduit for the passage of different proteins between the tip and the base. Smooth transport is ensured by the transport logistics of the system. \\
(iii) {\bf Tip:} The tip houses special proteins and machineries for sensing and probing the surrounding environment, interpreting the guiding cues and for adhering to the host cells. Besides, barring a few exceptions (like, for example, pili), the protrusions elongate by adding precursors at the tip.

\subsubsection{The need for communication between the protrusion and the cell body is universal}


Protrusions and cell body are effectively two functional modules of the cell. Usually protrusions lack the machineries for the synthesis and recycling  of structural proteins. It imports fresh proteins from the cell body and exports those discarded ones back to the cell body  for recycling. There are two common mechanisms which facilitate communication between these two functional modules and manage the exchange of proteins. 

(i) {\bf Intra-protrusion transport system:} In eukaryotic cells, the bidirectional transportation is facilitated by intra-protrusion transport system. Molecular motors are a crucial part of this dynamic arrangement. They walk on the cytoskeleton which form the core structure of the protrusion by consuming fuel (more precisely, hydrolyzing ATP molecules) \cite{howardbook,kolomeisky15,chowdhury13}. In microtubule based protrusions, kinesin and dynein motors are involved in the anterograde and retrograde trips respectively, whereas in the actin based protrusions, different family members of the myosin superfamily of motors run back and forth between the protrusion base and tip. 

(ii) {\bf Secretion system:} The bacterial protrusions lack such network of cytoskeletal based transport. Instead, the bacterial protrusions have machineries at their base for unfolding the proteins and secreting them into the hollow protrusions. These machines are powered by ATP. Inside the narrow conduit, the subunits diffuse to the tip. The bacterial protrusions, with the exception of pili, only import structural proteins and do not transport anything back to the cell. Pili take up structural proteins during the elongation and transport them back during retraction. In addition to the structure building proteins, the machinery secretes also some other proteins and toxins. 

In Table.\ref{tab-unity}, we have summarised the specific roles of these compartments in different long cell protrusions and the features of the intra-protrusion transport responsible for assembling and maintaining  them.  

\begin{figure}
\begin{center}
\includegraphics[width=0.7\textwidth]{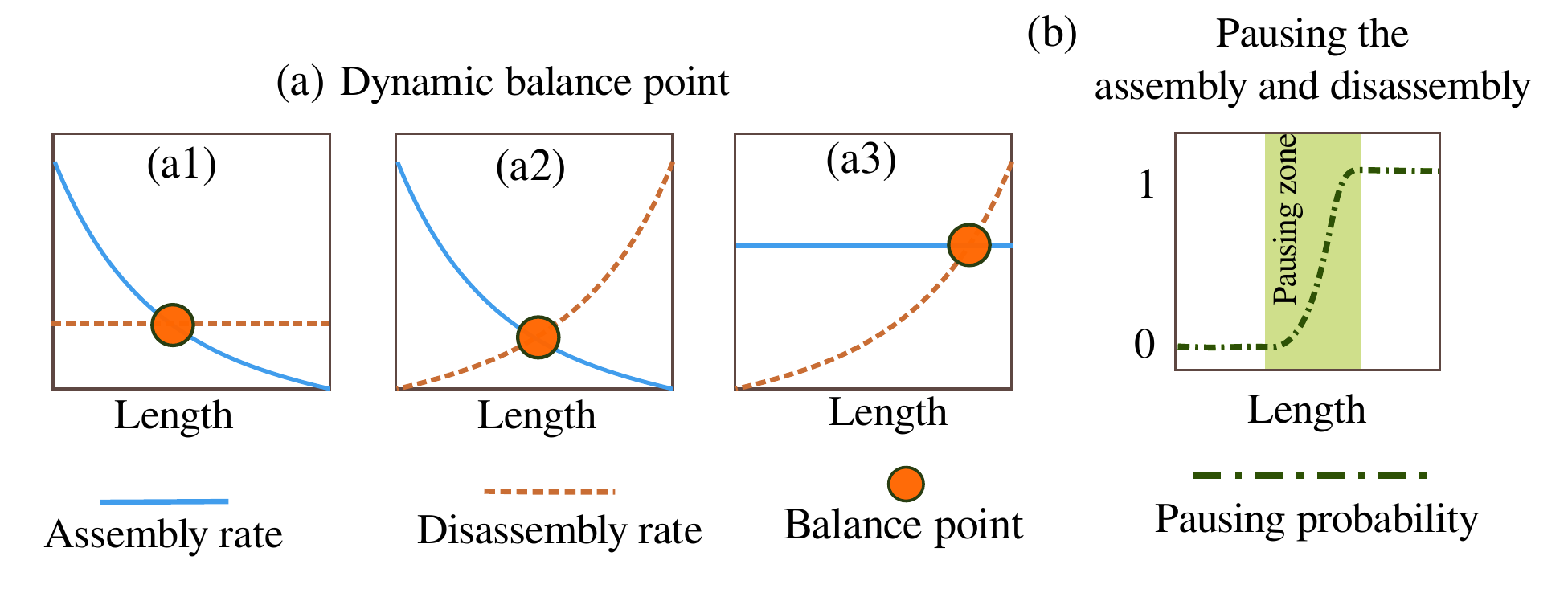}
\end{center}
\caption{ \textbf{Mechanisms of achieving a steady length} (see the text for details).}   
\label{fig-balance_pause}
\end{figure}

\section{Length control in long cell protrusion}
\label{sec-CellProtruLength}

For several different types of protrusions the steady-state length is determined by the distance between the parent cell body and another cell with which the tip of the elongating protrusion binds, thereby linking the two cells and resulting in the stoppage of further growth of the protrusion. The most common example of such protrusions, the axon of a neuron, will be considered only briefly, in part-III of this review, from the perspective of length control. Another class of protrusions can, in principle, continue to grow without ever attaining a steady length; filamentous fungi being the most prominent example of such cell protrusions. This type of cell protrusions will not be discussed further in this review. The general concepts associated with the mechanisms of length control in most of the other types of cell protrusions are reviewed in this section. From now onwards, the term ``precursor'' will be used to refer to the protrusion's structural proteins.

There are two primary modes by which the cells assemble protrusions of a controlled length and prevent them from growing forever. These modes are as follows:\\
(i) {\bf Dynamic balance point:} Some protrusions grow by adding fresh precursors and simultaneously shorten by removing precursors which are discarded due to ongoing turnover. For the protrusion to have a steady length, it is essential that either the assembly rate or the disassembly rate or both of them must be dependent on the protrusion length. As shown in Fig.\ref{fig-balance_pause}(a), a {\it balance point} emerges where the rates of these two opposing processes intersect and the corresponding length is steady state length of the protrusion. For example, eukaryotic flagella of {\it Chlamydomonas reinhardtii} \cite{marshall01} and various actin based protrusions \cite{orly15} belong to this category. \\
(ii) {\bf Pausing zone: } In some protrusions, the assembly and disassembly rates may or may not be length independent. But the protrusion ceases to grow or shorten  up on attaining a certain length. The pausing probability varies with protrusion length and displays a switch like behaviour as indicated in Fig.\ref{fig-balance_pause}(b). The zone of length over which the pausing probability switches its value is the {\it pausing zone} and the static steady length of the protrusion falls in this zone. Eukaryotic flagella of {\it Trypanosome brucei } \cite{bertiaux18} and bacterial flagellar hooks \cite{makishima01} are examples of such protrusions.\\

The primary difference between these two modes is that, in the first case the instantaneous length is dynamic even when the mean length attains a steady value but in the second case, both the mean and instantaneous length remain static in the steady state. In order to modulate the assembly and disassembly rates or the pausing probabilities, the cell must get continuous feedback about its  instantaneous length. Such feedback requires  special mechanisms for sensing the length. Controlled length also emerges because of the collective interaction among the multiple components inside the protrusion. 

In Table.\ref{tab-len_sensing}, we list the general mechanisms  which are used by the cell for sensing the length and assembling a protrusion of controlled length in the steady state. 
Different mechanisms are not necessarily completely distinct in the sense that they may have partial overlap of the nature of underlying processes.

\begin{longtable*} { |>{\centering\arraybackslash}m{0.7cm} | >{ \arraybackslash}m{3cm} |>{\arraybackslash}m{14cm} |}
\hline 
 & {\bf  Mechanisms for length sensing and length control} & {\bf ~~~~~~~~~~~~~~~ Description and comments} \\
\hline
1 & \parbox[r]{1em}{\includegraphics[width=1in]{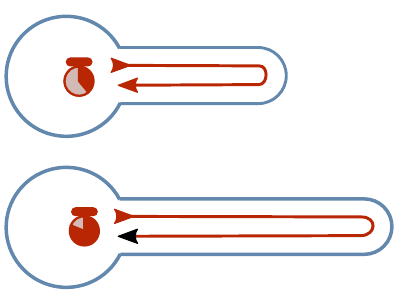}} & {\bf Time of flight based sensing:} Dynamic elements like molecular motors or cargo carried by these motors keep moving  between the base and the tip of the protrusion. As the protrusion elongates, the time of flight $T_\text{tof}$ from one end to the other \cite{folz19,rishal12,bressloff15} or the time of  one complete round\cite{lefebvre86,ludington15,patra20a,ishikawa17} is proportional to the length of the protrusion $L$ and inversely proportional to the velocity $v$ of these shuttling elements i.e $T_\text{tof} \propto \frac{L}{v}$. The concept of measuring the distance between these two points using the time of flight of sound waves was first proposed by Galileo.  
{\bf Protrusions:} Axon \cite{folz19,rishal12,bressloff15}, Eukaryotic flagellum (For: \cite{lefebvre86,ludington15,patra20a} Against: \cite{ishikawa17}).\\
\hline 
\\
\hline 
2 & \parbox[r]{1em}{\includegraphics[width=1in]{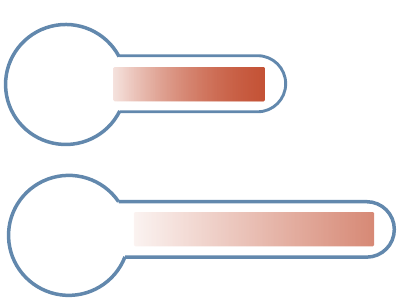}} & {\bf Gradient based sensing: } Length can be sensed and controlled by  gradients of   proteins \cite{toriyama10}, motors \cite{hendel18,chien17,fai19}  and precursors\cite{renault17}. Certain proteins are carried to the tip using active transport and they tend to diffuse back towards the base from the tip \cite{toriyama10, hendel18, chien17,fai19}. This sets a tip-to-base gradient. In certain protrusions the precursors diffuse to the tip and this sets a  base-to-tip gradient \cite{renault17,chen17}. Such gradients can control length by modulating the assembly and disassembly rates of the protrusion. The concentration of these elements at one of the ends is given by 
$C \propto \exp \left( -L(t)/\lambda \right)$ where $\lambda$ is the length scale of the gradient. {\bf Protrusions: } Neurites \cite{toriyama10}, Eukaryotic flagellum \cite{hendel18,chien17,fai19} , Bacterial flagellum \cite{renault17,chen17}.  \\
\hline
\\
\hline
3 & \parbox[r]{1em}{\includegraphics[width=1in]{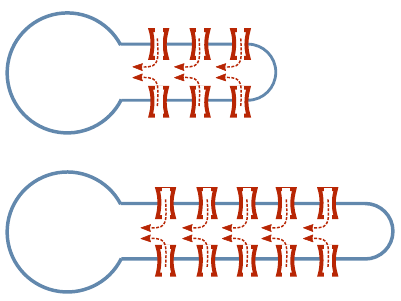}} & {\bf Ion current based sensing: } The number of ion channels present on the protrusion membrane is directly proportional to length and so is the total ion influx \cite{ludington15}. Hence, the amount of current received at the base is used as a length sensor and the current directly impacts the processes responsible for the elongation of the protrusion.  {\bf Protrusion:} (i) Eukaryotic flagellum \cite{johnson93,liang18, liang14, ludington15,besschetnova10} (ii) In stereocilia mechnotransduction current is essential for assembling and maintaining stereocilia of correct length \cite{hudspeth00}. (iii) In axon calcium current plays certain role in measuring axon length \cite{rishal19}. \\
\hline
\\
\hline
4 & \parbox[r]{1em}{\includegraphics[width=1in]{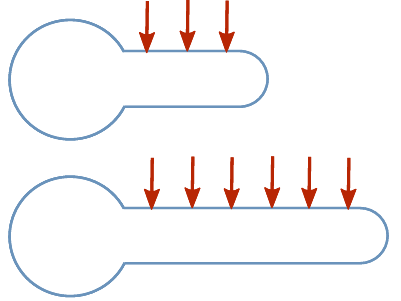}} & 
{\bf Shear force based sensing: } Certain protrusions are constantly subjected to mechanical shearing due to  the surrounding fluid. The shear stress is proportional to the length and the amount of stress the protrusion is subjected to could give an estimate of the length.\\
\hline
\\
\hline
5 & \parbox[r]{1em}{\includegraphics[width=1in]{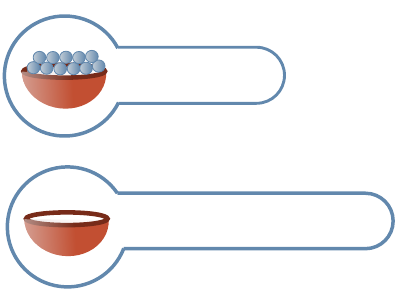}} & 
{\bf Limited components in pool: } In most protrusions multiple components collectively interact to  assemble the protrusion. If at least one of the structural  components is available in a limited quantity and is not replenished, continued depletion of the initial pool of that component by the protrusion growth can eventually stop its further growth after it attains a certain length. {\bf Protrusion:} (i) The stoichiometric amount of the base pilin controls the length of the heterotrimeric pilus.{\cite{mandlik08}}. \\
\hline
\\
\hline
6 & \parbox[r]{1em}{\includegraphics[width=1in]{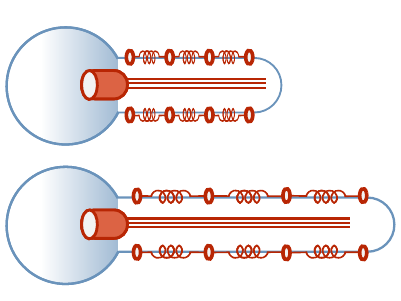}} & {\bf Membrane-cytoskeleton force balance: }  Length of the protrusion attains a steady value when the protrusive force (generated by the polymerization of cytoskeletal filaments) is balanced by the opposing restoring force (arising from membrane elasticity ).
 {\bf Protrusion:}  Actin based protrusion like stereocilia, microvili and filopodia \cite{orly14,orly15,gov06}.  \\
\hline
\\
\hline
7 & \parbox[r]{1em}{\includegraphics[width=1in]{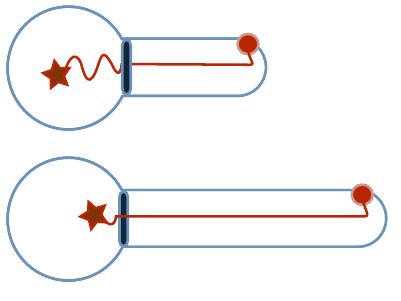}} & 
{\bf Measuring tape: } One end of a special protein is attached to the growing tip of the protrusion whereas the other end of the protrusion hangs loosely inside the cytoplasm. As the protrusion elongates, the chances of the loose end interacting with the base increases. This interaction is sufficient to stop the export of precursors for the further elongation of the protrusion.{\bf Protrusion:} Needle of Type-3 secretion system \cite{bergeron16}. \\ 
\hline
\\
\hline
8 & \parbox[r]{1em}{\includegraphics[width=1in]{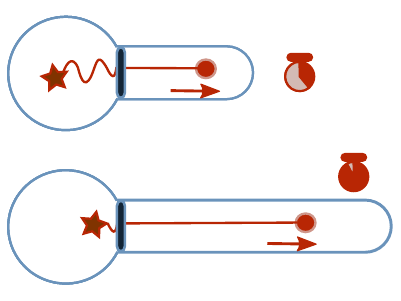}} & {\bf Secretion of molecular ruler: }  This is slightly different from the measuring tape mechanism in the implementation of the details. A ruler protein is secreted into the protrusion from time to time. One end of the protein slowly navigates the protrusion while the other end loosely hangs at the base.  As soon as the navigating end of the ruler protein reaches the  protrusion tip, the whole ruler protein is quickly released out of the protrusion.
 Longer protrusion means longer passage time and this increases the probability of the other loosely hanging end to interact with the base compartment and thereby stop further supply of precursors for elongation. {\bf Protrusion:} Hook of bacterial flagella\cite{erhardt11,wee15}. \\
\hline
\\
\hline
9 & \parbox[r]{1em}{\includegraphics[width=1in]{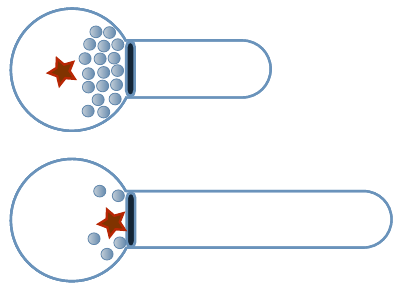}} & 
{\bf  Waiting room mechanism:} Some cells accumulate a pool of  structural components of the protrusion in a region (`waiting room') around a complex that would then serve as the base of the protrusion when it begins to grow. Such crowding at the base blocks the access of other kinds of proteins which could potentially stop the export of precursors from the base. However, as the pool is gradually depleted by the elongation of the protrusion, these blocking-capable proteins find access to the protrusion base and eventually stop the further supply of precursors. {\bf Protrusion:} Hook of bacterial flagella \cite{makishima01}   \\ 
\hline
10 & \parbox[r]{1em}{\includegraphics[width=1in]{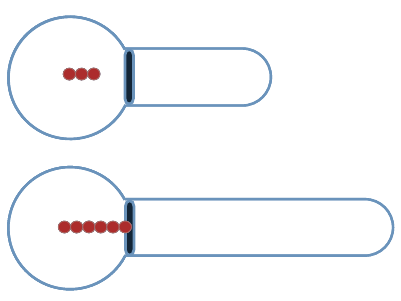}} & {\bf Coassembly of protrusion and  a subcellular basal structure : } A specific  sub-structure of the base and the protrusion can assemble simultaneously. Upon completion of the assembly of the former, the new basal sub-structure blocks the  supply of the components required for further elongation of the former, thereby deciding the final length of the protrusion. {\bf Protrusion:} Needle complex of the bacterial injectisome \cite{nariya16,lefebre14}. \\
\hline
\\
\hline
11 & \parbox[r]{1em}{\includegraphics[width=1in]{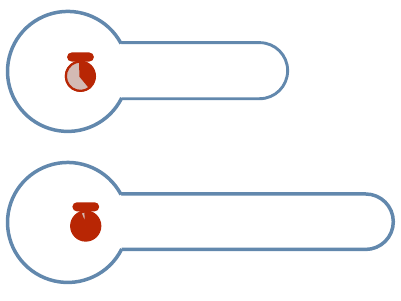}} & {\bf Time-keeper:} The cell sets aside a specific time interval for protrusion assembly. During this finite duration of time,  a protrusion can be assembled. As soon as this interval ends, the cell stops further changes of the length of the protrusion. Similar mechanism is proposed in the context of bacterial size control.  {\bf Protrusion:} Eukaryotic flagellum in {\it Trypanosome} \cite{bertiaux18}.\\
\hline
\\
\hline
12 & \parbox[r]{1em}{\includegraphics[width=1in]{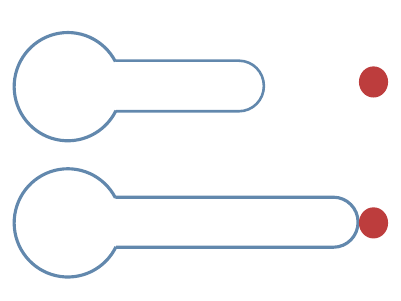}} & {\bf Reaching the target:} Certain protrusions just keep elongating  till the growing tip hits an external target. Hence, it is the distance between the external target and the  cell bearing the protrusion which govern the length of the protrusion.  {\bf Protrusion:} (i) Pollen tube is an example of such protrusion which doesn't cease growing till the tip establishes contact with the ovary for the delivery of sperm \cite{mizuta18}. (ii) Cytonemes and tunnelling nanotubes keep growing till they hit the target cell with which the host cell has to establish connection with \cite{yamashita18}. (iv) Axon \cite {luo05}. \\
\hline
\\
\hline
13 &  & {\bf Protrusions with no controlled length:} Certain protrusions, like filamentous fungi, keep growing if sufficient nutrition is available. {\bf Protrusion:} Filamentous fungi \cite{riquelme18,steinberg17}\\
\hline
\hline
\caption{\bf  Mechanisms of sensing and control of length of cell protrusions.} 
  \label{tab-len_sensing}
\end{longtable*}

\section{Length fluctuations of a single protrusion}
\label{sec-length_fluctuations}
\subsection{Mapping onto special types of stochastic processes}
\begin{figure*}
\includegraphics[width=0.9\textwidth]{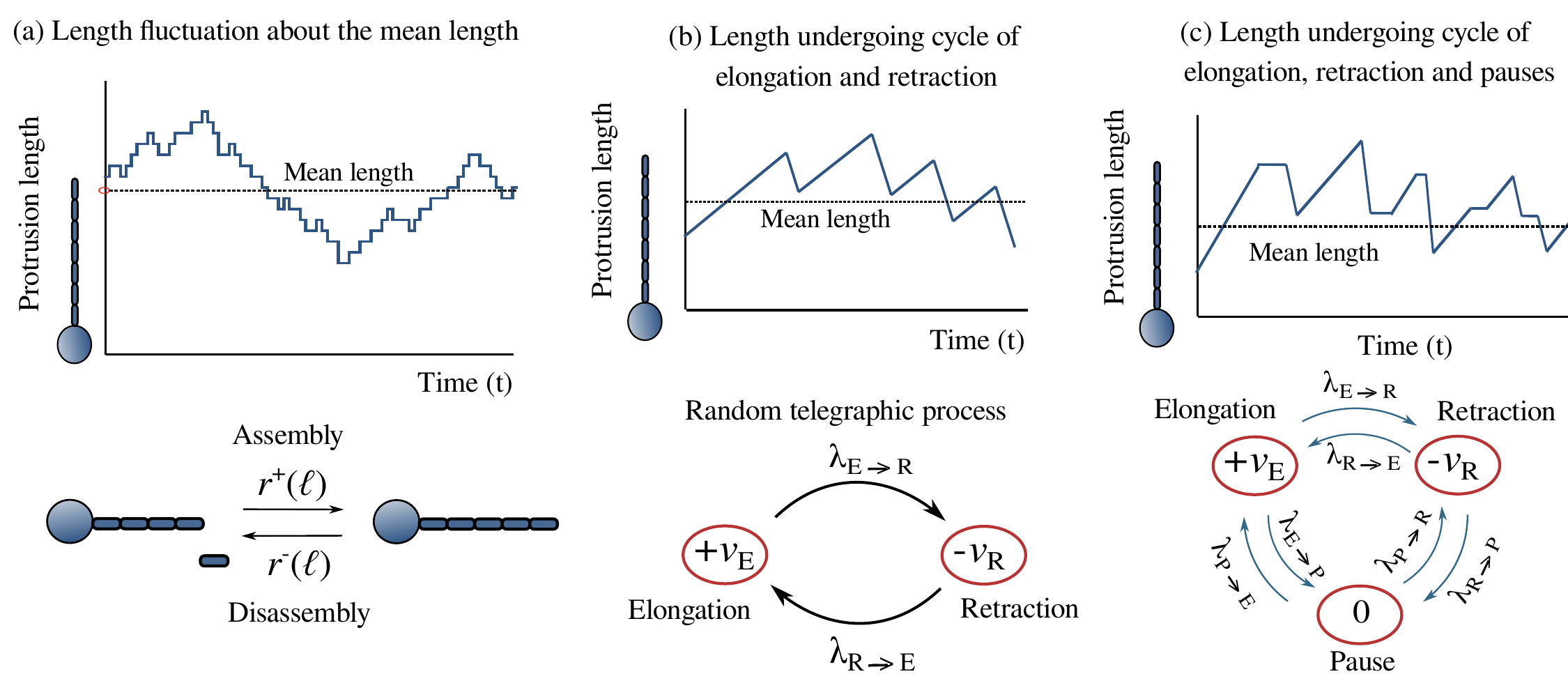}
\caption{{\bf Different trends of length fluctuations}:  (a) The length of  protrusions  fluctuate about the mean value because of ongoing incorporation of precursors and turnover of discarded components at the tip with length dependent rates  $r^+(\ell)$ and $r^-(\ell)$, respectively. (b) Certain protrusions keep elongating and retracting with velocities $v_E$ and $v_R$, respectively and the protrusion switches between  a strictly elongating and a strictly retracting phases  with the corresponding rates $\lambda_{E \to R}$ and $\lambda_{R \to E}$, respectively. \cite{kolesnik13} (c) Certain protrusions switch between an elongating and a retracting phase with in between pauses. } 
\label{len-fluc}
\end{figure*}

 Even at the balance point the protrusion length does not remain constant; it fluctuates around the steady mean value. In Fig.\ref{len-fluc}(a-b) we have shown schematically the typical temporal evolution of the instantaneous length for the two different cases. As elongation and shortening events are random, the protrusion length evolution can be treated as a stochastic process. Here we summarize how to deal with the length fluctuations with special classes of stochastic processes. \\
(i) {\bf Mapping onto Ornstein-Uhlenbeck process:} The length of certain  protrusions like eukaryotic flagellum \cite{marshall01} and stereocilia \cite{narayanan15}  fluctuate because of ongoing incorporation of precursors and turnover of discarded components at the tip even after the mean length attains a steady value.  In the generic case, the length evolution can be studied using a master equation with length dependent  rates  $r^+(\ell)$ and $r^-(\ell)$ of protrusion assembly  and disassembly, respectively. For such systems, the length fluctuations can be mapped onto an Ornstein-Uhlenbeck (OU) process (see Fig.\ref{len-fluc}(a)). A detailed study on length fluctuations of such protrusions was reported in ref. \cite{patra20b}.\\
(ii) {\bf  Mapping onto Telegraphic process:} Certain protrusions, like filopodia and neurites \cite{winans16}, are highly dynamic  because of the intrinsic dynamics of the constituent filaments like, for example, dynamic instability of microtubules. They keep elongating and retracting with velocities $v_E$ and $v_R$, resepctively.   In case the protrusion switches between  a strictly elongating and a strictly retracting phases (see Fig.\ref{len-fluc}(b)) with the corresponding rates $\lambda_{E \to R}$ and $\lambda_{R \to E}$, respectively, the length evolution could be best described as an asymmetric telegraphic process \cite{kolesnik13}. \\
(iii) {\bf Mapping onto a three state markov process:} Pilus, which is a  bacterial protrusion, keeps elongating and retracting with velocities $v_E$ and $v_R$ with in between pauses (see Fig.\ref{len-fluc}(c)) \cite{koch21}. The length dynamics can be described by a three state Markov process as shown in Fig.\ref{len-fluc}(c).


\begin{figure}
\includegraphics[width=1.0\textwidth]{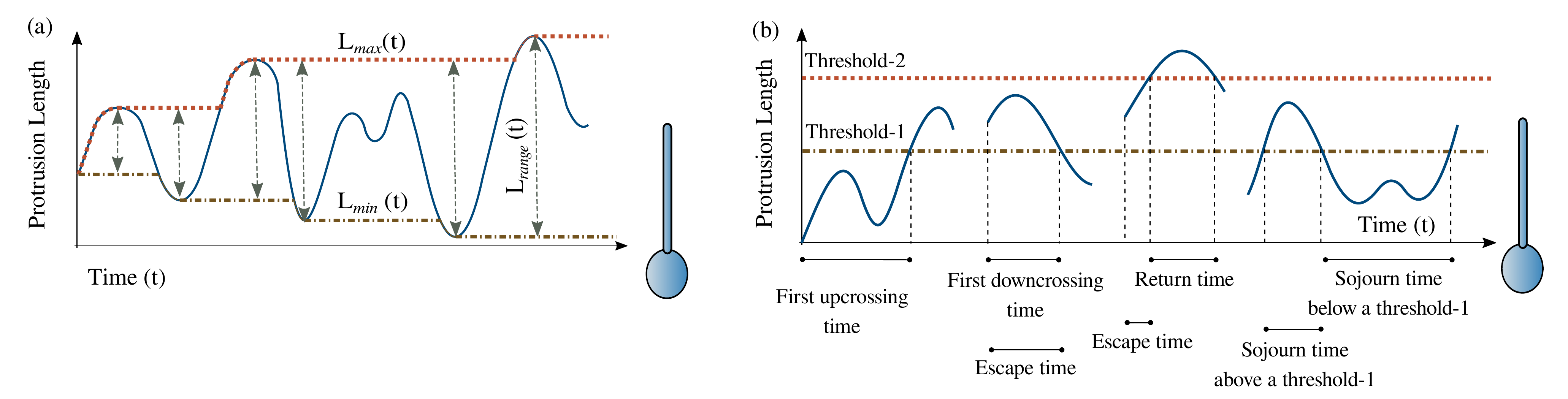}
\caption{Level crossing quantities for characterizing length fluctuations: (a) Characteristic length and (b) Characteristic time.}
\label{len-fluc-len-time}
\end{figure}

\subsection{Level crossing statistics}
Irrespective of the nature of trajectory (Fig.\ref{len-fluc}(a-b)) that the protrusion length follows, there are some common questions regarding the temporal fluctuations of the lengths of the protrusions. Protrusions whose length elongate and retract as shown in Fig.\ref{len-fluc}(b), are mostly involved in scanning and probing the environment for guiding cues. So, the maximum and the minimum length it can grow or shorten to in a finite duration and the range it can scan during its lifetime are important  characteristic lengths. Knowing these lengths are also important for those protrusions whose length fluctuate about their steady state mean value (Fig.\ref{len-fluc}(a-b)) because, for optimal performance, the length must lie between a narrow zone bounded by an upper threshold and a lower threshold. Hence, knowing the statistics of extreme length fluctuations is important for understanding how the cell responds to such extreme events (Fig.\ref{len-fluc-len-time}(a)). In addition to the important characteristic lengths, having the estimates of various  characteristic times is equally  important for a complete stochastic description of the length fluctuations. For example, the first upcrossing and downcrossing time to hit a particular threshold, the exit time to move out from a zone bounded by two thresholds, the sojourn time above and below a threshold are some of the important timescales (Fig.\ref{len-fluc-len-time}(b)). Note that   all the characteristic lengths and times shown in Fig.\ref{len-fluc-len-time}(a-b) are random variables. Their complete description  requires either their distribution or all the moments of each of those distributions. Interested readers are referred to the works of Syski \cite{syski92}, Rice \cite{rice44,rice45}, Stratonovich \cite{stratonovic81} and Masoliver \cite{masoliver14} where various techniques of stochastic processes are presented which can be applied to length fluctuation of long cell protrusions. These statistical quantities that characterize the random level crossing process  have been estimated recently for eukaryotic flagella. The techniques summarised there are applicable to all such protrusion whose length fluctuations are described by Ornstein-Uhlenbeck process\cite{patra20b,bauer20}.

\section{Protrusion loss and regeneration}
\subsection{Causes of protrusion loss { and cost-benefit analysis}}
\begin{figure}
\includegraphics[width=0.8\textwidth]{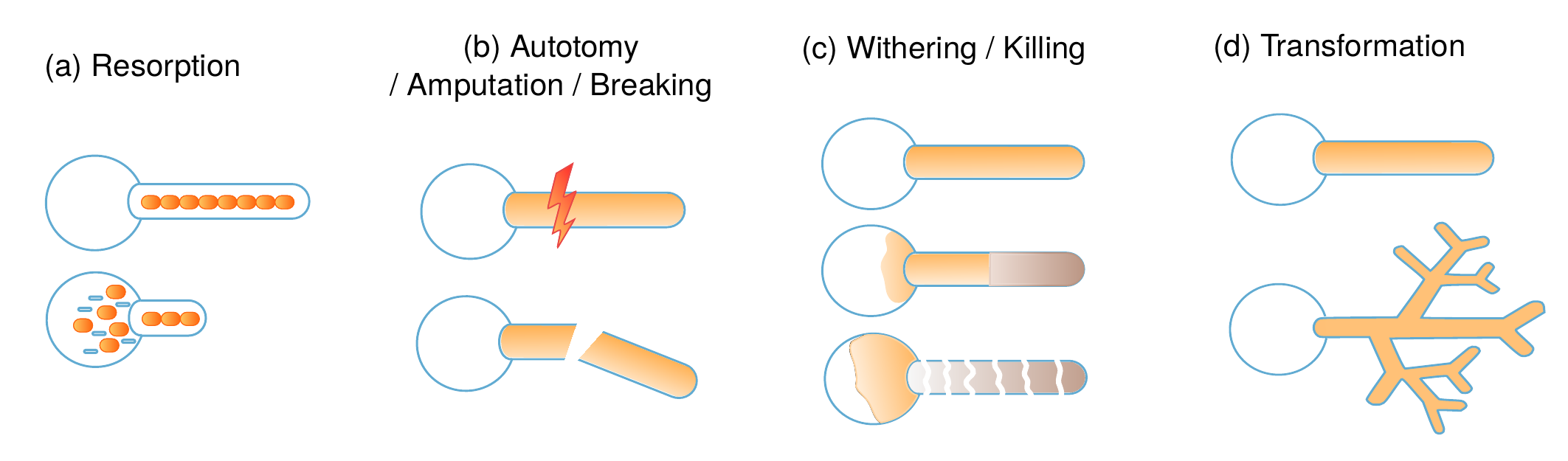}
\caption{{\bf Modes of protrusion loss}:  (a) resorption (b) autotomy or amputation (c) withering and (d) transformation. }.
\label{fig-protrusion_loss}
\end{figure}

A cell can lose its protrusion voluntarily or involuntarily by resorption, autotomy, amputation or eliminate it by withering it. { Once a protrusion-like appendage is lost, proper biological functions of the cell are affected in the absence of the services of the missing structure (e.g. decreased locomotory function or sensory perception)}. Here we look at these processes one by one.\\
(i) {\bf Resorption }: A cell can resorb its protrusions voluntarily as shown in Fig.\ref{fig-protrusion_loss}(a). Protrusions involved in surveying their surroundings undergo a cycle of elongation and resorption;  resorption is a part of the lifestyle and a functional necessity for  the cell (example: neurites \cite{winans16}). Unlike these dynamic protrusions which undergo retraction multiple times, some other protrusions get permanently resorbed during a certain phase of the cell cycle and the  daughter cells acquire  new copies of the protrusion after cell division (example: eukaryotic flagella \cite{quarmby05}). Protrusions sometimes act as reservoirs of membrane and cytoskeletal monomers and hence the cell resorbs such protrusions whenever it needs the supply from these reservoirs (example: microvilli \cite{figard16}). In order to  survive sudden environmental stress or fluctuations in environmental factors like pH or temperature,  a cell can retract a protrusion into the cell body just like a tortoise or a snail can pull back their head and neck inside their shell. Hence  resorption is a functional necessity and performed often voluntarily by the cell. Resorption could  also be initiated by stopping the supply of precursors, or by exporting depolymerase to the tip or by capping the tip and preventing its elongation. A key point is that the structural components of a resorbing protrusion are not lost by the cell. During this process all the components of the protrusion are collected in the cell body and these can be reutilised while regenerating the protrusion.\\
(ii) {\bf Autotomy}: Another mode of losing the protrusion voluntarily is by autotomy where the cell sheds its protrusion by severing it from the base as shown in Fig.\ref{fig-protrusion_loss}(b). The most commonly recognized cause for autotomy in natural populations is to escape from the entrapment of the predator. Adverse environmental factors like unfavourable pH leads to deflagellation in {\it Chlamydomonas reinhardtii} \cite{quarmby04}. Deciliation in ciliated microorganisms is crucial for the progression of cell cycle \cite{gogendeau20}. During a famine like situation when  there is dearth of nutrition, shedding the protrusion reduces the metabolic cost needed for maintaining the protrusion and increases the chances of survival of the cell. For example, bacteria autotomize their flagella in medium lacking nutrients \cite{zhuang20,zhu20}.  Autotomy  can increase the chance of survival by facilitating escape, for example by enabling the individual to avoid entrapment. But , in this process, unlike resorption, the cell suffers loss of materials that constitute the protrusion. Nevertheless, autotomy also provides a predictable wound plane and can minimize fluid loss and cell damage, thus reducing the cost of injury. \\
(iii) {\bf Amputation and breaking off}: Sublethal predation could lead to the amputation of a protrusion. Amputation of protrusions can also be performed  in controlled experiments by shining a laser beam which results in a precise cut (eukaryotic flagella \cite{ludington12}, axon\cite{ruth08}, bacterial flagella \cite{paradis17}). Crushing the protrusion is an outdated technique now \cite{ruth08}.  Moreover, those protrusions which are subjected to sustained shear stress from the circulating fluid medium, can loose a part of the protrusion due to breaking \cite{paradis17}.  Structure loss can lead to local cell damage and loss of body constituents. In some situations, injury costs can be almost trivial whereas others can be life threatening for the cell. \\
(iv) {\bf Slow death}: In certain cases, the cell kills the protrusions by either cutting off the supply of materials essentials for maintaining  the protrusion (example: {axon \cite{luo05}}) or simply by  sucking the cytoplasm and/or  other essential components back into the base (example: root hairs \cite{distefano17}) thereby starving the protrusion to death. In these cases, the protrusion slowly withers and then falls off from the cell as depicted schematically in Fig.\ref{fig-protrusion_loss}(c).\\
(v) {\bf Transformation:} Another way of losing a protrusion with a specific structure and function is by transforming its structure or/and function as shown schematically in Fig.\ref{fig-protrusion_loss}(d). Certain transient protrusions like neurites transform into more permanent structures like dendrites and maintain this new identity throughout their lifetime \cite{dotti88}. In certain cases, filopodia  get converted to cytoskeletal bridges \cite{yamashita18}.

\subsection{Regeneration of protrusion and cost-benefit analysis}

Before the process of regeneration of a cell protrusion can begin, the length-sensing mechanism must give feedback to the cell body about the change in the length of the  affected protrusion. The cell is expected to transport the materials required for the regeneration at the appropriate location at the appropriate rate. If the precursor pool at the base is inadequate to supply the material, the cell needs to upregulate the synthesis of these components unless such synthesis is blocked by other intracellular signals or suppressors. In cells where the fresh synthesis of precursors is inhibited, the cell may not be able to regenerate the protrusion to its full original length \cite{rosenbaum69}.  Amputation may cause some local damage at the tip of the surviving segment of a protrusion because of which incorporation of fresh subunits at the damaged tip may be difficult or impossible. In such cases, in spite of sensing the amputation and even if enough supply of precursors may be available, regeneration of the amputated appendage may not take place  \cite{paradis17}.

Let us now carry out a cost-benefit analysis of regeneration. \\
(i) {\bf Energy and material allocation cost:} regeneration can be energetically expensive and the energetic burden of regeneration can adversely affect other body functions. These effects are exacerbated if structure loss results in a significant loss of energy stores (e.g. ATP) or reduces the ability to provide fresh supply of energy (e.g., loss of mitochondria) or that of materials (e.g., loss of ribosomes).\\
(ii) {\bf  Operational cost because of low fidelity:} If the regenerated structure is an imperfect copy of the original,  for example shorter or longer than the original, it may not perform its biological function fully satisfactorily. In such situations, the cell must bear a permanent or recurring operational cost arising from the low fidelity of the regenerated appendage. \\
(iii) {\bf Benefit of structure replacement:} Unless  the operational cost arising from low fidelity of regenerated protrusion is very high, the benefit of regeneration outweighs the cost of operation without the service of the protrusion. The lifetime benefit of structure replacement also depends on the age of the cell at the time of suffering the injury and its expected longevity.

\subsection{Condition for regeneration and the nature of regenerate}

The most important conditions for regeneration after autotomy, amputation or breaking are the following: \\
(i) {\bf Estimating the extent of loss }: The cell must sense that it has suffered loss. After suffering a partial loss, the primary goal of the cell is to estimate the extent of loss. In case of long protrusions with linear geometry, the cell has to know the current length of the intact protrusion, so that while rebuilding the lost part it has to supply precursors according to the corresponding needs. Only certain length sensing mechanisms may be successful in measuring the length of the shorter protrusion and precisely add the rest of the part so that the protrusion regenerates back to its original length. Some protrusions attain stable length because no more precursors are supplied for further elongation due to permanent conformational or biochemical change of the base. Regenerating those protrusions whose length control solely depends on the base compartment and does not involve taking feedback of length will find it difficult to regenerate.\\
(ii) {\bf Ability to resynthesize the precursors}: In cells where the fresh synthesis of precursors is inhibited, the cell may or may not be able to regenerate the full protrusion \cite{rosenbaum69}.\\
(iii) {\bf Ability of protrusion to grow}: Amputation of certain protrusions may cause such local deformation of the exposed tip that prohibits incorporation of fresh precursor there. In such cases, regeneration is impossible, even if the cell has the ability to estimate the extent of the loss and has capacity to supply fresh precursors \cite{paradis17}.

Loss and regeneration of various protrusions are summarised in a tabular form in Table.\ref{tab-loss_regen}.

\begin{longtable*} 
{ |>{\centering\arraybackslash}m{1cm}|>{\centering\arraybackslash}m{8.5cm} | >{ \arraybackslash}m{8cm} |}
\hline 
{\bf Sl.no} & {\bf  Mode and reason of losing protrusion} & {\bf Condition and mechanism of regeneration} \\
\hline
\multicolumn{3}{|c|}{\bf Eukaryotic flagellum and cilium}\\
\hline
1.1 & Complete resorption prior to cell division \cite{rosenbaum69}. & Regeneration by ciliogenesis in the next cell cycle \cite{rosenbaum69}.\\
\hline
1.2 & Partial resorption of the intact flagellum during the elongation of the other flagellum \cite{rosenbaum69,he19,ludington12}. & Regeneration after the length equalization of  the shortening and the elongating flagellum  \cite{rosenbaum69,he19,ludington12}.\\
\hline
1.3 & Shedding the complete flagellum by deflagellation for escaping
the predator or in absence of nutrients \cite{rosenbaum69,quarmby04}. & Regeneration subjected to the availability of precursors in the pool \cite{rosenbaum69,quarmby04}.\\
\hline
1.4 & Selective amputation of one of the flagellum of a biflagellate \cite{rosenbaum69,ludington12}. & Regeneration of the amputated flagellum partially at the cost
of shortening of the intact flagellum and the synthesis of new
precursors \cite{rosenbaum69,ludington12}.\\
\hline 
\multicolumn{3}{|c|}{\bf Neurites and axon of a neuron}\\
\hline
2.1 & Neurites transform into dendrites and axon \cite{dotti88}. & \\
\hline
2.2 & Amputating the axon of a nascent neuron whose neurites have
not transformed into dendrites by crushing \cite{dotti88,toriyama10}. & The longest protrusion (among all the neurites and the amputated axon), and rarely the second longest protrusion, converts into axon \cite{dotti88,toriyama10}.\\
\hline
2.3 & Amputation of the axon of a mature neuron embedded in a
network \cite{ruth08}. & Conditional regeneration: If the length of the amputated axon exceeds a critical length, axon retains its identity and
regenerates. If the length of the amputated axon is shorter than one of the dendrites, the latter 
converts to axon and the former converts to a stump \cite{ruth08}.\\
\hline
2.4 & Elimination of axon which overshoots the target and the axon
collaterals which do not reach target or make the correct
connections \cite{luo05}. & \\
\hline
\multicolumn{3}{|c|}{\bf Filopodia}\\
\hline
3.1 & The filopodium retracts by collapsing suddenly. & After retraction, it again regenerates and achieves a stable length.\\
\hline

\multicolumn{3}{|c|}{\bf Stereocilia}\\
\hline

4.1 & Partial retraction of stereocilium on reduction of  mechanotransducer current  \cite{ortega17,hudspeth00}. & Regeneration on the restoration of  the current \cite{ortega17}. \\
\hline
4.2 & Breaking off or uprooting of stereocilia by ultrasonic sounds \cite{knalltrauma}. & Fail to regenerate\cite{knalltrauma}.\\
\hline
\multicolumn{3}{|c|}{\bf Microvilli}\\
\hline
5.1 & Microvilli are reservoirs of membrane. They retract to supply to membrane \cite{figard16}. & Microvilli regenerate by reutilising the  actin monomers accumulated  during retraction \cite{figard16}. \\
\hline
\multicolumn{3}{|c|}{\bf Root hairs}\\
\hline
6.1 & Root hair retraction in nutrient deficiency \cite{hogg11}. & 
Cell death due to cytoplasm retraction of condensation \cite{hogg11}.  \\
\hline
\multicolumn{3}{|c|}{\bf Filamentous fungi}\\
\hline
7.1 & Fungi are preys to a variety of animal predators, including
fungivorous nematodes and insects which cause injury \cite{riquelme18}. & Respond to injury by sealing the pore to prevent the loss of
cytoplasmic content. It is followed by the extension of thinner
filamentous hyphae for a brief period and formation of
fruiting body at these sites of injury \cite{riquelme18}.\\
\hline
\multicolumn{3}{|c|}{\bf Pollen tube}\\
\hline
8.1 & Premature bursting of pollen tube tip before reaching the female gamete \cite{li18}. & \\
\hline
\multicolumn{3}{|c|}{\bf Bacterial flagella}\\
\hline
9.1 & Breaking of flagellar filament by mechanical shear while
swimming \cite{paradis17}. & Regrowth with a length dependent rate which is facilitated by a formation of cap like structure at the broken tip \cite{paradis17}.\\
\hline
9.2 & Amputation of filament by laser pulse which cause
local damage to the flagellum \cite{paradis17}. & No regrowth \cite{paradis17}.\\
\hline
9.3 & Flagellar ejection induced by nutrient starvation \cite{ferreira19,zhu20,zhuang20}. & Regrowth later upon  availability of sufficient nutrients \cite{ferreira19,zhu20,zhuang20}.\\
\hline
\multicolumn{3}{|c|}{\bf Pili and fimbria}\\
\hline
10.1 & Retraction by disassembling monomers from base for enabling DNA and phage uptake, twitching motility, pulling the bacteria towards sites of
adhesion and making intimate contact with the host cell surface, for coaggregation and colonization \cite{burrows05}. & Regeneration by reutilising the monomers accumulated at the base during retraction \cite{burrows05}.\\
\hline
\caption{Loss and regeneration of long cell protrusions}
\label{tab-loss_regen}
\end{longtable*}

\section{Length control and coordination in cells bearing multiple copies of protrusions}

\subsection{Challenges related to the length control of multi-protrusion cell}
From the perspective of length control, there are numerous interesting questions in the context of cells bearing multiple copies of the the same protrusion. 
\subsubsection{Protrusions with different steady length}
Certain cells bear multiple protrusions of same type that, however, have different lengths in the steady state. For understanding how the cell establishes and maintains the different lengths, it is necessary to experimentally determine (a) the relative distribution of cystoskeletal components and other length controlling proteins (like depolymerase, cappping proteins, etc) in different protrusions, and (b) the role of collective transport in establishing and maintaining those relative distributions.  A classic example is  {\it Giardia} each of which bears four pairs of flagella where different pairs have different lengths while the two members of each pair have approximately equal length. It has been shown experimentally that the different amount of depolymerases at the tips of different pairs  leads to the length difference \cite{mcinally19}.  

\subsubsection{Different dynamicity of different protrusions}
During certain stages or throughout their existence, a group of protrusions may elongate while the others retract at the same time. It may also happen that the length of certain protrusions remain unchanged during the elongation and retraction of the others. How does the cell coordinate the dynamicity of its multiple protrusion simultaneously is an open question. These questions are relevant, for example, in the context of neurites of a nerve cell \cite{toriyama10} and flagella of monoflagellated cell (during the multiflagellate stage) \cite{patra21}.

\subsubsection{Distribution rules during cell division}
One challenge for the cells with multiple protrusions is how to distribute their protrusions among the daughter cells post cell division. Here we list certain general rules:\\
(i) {\bf Replication followed by distribution:} Prior to cell division, the cell doubles the copies of its protrusion and distribute them equally among the daughter cells \cite{wetherbee88,heimann89}. \\
(ii) {\bf Distribution followed by de novo synthesis:} The cell first distributes the existing protrusions among the daughter cells and then the daughter cells assemble the extra protrusions from scratch \cite{aizawa98}.\\
(iii) {\bf Removal followed by de novo synthesis:} Certain cells get rid of the their existing protrusion prior to cell division and then the daughter cells synthesize the required number of protrusions from scratch \cite{rosenbaum69}. 

In case, the steady state length of the protrusions are unequal, it is a challenge from the perspective of length control how this length asymmetry is inherited by the daughter cells.

\subsection{Correlations in length fluctuations for intra-cell inter-protrusion communication}
The correlations in length fluctuations can be used for probing the nature and consequences of communications between different protrusions of the cell. For the numerical computation of the correlations, we begin with the following definitions: suppose, the total number of realizations recorded is $n$. Let $L_j^i(t)$ and $L_k^i(t)$ denote the length of protrusions labelled by the indices $j$ and $k$, respectively, at time $t$ in the $i^{th}$ realization. The instantaneous \textit{mean} lengths of these two protrusion are defined by 
\begin{equation}
\langle L_j(t) \rangle=\frac{\sum_{i=1}^n {L_j}^i(t)}{n}, ~~~ \& ~~~ \langle L_k(t) \rangle=\frac{\sum_{i=1}^n {L_k}^i(t)}{n},
\end{equation}
while the corresponding \textit{variances} are given by
\begin{eqnarray}
Var(L_j)= \bigg{[}\frac{1}{n-1}{\sum_{i=1}^n(\langle L_j(t) \rangle - {L_j}^i(t))^2}\bigg{]}^{1/2} ~~~~ \& ~~~~
Var(L_k)= \bigg{[}\frac{1}{n-1}{\sum_{i=1}^n(\langle L_k(t) \rangle - {L_k}^i(t))^2}\bigg{]}^{1/2}.  
\end{eqnarray}
and the \textit{covariance} $Cov(L_jL_k)$ is given by
\begin{eqnarray}
\frac{1}{n-1}\bigg{[}{\sum_{i=1}^n(\langle L_j(t) \rangle - {L_j}^i(t))(\langle L_k(t) \rangle - {L_k}^i(t))}\bigg{]}^{1/2}~.
\end{eqnarray}
In terms of \textit{variances} and  \textit{covariance}, 
the \textit{correlation} is defined as 
\begin{eqnarray}
Corr(L_jL_k)=\frac{Cov(L_jL_k)}{Var(L_j)Var(L_k)};
\label{corr_flagella}
\end{eqnarray}
and it gives a quantitative measure of the correlation in length fluctuations of  protrusion $j$ and protrusion $k$.

The sign and the magnitude of the correlations give crucial information about the nature of communication and coordination among the multiple protrusions of a cell as inferred in the context of eukaryotic flagellum \cite{patra21, bauer20, patra20a}.  

\part{Eukaryotic flagellum and cilium }
Eukaryotic flagella and cilia are hair like projections which emerge from the surface of the cell. Cilium and flagellum are often grouped together because of their identical internal structure and anatomy. From now onwards we will   use these two terms interchangeably. However, the flagellum referred in this part should not be confused with the bacterial flagellum (which will be discussed in Part-III). Flagella and cilia are present in a vast type of different cells starting from unicellular microorganisms to highly evolved multicellular mammals like human beings. However, the number of copies per cell, the location of appendages on the cell and their beating pattern may differ from one cell type to another.  In spite of this diversity, there is universality in their internal structure  and the transport logistic which  support the processes of assembling, maintaining and disassembling the flagella. In this part, we review the mechanisms of length control of eukaryotic flagellum, interesting trends of flagellar growth and regeneration during different stages of length control, etc.

From the perspective of organelles size control,  what makes flagella more interesting than many other organelles is not only the one-dimensional nature of the problem but also their highly dynamic lengths. The lengths of flagella change with time in sync with the cell cycle. Any deviation from this temporal dependence  can adversely affect not only the cell division but also the speed of swimming and the efficiency of circulating extracellular fluids  \cite{cross15,quarmby05}. Even when their growth is complete, the flagellar structure remains highly dynamic because each of the flagella continues to incorporate new proteins to make up for the high ongoing turnover, thereby maintaining a steady balance of the elongation and shortening\cite{marshall01}. How a specific cell maintains this balance at a particular length of a flagellum is one of the challenging open questions in the context of length control of a single flagellum. \\
\begin{longtable*} 
{ |>{\centering\arraybackslash}m{3.5cm}>{\centering\arraybackslash}m{3cm} | >{ \arraybackslash}m{10cm} |}
\hline 
{\bf  Number of  flagella} & { } & {\bf Example} \\
\hline
& & \\
1  & Monoflagellate & Sperm, {\it Leptomonas pyrrhocoris} \cite{he19}, {\it Pedinomonas tuberculata}\cite{heimann89}, {\it Pseudopedinella elastica} \cite{heimann89}, {\it Monomastix spec }\cite{heimann89}, {\it Peranema trichophorum} \cite{chen50}, {\it Trypanosome brucei} \cite{bertiaux18}  \\
& & \\
\hline
& & \\
2  & Biflagellate &  {\it Chlamydomonas reinhardtii}\cite{marshall01}, {\it Volvox carteri} \cite{coggin86}, {\it Nephroselmis stein} \cite{melkonian87}, {\it Spermatozopsis} \cite{schoppmeier03}  \\
& & \\
\hline
& & \\
4 & Quadriflagellate & {\it Tertraselmis} \cite{melkonian79}, {\it Tritrichomonas foetus} \cite{lenaghan14} \\
& & \\
\hline
& & \\
5 & Pentaflagellate & {\it Trichomonas vaginalis} \cite{petrin98}\\
& & \\
\hline
& & \\
8 & Octoflagellate & {\it Pyramimonas octopus, Giardia} \cite{dawson10,lujan11}\\
& & \\
\hline
& & \\
16 &  Hexadecaflagellate & {\it Pyramimonas cyrtoptera}  \cite{daugbjerg92}\\
& & \\
\hline
\caption{Example of mono- and multi-flagellated cells}
\end{longtable*}

Multi-flagellated microorganisms are excellent candidates for controlled experimental studies of the mechanisms of  length control of a single membrane bound organelle and coordination and communication among the multiple flagella of the same cell. The lengths of the flagella are thought to be adapted to their function in the respective cells \cite{bottier19}. The number and length of flagella vary from one species to another. For example, mammalian sperm are the classic examples of monoflagellates (cell with a single flagellum) whereas {\it Tetraselmis} is a quadriflagellate \cite{melkonian79}, {\it Pyramimonas octopus} and {\it Giardia} \cite{dawson10,lujan11} are octoflagellates \cite{moestrup87} while {\it Pyramimonas cyrtoptera}, to our knowledge, is the only example of unicellular eukaryotes with 16 flagella (i.e., a hexadecaflagellate) \cite{daugbjerg92}. The positions and lengths  of the flagella can also vary widely.  For example, the lengths of the four pairs of flagella of {\it Giardia} vary significantly although the two members of each pair have roughly the same length. Another interesting feature is that the length and number of the flagella on a cell can vary  as the cell enters different stages of the cell cycle. For example, the monoflagellates becomes transiently biflagellated \cite{heimann89,he19,wheeler11} or triflagellated \cite{heimann89} and the biflagellates becomes transiently quadriflagellated \cite{melkonian79,schoppmeier03,wetherbee88}. What makes biflagellated and multiflagellated cells even more interesting than monoflagellates is their abilities to coordinate the dynamics of lengths of different flagella. At the same moment, different flagella show different dynamics and this indicates that a certain mechanism must be there which facilitates communication among the multiple flagella of the cell \cite{heimann89,he19,wheeler11,melkonian79,schoppmeier03,wetherbee88}. 
 Finally, comparison of the lengths, numbers, positions of the wild type cells with the corresponding mutants of various types give further insight into the mechanisms of length control.

\section{Elongation and shortening of flagella: biophysical phenomena} 

\begin{figure}[h]
\includegraphics[width=0.98\textwidth]{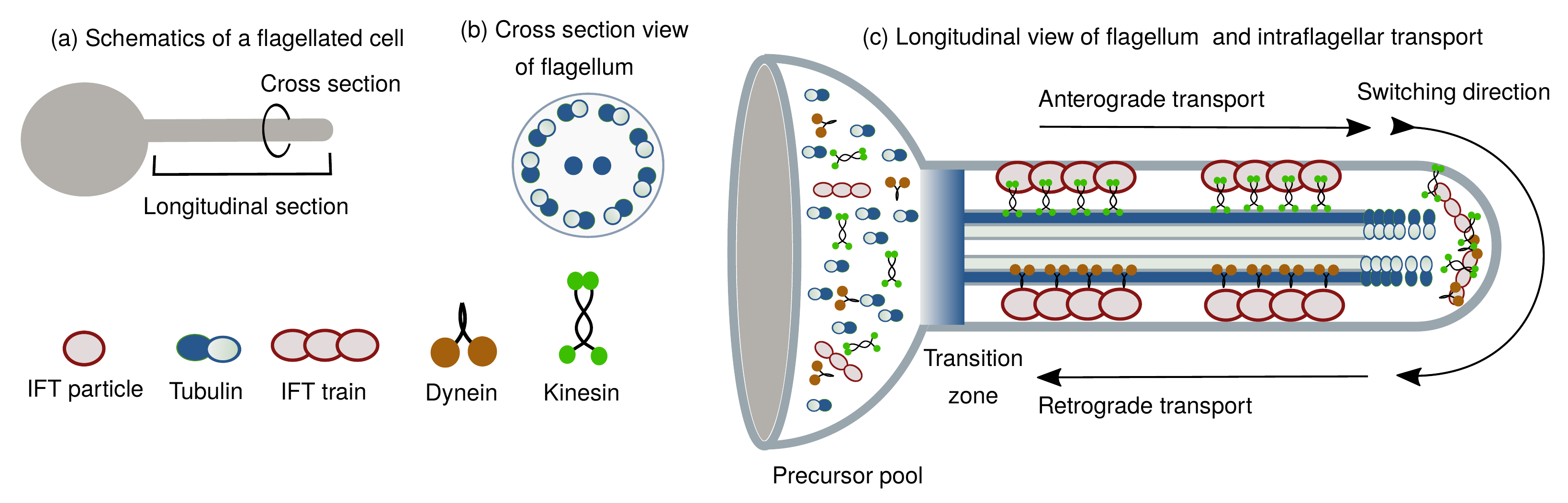}
\caption{Eukaryotic flagellum: internal structure and IFT transport}
\label{flagellum_intro}
\end{figure}

\subsection{Internal structure of eukaryotic flagella}
These organelles consist of an axonemal complex that is assembled on a basal body and projects out from the cell surface. The major structural component of all axonemes is microtubule (MT) each of which is essentially a tubular stiff filament  made of a hierarchical organization of tubulin proteins. Inside the cylindrical flagellum nine doublet MTs, arranged in a cylindrically symmetric fashion, extend from the base to the tip of the protrusion. Most axonemes have a 9+2 arrangement of MTs (see Fig.\ref{flagellum_intro}(a-b)), where nine outer doublets surround a coaxial central pair. In many immotile cilia, axonemes are said to have a 9+0 arrangement of MTs because they lack the central pair.  

\subsection{Intraflagellar transport: cargo, vehicle and motor}

Proteins are synthesized in the cell body, and not in the flagella. Therefore, the ciliary structural proteins are transported from the base to the tip of each flagellum by a motorized transport system \cite{chowdhury13,kolomeisky15}; this phenomenon is called intraflagellar transport (IFT) \cite{kozminski93,kozminski12,rosenbaum02,lechtreck15,prevo17}. The discarded structural components of flagellum released from the flagellar tip region are transported back to its base also by IFT. 
Such cargo transport plays a crucial role not only in the growth, but also in the maintenance and shrinkage of wide varieties of long protrusions of cells, including flagella. Obviously, regulation of the cargo transport regulates the 
rates of growth and decay which, in turn, determine the overall dynamics of the length of a flagellum. 
The crucial role of IFT in the construction of a growing flagellum was also established experimentally by demonstrating disruption of flagellar growth upon disruption of IFT \cite{pazour00,rosenbaum02} (see Fig.\ref{flagellum_intro}(c)). In more recent times, direct evidence in support of the transport of structural proteins as cargo of IFT trains have been reported \cite{wren13,craft15}.

Transport of various types of molecular and membrane-bound cargoes in eukaryotic cells is carried out by molecular motors that are driven along filamentous tracks \cite{chowdhury13,kolomeisky15}. A MT serves as a track for two `superfamilies' of molecular motors, called kinesin and dynein, which move naturally in opposite directions by consuming chemical fuels. 
IFT particles, which are multi-protein complexes at the core of the IFT machinery, operate essentially as the ``protein shuttles of the cilium'' \cite{lechtreck17}. Powered by molecular motors, the IFT trains cycle between the flagellar tip and base \cite{iomini01,buisson13}. During each leg of their journey the IFT trains remain constrained in the narrow space between the outer surface of the axoneme and the inner surface of the ciliary membrane.  The IFT particles switch their direction of movement only at the base and the tip of the flagellum. This indicates the plausible existence of a regulatory mechanism for differentially activating and inactivating the appropriate IFT motors at the base and tip to facilitate the directional switching.  However, neither the mechanism of this regulation nor the number of motors per IFT train is well known. 

The molecular components of the IFT machinery have also been catalogued in detail \cite{cole09,hao09}. 
Broadly,  four different types of proteins perform crucially important distinct functions in IFT:\\
(a) Axonemal proteins (mainly tubulins) and other {\it structural proteins} are transported as {\it cargoes} within flagella.\\ 
(b) The {\it vehicles} for the transport of these cargoes are a special type of proteins called {\it IFT particles} \cite{wren13,craft15}. Because of their superficial similarities with cargo trains hauled along railway tracks, chain-like assemblies formed by IFT particles are called IFT trains  \cite{pigino09}. Not all IFT particles are loaded with cargo before they begin their journey.  \\
(c) Both the empty and loaded IFT particles are hauled along the narrow space between the axoneme and the ciliary membrane by {\it motor proteins} that walk along the MT tracks. Since the number of motors per IFT train is not known, movement of the motors are not described explicitly in some models \cite{patra20a}; instead, the stochastic movement of the IFT trains along the MT tracks are described in terms of kinetic equations.\\
(d) Special {\it flagellum stabilizing or destabilizing} proteins can control the length of a flagellum. For example, a stabilizing protein can cap the flagellar tip by stopping its further elongation or shortening. On the other hand, a flagellum can be destabilized by driving active depolymerization of its filamentous constituents (like microtubules) by depolymerases.

One natural question is: why is IFT required in fully grown flagella? This mystery was unveiled when it was observed that there is an ongoing turnover of axonemal proteins at the tip of a fully grown flagellum. Unless the discarded material is removed from the flagellar tip and replenished by fresh supply of these proteins in a timely manner the fully grown flagellum cannot continue to maintain its length. Thus, in a fully grown flagellum it is the dynamic balance between the rates of growth and decay that maintains the average length at a stationary value \cite{marshall01}.

\subsection{Ciliogenesis, resorption, deflagellation, amputation and regeneration}
The process of assembling fresh flagella (and cilia) in a new born cell  is known as {\it ciliogenesis}. A fully grown flagellum of a wild type cell has a length that is convenient for its biological function and has been selected in the course of Darwinian evolution. For example, each of the flagella of the biflagellated green algae cell of {\it Chlamydomonas reinhardtii}  is approximately 12 $\mu$m long. 
 However, mutants cells can have longer or shorter flagella. In certain flagellated cells, the flagella are gradually retracted into the cell prior to the cell division \cite{bloodgood74}. This phenomenon, usually referred to as ``{\it resorption}'', just like the phenomenon of {\it ciliogenesis}, depends crucially on IFT. 

Flagellar disassembly \cite{liang16} via resorption should be distinguished from ``{\it deflagellation}'' (also known as deciliation, flagellar excision, flagellar shedding or flagellar autotomy) \cite{quarmby04}.  In the latter process, in response to wide varieties of stimuli, like heat shock or mechanical strain, the axoneme is severed resulting in a complete detachment of the flagellum from the cell body. Deflagellated cells can regenerate their flagella when stress causing stimulus disappears. For example, the flagella regain their original full length in about 90 minutes in deflagellated wild type {\it Chlamydomonas reinhardtii} cells after stress is removed.

The dynamics of flagella is interesting also from the perspective of regeneration \cite{johnson93}. It is worth emphasizing that in the regeneration of flagella a cell replaces its own severed parts; it does not require formation of new {\it Chlamydomonas reinhardtii} cells from undifferentiated precursor cells. In this respect, regeneration of cellular protrusions, like flagella and axons, differs from the cell replacements in injured tissues.
The cooperation between the dynamics of the two flagella of a {\it Chlamydomonas reinhardtii} is displayed most vividly during regeneration of an amputated flagellum \cite{ludington12}, or during the ciliogenesis of extra flagella in the monoflagellated or biflagellated cells prior to the cell division \cite{heimann89,he19,wheeler11,melkonian79,schoppmeier03, wetherbee88}.

\subsection{Ciliogenesis in monoflagellates and multiflagellates}
The biflagellate {\it Chlamydomonas} has been used extensively in the experimental studies of the mechanism of the  genesis of a single flagellum. This approach worked quite well mostly because, during ciliogenesis in a {\it Chlamydomonas} cell,  both the flagella elongate simultaneously at the same rate. However, a complete understanding of the genesis of the flagella of a cell requires a holistic approach that would explain how a cell controls the lengths of all of its flagella simultaneously. Here, we summarize various mechanisms of flagellar development and distribution in monoflagellates as well as in multiflagellates.

\subsubsection{Ciliogenesis in monoflagellates }
Monoflagellated cells bear a single flagellum during the longest phase of cell cycle i.e, the interphase. However prior to the cell division it becomes multiflagellated  and,  after completion of the division, the daughter cells inherit one flagellum each which are either fully or partially grown. There are are two mechanism by which ciliogenesis and cell division proceed hand in hand in the monoflagellates : \\
{\bf Mechanism I} : The monoflagellated cell becomes transiently biflagellated by growing one additional flagellum prior to the cell division. During the elongation of the new flagellum, the length of the older flagellum either remains constant, as seen in {\it Pedinomonas tuberculata} \cite{heimann89}, {\it Trypanosome} \cite{bertiaux18}, or it may undergo partial resorption, as seen in {\it Leptomonas pyrrhocoris} \cite{he19}. In case the older flagellum shortens, it never becomes shorter than the new elongating flagellum \cite{he19}. Each daughter cell receives one flagellum upon cell division. Either the flagellum is fully grown or it continues elongation even after being inherited by the daughter cell  till it attains its normal full length in the interphase. (see Fig.\ref{Ciliogen_cycle}(a1)) \\
{\bf Mechanism II} : The monoflagellated cells like {\it Pseudopedinella elastica} and {\it Monomastix spec} become  transiently triflagellated by growing two additional flagella prior to the cell division \cite{heimann89}. During the elongation of the pair of new flagella the old flagellum shortens. Prior to the cell division, the cell gets rid of the old flagellum either by complete resorption ({\it Pseudopedinella elastica}) or by deflagellation of the flagellum which has undergone partial resorption ({\it Monomastic spec}). 
Thereafter, the mother cell remains biflagellated for a brief period till complete division which results in the two daughter cells each inheriting one of the two new elongating flagella \cite{heimann89} (see Fig.\ref{Ciliogen_cycle}(a1)).

\subsubsection{Ciliogenesis in biflagellated isokont}

\begin{figure}
\includegraphics[width=1.0\linewidth]{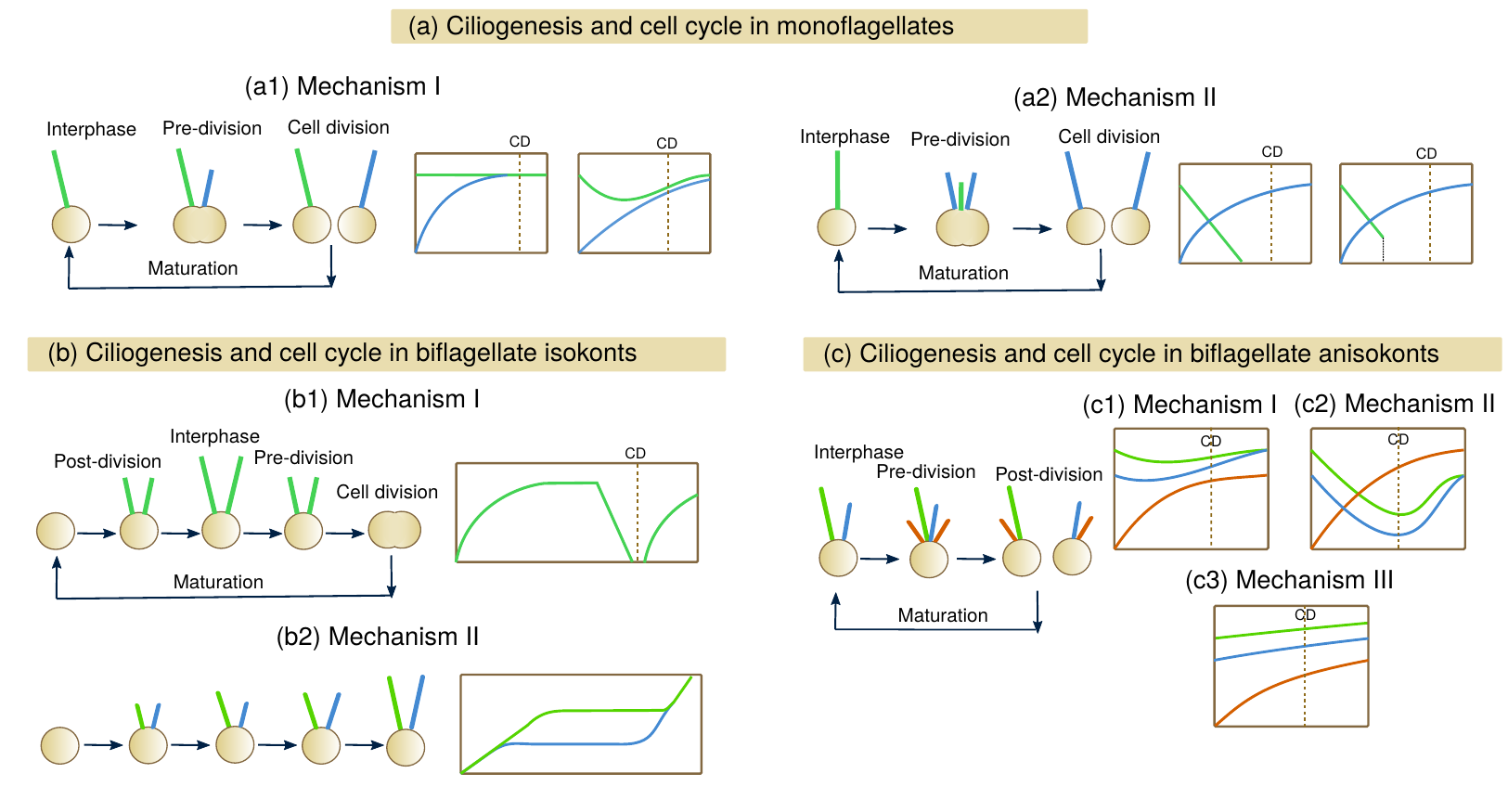}
\caption{{ \bf Ciliogenesis and cell cycle} in (a) monoflagellates (b) biflagellate isokonts and (c) biflagellate anisokonts. (a1) In a monoflagellated cell, flagellum replication followed by the distribution of flagella among the two  daughter cells. The old flagellum (denoted in green colour) may or may not shorten partially during the elongation of the new flagellum (denoted in blue colour). (a2) In a monoflagellated cell, a pair of new flagella (denoted by blue colour) elongate while the old flagellum (denoted in green colour) is either resorbed completely or removed by deflagellation after partial retraction. Later the new pair of flagella is distributed among the two daughter cells. (b1) In isokont biflagellated cell,
two flagella of equal length retract prior to the cell division of the mother cell and, after division, a new pair of flagella grow in each of the daughter cells. During the elongation, the two flagella may or may not elongate together. (b2) In anisokont biflagellated cell, an additional pair of flagella elongate prior to the cell division and later each daughter cell inherits one old (denoted in blue and green color) and one new flagellum (denoted in red color). During the elongation of the new pair of flagella, the pair of old flagella may either shorten partially or keep growing.} 
\label{Ciliogen_cycle}
\end{figure}

Biflagellated cells which have two flagella of equal length are referred to as isokonts. The two most widely studied examples are {\it Chlamydomonas reinhardtii} and {\it Volvox carteri}. Here we mention the two identified mechanisms by which the cells assemble their flagella.\\
{\bf Mechanism I }: As we have already discussed, prior to the cell division, a {\it Chlamydomonas}  cell resorbs both of its equilength flagella simultaneously \cite{rosenbaum69}.  Then,   after cell division, each of the daughter cells reassemble a pair of  flagella that grow simultaneously attaining, eventually, the same steady length as that of their mother cell (see Fig.\ref{Ciliogen_cycle}(b1)).\\
{\bf Mechanism II }: In {\it Volvox carteri} both the flagella start to elongate together. However, after a certain period, one of the flagella ceases to grow while the other flagellum keeps elongating. After a certain while, the growing flagellum also ceases to grow. Somewhat later, the shorter flagellum resumes growth, and when its length becomes equal to that  of the stalled flagellum, the stalled flagellum  resumes growth and both the flagella grow together till they attain their normal length in the interphase \cite{coggin86} (see Fig.\ref{Ciliogen_cycle}(b2)).

\subsubsection{Ciliogenesis in biflagellated anisokont}

Anisokonts are cells which bear flagella of unequal length.  Just as the monoflagellates become transiently biflagellated prior to cell division, the biflagellated anisokonts become quadriflagellated before division into two daughter cells. During the cell division, each daughter cell receives an old flagellum  and a newly assembled flagellum which may or may not have reached their steady state full length. There are three mechanisms by which these anisokont cells undergo  ciliogenesis.\\
{\bf Mechanism I }: Two new flagella which emerge prior to the cell division make the mother cell transiently quadriflagellated. The new pair of flagella keep growing whereas the older pair of flagella keep shortening. Upon division of the mother cell, each daughter cell inherits one new growing flagellum and an old shortening flagellum. The old flagellum does not shorten fully but regrow and attains a shorter steady state length whereas the new flagellum elongates to become the new longer flagellum. This kind of dynamics is seen in {\it Epiphyxis pulchra} \cite{wetherbee88}. (see Fig.\ref{Ciliogen_cycle}(c1))\\
{\bf Mechanism II }: The mother cell becomes quadriflagellated. Each daughter cell inherits one older flagellum and a newly assembled flagellum. The older flagellum is longer than the new flagellum. During the elongation of the shorter flagellum, the longer flagellum partially retracts but finally both attain their steady state interphase length. The longer one is the older flagellum. This kind of dynamics is seen in {\it Nephroselmis olivacea } \cite{melkonian87} (see Fig.\ref{Ciliogen_cycle}(c2))\\
{\bf Mechanism III }: Both the pair of old flagella and the pair of newly assembled flagella prior to the cell division keep elongating. Even in the new daughter cell, which inherits one older and  one newly assembled flagellum, the pair of flagella keep elongating. A particular flagellum may take 3 cell cycles to completely grow to its full steady state length and it ceases to grow after that and only the shorter one elongates. In a given cell, the longer flagellum is older than the shorter flagellum.  This kind of trend is seen in {\it Spermatozopsis} \cite{lechtreck97} (see Fig.\ref{Ciliogen_cycle}(c3)).

\subsubsection{Ciliogenesis in octoflagellates}

{\it Giardia} is an octoflagellated cell which is drawing a lot of attention recently {\cite{lujan11,patra22}}. It is an interesting species to understand how the size control of multiple  appendages is achieved by the cell at the same time. The four pairs of flagella are referred to as caudal, ventral, anterior and posterior as depicted in Fig.{\ref{Giardia}(a)} .

Unlike the other flagellates, the axoneme of  {\it Giardia} is very long. The length of the cytoplasmic bound and membrane bound axoneme of a particular flagellum is comparable (length compared in Fig.{\ref{Giardia}(b)}). The IFT particles perform diffusive motion on the cytoplasmic part. They are assembled as trains at the flagellar pore and perform directed motion in the membrane bound part. The question is, whether the cell measures the length of the membrane bound axoneme only or the length of the whole axoneme. In a recent paper, the authors claimed that the pair of caudal flagella is the shortest \cite{mcinally19}. They actually claimed that the amount of depolymerase at the tip of caudal flagella is maximum as compared to that in the other flagella. As a result, the membrane bound length of the caudal flagellum is the shortest as compared to the other flagella. 

{\bf Cell cycle and flagella distribution:} Another interesting feature of {\it Giardia} is the transformation of flagella during cell division. As shown schematically in Fig.{\ref{Giardia}(c)}, two daughters receive different combination of flagella, transform one type of flagella into another type and  synthesize the extra  flagella from the scratch so that at the end each daughter could have 8 flagella. The caudal flagella is the oldest among all. So according to the age-length relation, the longest is the oldest (if we consider the total length) or the shortest is the oldest (if we consider the membrane bound region length only).    

\begin{figure*}[htbp] 
\begin{center}
{\bf Figure NOT displayed for copyright reasons}.
\end{center}
\caption{{\bf Giardia cell with four pairs of flagella:}  (a) Axonemes initiate at the basal body (BB) and  protrude the cell membrane at the flagellar pore (fp). The median body (MB) and the ventral disc (VD) are microtubule-based structures. (Fig 1(a) of  ref.\cite{mcinally19}).  (b) Independent IFT particles diffuse in the region of axoneme which remains exposed in the cytoplasmic region. In the membrane bound region, the assembled IFT particles perform directed motion. (Fig.2 and Fig.1(b) of ref.\cite{patra22}). (c) Flagellar distribution followed by transformation during cell division of Giardia cell amond two daughter cells (Fig.3(a) of ref.\cite{patra22}). }
\label{Giardia}
\end{figure*}

{\bf Encystation: } During a certain phase of its cell cycle, the {\it Giardia} undergoes encystation during which an envelope of cyst covers the entire cell. So prior to this, the cell resorbs its membrane bound flagella. But even when they are resorbed, the cytoplasmic bound axoneme is present and it continues to beat. After excystation, the cell emerges from the cyst shell and undergo cell division by  longitudinal binary fission. The question is how the cell can have two different balance point for the same axoneme?

\section{Models for length control of a flagellum: ciliogenesis and resorption}
\label{sec-flag-len-cont}
Flagella of optimum length are essential for proper swimming through the medium \cite{heddergott12} and adhesion to the host cells \cite{dawson10}. Using an excellent experiment setup, it was demonstrated \cite{heddergott12} that  20 $\mu$m long flagella is essential for navigation of {\it Trypanosome}  through the blood that contains blood cells of fixed  diameter. But cells with altered flagellar length find it difficult to move through the obstacles and navigate the medium. In a recent work, it has been shown that the length alterations caused by mutations need not be lethal but, instead, cause {\it Chlamydomonas} to alter its swimming pattern \cite{bauer20}. Just like the flagellar length, length of cilia is also optimally controlled for mucus clearance \cite{juan20} and decoding the signals \cite{challis15}. Any length changing mutation could either be lethal or force the cell to adapt accordingly.

\subsection{Length control of a single flagellum}    
Recall that the two main modes of length control of long cell protrusions that we summarized in section \ref{sec-CellProtruLength} are (a) dynamic balance point, and (b) pausing zone. It was experimentally established many years ago that flagellar length is controlled by the mode (a), namely by balancing of the rates of assembly and diasassembly. This principle requires that at least one of the two rates must be dependent on the flagellar length. The natural question is: how does the cell sense the instantaneous length of a flagellum and accordingly down-regulate or/and up-regulate the assembly rate or/and the disassembly rate with the changing flagellar length? In table \ref{tab-len_sensing} of section \ref{sec-CellProtruLength} we summarized the known mechanisms of sensing and regulation of lengths of long cell protrusions. Not all the different flagella in the same cell type or flagella in different types of cells are expected to be sensed and controlled by a single unique mechanism. Even the experimentally measured data for ciliogenesis in a given cell can be accounted for, at present, by more than one mechanism of length control, as we review in this section, although some of the mechanisms listed in table \ref{tab-len_sensing} can be easily ruled out for flagellar length control. 
A brief description of all the mechanisms that have been used in various attempts to explain flagellar length control, as well as their successes and criticisms, are provided in Table. \ref{flag_len_cont_table}. It may turn out that the actual length control mechanism in a specific example is a combination of some of these mechanisms.

\begin{figure}[htbp]
\begin{center}
{\bf Figure NOT displayed for copyright reasons}.
\end{center}
\caption{ {\bf Evolution of flagellar length predicted by different models:} (a) Balance point model by Marshall and Rosenbaum (Fig.8(C) of ref.\cite{marshall01}) (b) Diffusing motors as rulers by Hendel et al (Fig.2(B) of ref.\cite{hendel18} (c) Time-of-flight based length sensing by Ishikawa and Marshall. (Fig.2(A) of ref.\cite{ishikawa17}) (d) Differential loading and time-of-flight by Patra et al(Fig.14(a) of ref.\cite{patra20a}). }
\label{fig_math_model}
\end{figure}

Here we describe, in little more detail, the mechanisms that have been proposed mainly in the specific context of flagellar length control in the biflagellated {\it Chlamydomonas reinhardtii} cell,  a green algae. Nevertheless, we discuss the topic from a broader perspective using a language that would help in deeper connection with flagellar length control mechanisms also in other flagellated cells.

Recall that there are three different types of important proteins that together control flagellar length, namely, the IFT trains (and the motors that pull them) that serve as the transport vehicles, the structural constituents of the flagellum that are carried as cargo by the transport vehicles, and the proteins that accumulate at the distal tip where they stabilize or destabilze the flagellum. Therefore, there are  at least three ways to control the assembly and disassembly of a flagellum.\\
(i) {\bf Control through IFT trains (transport vehicles) and motor proteins (engines):} In this mechanism it is assumed that the tip undergoes continuous turnover with a disassembly rate which is independent of flagellar length and the assembly rate is a function of the flux of IFT trains which carry precursors for elongating the flagellum. Certain class of models are based on the premise that flux of IFT trains reaching the tip decreases with the increasing flagellar length. Such control could  be implemented in several different ways: (a) by limiting the number of trains shuttling inside the flagellum \cite{marshall01}, or (b) by gradually reducing the entry of IFT trains into the flagellum with its growing length responding to the feedback about the instantaneous length \cite{wren13}. The feedback could be based on a length-sensing mechanism that exploits either a timer associated with the IFT particles \cite{wren13} or currents through the calcium ion channel embedded in the ciliary membrane \cite{liang14,liang18} or a concentration gradient of specific proteins along the flagellar length \cite{chien17}. The three models of flagellar length control based on this are the following: 
\begin{itemize}

\item {\bf (a) Balance-point model: length-dependent assembly rate}:
The balance-point model is actually not a single model but one of the two general scenarios that we presented in section \ref{sec-CellProtruLength} and three distinct ways of realizing it were depicted in Fig.\ref{fig-balance_pause}(a). The balance-point model proposed by Marshall and Rosenbaum \cite{marshall01}, is a physical realization of the scenario shown on the leftmost panel of Fig.\ref{fig-balance_pause}(a). It is built upon two key assumptions: (1) The disassembly-rate $d$ is independent of the flagellar length, and (2) The assembly rate decreases with increasing length of the flagellum. The length-dependent rate of assembly is obtained by, first, assuming that the number of IFT particle proteins per flagellum is independent of flagellar length. The time taken by an IFT train to complete a single trip is $2L(t)/v$ where $v$ is its velocity. Therefore, the flux (also called current) of the IFT trains is $N/(2L/v)$ where $N$ is the number of IFT trains shuttling inside the flagellum. The time-dependence of the flagellar length $L(t)$  is governed by the equation
\begin{eqnarray}
\frac{dL(t)}{dt}=\underbrace{\frac{\alpha v N }{2L(t)}}_{\substack{  \text{Assembly rate is inversely } \\ \text{proportional to length}}} - d ~~~~~(\rm {\text{Original balance point model}}) \nonumber \\
\label{eq-MarshallRosen}
\end{eqnarray}
where  $\alpha$ is a constant of proportionality. Note that the assembly rate, which is the first term on the right hand side of (\ref{eq-MarshallRosen}), is inversely proportional to the length of the flagellum. From Eq.(\ref{eq-MarshallRosen}) we find the steady-state flagellar length to be
\begin{eqnarray}
L_{ss}^{(BP)}=\frac{\alpha v N }{2d}.
\end{eqnarray} 
The increase of flagellar length with time, as predicted by the balance point model, is shown Fig.\ref{fig_math_model}(a).

\item {\bf (b) Length sensing by diffusion ( diffusing motor as a ruler)}: 
In an alternative scenario \cite{chien17, hendel18} the kinesin motors walk actively, fuelled by ATP hydrolysis, pulling IFT trains only up  to the flagellar tip. But, after delivering cargo at the tip, these motors unbind  from the IFT particles and return to the base by diffusion and are then re-used for pulling new IFT trains from the flagellar base to the tip. For simplicity of calculation, let us make three assumptions: (1) Since the active transport time of a kinesin motor from the flagellar base to the tip is negligibly small compared to its passive diffusion time from the tip to the base one can make the approximation that upon arrival at the base each kinesin motor is transported instantaneously back to the tip \cite{hendel18}. (2) A kinesin motor begins its return journey to the base immediately after reaching the flagellar tip. (3) No diffusing kinesin rebinds with the axonemal microtubule tracks within the flagellum. 

Thus, this scenario effectively models the shuttling of the motors as one-dimensional diffusion with a constant source of free motors at the flagellar tip and a sink at the flagellar base. The diffusion time of the kinesins from the tip to the base can provide feedback to the cell as to the length of the flagellum. The diffusion time $t_d$ is given by $t_D = L^{2}/(2D)$ where $D$ is the diffusion coefficient of the motors in the flagellum. Let $\delta L$ be the increment of flagellar length when a motor reaches the tip and suppose $N$ is the number of diffusing motors. Then, in the time interval $t_d=L^2/(2D)$ the $N$ motors add a length $N \delta L$ to the tip. Thus the rate of growth of the flagellum is given by 
\begin{eqnarray}
\frac{dL(t)}{dt}=\underbrace{\frac{N \delta L}{L^2/(2D)}}_{\substack{ \text{Assembly rate is inversely }\\ \text{proportional to square of length}}} - d ~~~~~(\rm{ \text{Diffusive ruler model }})
\end{eqnarray}
where  $d$ is the constant rate of disassembly. Consequently, the steady-state length of a flagellum in this model is 
\begin{eqnarray}
L_{ss}^{(DR)} =\biggl(\frac{2 D N \delta L}{d}\biggr)^{1/2}
\end{eqnarray}
The growth of a flagellum, as  predicted by the model of Hendel et al \cite{hendel18}, is shown Fig.\ref{fig_math_model}(b).

The above model was improved \cite{fai19} by relaxing the assumption of instantaneous transfer (infinite velocity) of the kinesin motors from the flagellar base to the tip. The flux of the kinesin motors can be expressed as 
\begin{equation}
J = k_{on} (N-N_{b} - N_{d}) 
\label{eq-Fai1}
\end{equation}
where $N$ is the total number of motors while $N_{b}$ and $N_{d}$ are motors engaged in ballistic and diffusive movement in the flagellum, leaving the remaining $N-N_{b} - N_{d}$ motors free in the pool at the flagellar base. 
The anterograde flux of the motors (from base to tip) is  $J={\bar{c}}_{b} v$ where  ${\bar{c}}_{b}$ is the  average concentration of the motors executing ballistic motion. Since ${\bar{c}}_{b}=N_{b}/L$, we have  $J=v(N_{b}/L)$, i.e., 
\begin{equation}
N_{b} = \frac{JL}{v}.
\label{eq-Fai2}
\end{equation}
In the steady state the magnitude of the current in both the anterograde and retrograde directions must be equal. Under such a situation one gets \cite{fai19}
\begin{equation}
N_{d} = \frac{JL^2}{2D}. 
\label{eq-Fai3}
\end{equation}
Substituring (\ref{eq-Fai2}) and (\ref{eq-Fai3}) into (\ref{eq-Fai1}) we get 
\begin{equation}
J = \frac{k_{on}N}{1+[k_{on}L/v]+[k_{on}L^2/(2D)]}.  
\label{eq-Fai4}
\end{equation}
Fai et al.\cite{fai19} wrote that $dL/dt = \gamma J (T-L) - d$ where $T$ is the total number of tubulin monomers out of which $L$ constitute the flagellum leaving $T-L$ monomers freely available in the pool. Then, using the expression (\ref{eq-Fai4}) for $J$, one gets \cite{fai19} 
\begin{equation}
\frac{dL}{dt} = \frac{\gamma k_{on}N}{1+[k_{on}L/v]+[k_{on}L^2/(2D)]} (T-L) - d.
\label{eq-Fai5}
\end{equation}
where $k_\text{on}$ is the injection rate constant, $v$ is the IFT speed, $D$ is the diffusion constant, $T$ is the size of tubulin pool, $d$ is the disassembly rate and $\gamma$ is a constant. The steady state flagellar length is given by
\begin{equation}
L_\text{ss}=\left( \frac{D}{v}+\frac{\gamma D M}{d}  \right) \left(-1+\sqrt{1+\left(\frac{2dT}{\gamma M D}\right) \frac{1-d/\gamma K_\text{on}MT}{(1+d/\gamma M v)^2}} \right)~,
\end{equation}
where $M$ is the total number of motors.

\item {\bf (c) Length-sensing by time-of-flight: }Next we consider an alternative scenario where the instantaneous length of the flagellum is sensed using a time of flight (ToF) mechanism \cite{ishikawa17}; this feedback decides the probability of undertaking a round-trip journey by a cargo carrying IFT train in the flagellum.

Let $v_{a}$ and $v_{r}$ denote the velocities of an IFT train during its anterograde and retrograde journeys, respectively, during a round-trip journey. Suppose the train spends a time $\tau$ after reaching the flagellar distal tip by anterograde transport and before beginning its retrograde  journey. 
Either each of the IFT train itself or a molecule bound to is assumed to have two possible chemical (or conformational) states $S_1$ and $S_2$ such that the state $S_1$ makes a spontaneous irreversible transition to the $S_2$  with a rate (more precisely, probability per unit time) $k$. We call it a timer for the reason that will be clear soon. The timer is assumed to be in the state $S_1$ when it begins its anterograde journey from the flagellar base. 
This model postulates that the train that has just returned to the flagellar base will undertake another round-trip with its cargo only if it is still in the state $S_1$ when it returns to the same location at the end of its retrograde journey. The probability of the latter event is  $\exp[-k t_{ToF}]$ where the total ToF is $t_{ToF} = \frac{L(t)}{v_a}+\frac{L(t)}{v_r}+\tau$. Therefore, in this model  
\begin{eqnarray}
\frac{dL(t)}{dt}=\underbrace{A(T-2L(t)) ~e^{-k(\frac{L(t)}{v_a}+\frac{L(t)}{v_r}+\tau)}}_{\substack{  \text{Assembly rate falls } \\ \text{exponentialy with length}}}-d ~~~~~({\rm \text{Time of flight model }})
\end{eqnarray}
where $A$ is a constant of proportionality, $T$ is the total pool size of tubulins  and $d$ is the length-independent disassembly rate. The equation is solved numerically for getting the  steady-state flagellar length.  Flagellar length  increase with time, as predicted by the model of Ishikawa and Marshall  \cite{ishikawa17}, is shown Fig.\ref{fig_math_model}(c).

\end{itemize}

(ii) {\bf Length control through differential loading of precursor proteins (transported cargo)}: The mechanisms described above in the categories (i) are based primarily on the regulation of IFT proteins and/or  the motors that pull them. So, now we consider mechanisms of controlling the assembly rate is by controlling the amount of precursors that are carried as cargo by the individual IFT trains. 

In this model where the total  flux of IFT trains per flagellum remains a constant independent of the flagellar length. Instead, the rate of supply of precursors is reduced with the increasing flagellar length by reducing the fraction of IFT trains that are actually loaded with the precursor cargo at the flagellar base at the beginning of a round-trip journey in the flagellum. This is the {\it principle of  differential loading}. Its implementation requires a length sensing mechanism \cite{wren13}. A time of flight (ToF) mechanism would be adequate for this purpose \cite{wren13,ishikawa17}. Combining the above ideas, a model was proposed by Patra et at \cite{patra20a}. 

The coupled set of equations which govern the evolution of the flagellar length $L(t)$ and population $n(t)$ of tubulins in the basal pool are 
\begin{eqnarray}
\frac{d L(t)}{dt} &=& \underbrace{\frac{n (t)}{n_\text{max}}J\Omega_e \exp \left(  -\frac{2k L(t) }{v} \right)}_{\substack{ \text{Assembly rate exponentially falls with length}}} -(1-\rho)^2 \Gamma_r ~~~~~({\rm \text{Differential loading with time of flight }})
\end{eqnarray}
and
\begin{eqnarray}
\frac{d n (t)}{dt}&=&\omega^+ \left( 1-\frac{n (t)}{n_\text{max}} \right) - \omega^-  n (t) 
\label{flagella-rate-eq}
\end{eqnarray}
where  $\Omega_e$ is the probability of incorporation of precursors at the tip, $\Gamma_r$ is the rate of removal of precursors from the tip, $\rho$, $v$ and $J$ are the number density, velocity and number flux of the IFT trains, $k$ is the flipping rate of the timer and  $n_\text{max}$ is the capacity of the pool, $\omega^+$ and $\omega^-$ are the rate of synthesis and degradation of precursors. Hence, the steady state flagellar length is 
\begin{equation}
L^{DL}_\text{ss}=\frac{v}{2k}\ell og \left( \frac{n_\text{ss}}{n_\text{max}} \frac{J \Omega_e}{(1-\rho)^2 \Gamma_r} \right)~.
\end{equation}
The temporal evolution of flagellar length, as predicted by the model of Patra et al \cite{patra20a}, is shown Fig.\ref{fig_math_model}(d).

Note that eq.(\ref{flagella-rate-eq}) can be expressed as 
\begin{eqnarray}
\frac{d L(t)}{dt} = k_{1}~  C_{p}~ T(L) - k_{2} 
\label{rate_l2}
\end{eqnarray}
with $k_{1}=J \Omega_{e}$, $C_{p} = n (t)/n_\text{max}$, $T(L) = e^{-2k L(t)/v}$ and $k_{2} = (1-\rho)^2 \Gamma_r$. The form (\ref{rate_l2}) looks exactly like the eq.(1) in the supplementary information of ref.\cite{ludington12}. However, the crucial difference between (\ref{rate_l2}) and eq.(1) in the supplementary information of ref.\cite{ludington12} is that $T(L)$ in (\ref{rate_l2}) is given by a mathematical expression that follows naturally from the ToF mechanism whereas it was treated as a phenomenological parameter in ref.\cite{ludington12}. \\

(iii) {\bf Length control through tip stabilizing / destabilizing  proteins:} The fourth player among the proteins that play important roles in flagellar length control are those that significantly alter the rates of polymerization and depolymerization kinetics of the tip of the underlying axoneme thereby influencing the rates of assembly and disassembly of the flagellum. MT capping proteins or locking proteins like FLAM8 which affect both the polymerization and the depolymerization rates \cite{bertiaux18} and the depolymerase KIF13 which affect depolymerization rates are examples of two such proteins \cite{mcinally19}. 
\begin{itemize}

\item {\bf (a) Length control by depolymerase:} In case of {\it Giardia}, the depolymerase accumulate in a length-dependent fashion at the tip \cite{mcinally19}. The concentration profile of kinesin-13 depolymerases inside the flagellum is given by
\begin{equation}
c(L)=c_\text{init}\frac{e^{L/\lambda}}{1+ve^{L/\lambda}}
\label{giardia_cl}
\end{equation}
which is obtained by solving the transport equation of kinesin-13 mediated by IFT trains and pure diffusion. In the above equation, $c_\text{init}$ is the initial concentration when all the kinesin-13 is at the flagellar base, $v$ is the ratio of the volume of a stretch of flagellum of length $\lambda$ and the volume of the cellular reservoir from which kinesin-13 is taken up during ciliogenesis.

The rate of assembly is proportional to the free tubulin population ($N-L$) in the basal pool and the flux of IFT trains $I$ whereas the rate of disassembly is proportional to the concentration of kinesin-13 at the tip given by $c(L)$ in eq.(\ref{giardia_cl}). The evolution of flagellar length can be written as 
\begin{equation}
\frac{dL}{dt}=I(N-L)-k_- c(L) ~~~~~({\rm \text {Length control by depolymerase}})
\end{equation}
where $k_-$ is a constant. The steady state occurs at a length $L^{DP}_{ss}$ which satisfies
\begin{equation}
L^{DP}_{ss}=N-\frac{k_-  c(L^{DP}_{ss})}{I}~.
\end{equation}

\item {\bf (b) Length control by capping or locking proteins:} In {\it Trypanosome} flagellum, a locking protein known as FLAM8 prevents further polymerisation and depolymerisation of the flagellar tip \cite{bertiaux18}.  FLAM8 is available abundantly at the tip of a long matured flagellum, but a very low amount of it is present at the tip of a short new flagellum. However, we are not aware of any model that describes how the population of FLAM8 increases with increasing length of the flagellum and when it caps the flagellar tip.

\end{itemize}


\begin{longtable*} {|>{\centering\arraybackslash}m{0.5cm}|>|>{\centering\arraybackslash}m{7cm}|>{\centering\arraybackslash}m{10cm}|}
\hline 
&\multicolumn{2}{c} {} \\
1&\multicolumn{2}{c} {\bf Control through IFT trains and motor proteins}\\
&\multicolumn{2}{c} {}\\
\hline
1.1 & {\bf Original balance point:} Fixed number of trains shuttle inside the flagellum. As the length increase, inter train distance increase which lead to decrease in flux of IFT at the tip. Flagellar tip undergoes continuous turnover which leads to constant disassembly rate. Controlled length emerges when the opposing rates balance each other.
 & \begin{itemize}
\item  Evidence (for):  Model proposed by Marshall \& Rosenbaum  with experimental evidence \cite{marshall01}.
\item Criticism :  (i) Dentler observed the number of IFT trains shuttling inside the flagellum increase with increasing  flagellar length and flux remains unchanged \cite{dentler05}. (ii) A large pool of IFT particles is maintained and only 20\% of the IFT particles of reservoir participate in transport (\cite{silva12}). 
\item Other comment: First model for flagellar length control.
\end{itemize}  \\

\hline
1.2 & {\bf Revised balance point model:} Fixed number of IFT particles shuttle inside the flagellum. However, the length of the IFT trains, which are formed by arranging the IFT particles in linear arrays, is inversely proportional to the flagellar length. Splitting of longer trains into shorter trains leads to increase in number flux with increasing flagellar length but decrease in the mass flux because of the shorter trains.   It was assumed that some unidentified component of the IFT machinery is present in limited amount. As the flagellum elongates, smaller amounts of this protein come back to the flagellar base via retrograde transport, which results in the production of smaller anterograde IFT trains. & 
\begin{itemize}
\item Proposed by Engel et al \cite{engel09} in response to Dentler’s criticism \cite{dentler05}. 

\item Supported by Vannucini et al. \cite{vannuccini16} who observed that the length of trains shuttling inside the flagellum is inversely proportional to flagellar length.

\item Criticism: Length control mechanism which is responsible for splitting the longer trains into shorter ones is not known. 

\item Evidence against : Stepanek \& Pigino \cite{stepanek16} observed  long trains in fully grown flagella which are mostly stationary.

\end{itemize} \\

\hline

1.3 & {\bf Time of flight mechanism controlling the entry of trains:}  IFT train itself or a molecule bound to it  has two possible chemical (or conformational) states $S_1$ and $S_2$ such that the state $S_1$ makes a spontaneous irreversible transition to the $S_2$  with a rate  $k$. The timer is assumed to be in the state $S_1$ when it begins its anterograde journey from the flagellar base.  This model postulates that the train that has just returned to the flagellar base will undertake another round-trip with its cargo only if it is still in the state $S_1$ when it returns to the same location at the end of its retrograde journey. & \begin{itemize}
\item Proposed by Lefevbre \cite{lefevbre09} and analysed by Ludington et al \cite{ludington15}.

\item Supported by (i) Bhogaraju et al \cite{bhogaraju11}  observed IFT27 acts as a switch that flips between an active and inactive state.

\item Negated by Ishikawa and Marshall \cite{ishikawa17} who observed that on slowing down the IFT train by mutating the dynein increased the import of IFT trains instead of decreasing the import. However, Patra et al \cite{patra20a} reinterpreted the results of  Ishikawa and Marshall  \cite{ishikawa17} and concluded that higher amount of IFT trains on the anterograde track are seen due to high density of anterograde traffic caused by the slow down of  retrograde traffic.
\end{itemize} \\

\hline

1.4 & {\bf Diffusion ruler:} Kinesin motor proteins which are responsible for the anterograde trip of IFT trains, diffuse back to the base after reaching the tip. The retrograde flux decreases with increasing length and this limits entry of IFT trains & \begin{itemize}
\item Proposed by Chien et al \cite{chien17} who observed that the kinesin diffusing back to the base from tip controls the loading of trains. \item Hendel et al \cite{hendel18} developed a theoretical model to complement the finding by Chien et al \cite{chien17} and showed it to be an effective mechanism for length control. 
\end{itemize}\\
\hline

1.5 & {\bf Length sensing based on ciliary electric current:} Current produced by the ion channels located on the ciliary membrane is proportional to ciliary length. The current  blocks the entry of IFT trains into the cilium. & \begin{itemize}
\item Proposed by Besschetnova et al \cite{besschetnova10}  and theoretically analysed by Ludington et al \cite{ludington15}.

\item Supported by Liang et al \cite{liang14,liang18} who observed that  kinesin-II undergo phosphorylation due to the calcium current with increasing flagellar length and this limits entry of trains. 

\end{itemize} \\

\hline
&\multicolumn{2}{c}{}\\
2 & \multicolumn{2}{c} {\bf Control through the differential loading of precursor proteins}\\
&\multicolumn{2}{c}{}\\
\hline
2.1 & {\bf Differential loading of cargo:} Amount of tubulin per IFT train decrease with the increasing flagellar length. Length is sensed using a time of flight mechanism and this controls the loading of precursor into the IFT trains at the base.
& \begin{itemize}

\item Wren et al \cite{wren13} proposed differential cargo loading as an alternative to balance point model \cite{marshall01} and provided experimental evidence for it as well. 

\item Bhogaraju et al \cite{bhogaraju11} and  Huet et al \cite{huet14} proposed that IFT27 is the timer that changes confirmation and modulates the binding affinity to different cargo.

\item Through their model, Patra et al \cite{patra20a}  demonstrated that differential loading combined with a time of flight mechanism could be efficient in controlling flagellar length and could explain related  phenomena in monoglagellates \cite{patra21} and biflagellates \cite{patra20a}.
\end{itemize}
\\
\hline 
&\multicolumn{2}{c}{}\\
3 & \multicolumn{2}{c} {\bf Control through tip stablizing or destabilizing
 proteins}\\
 &\multicolumn{2}{c}{}\\
\hline
3.1 & {\bf Length dependent accumulation of depolymerase KIF13} & While exploring the mechanisms underlying the four different steady state flagellar lengths in the octoflagellate Giardia, McInally observed that the amount of depolymerase KIF13 at each of the tips is inversely proportional to the flagellar length \cite{mcinally19}. \\
\hline
3.2& {\bf Grow and lock model:} After the initiation of ciliogenesis, a locking or capping protein gets activated at the tip after a fixed duration of time which inhibits the flagellum from further elongation and retraction. & Bertiaux et al \cite{bertiaux18} observed that the tip of flagella of steady length cease to elongate or shrink  and become stable. A protein called FLAM8 is responsible for the stabilization. FLAM8 is sent to the  tip / or gets activated at the tip after a certain interval of time. Therefore, if the rate of elongation is slowed by slowing the intraflagellar transport, the tip of the slowly elongating flagellum is locked by FLAM8 after a fixed interval of time starting from the initiation of the ciliogenesis and hence, a shorter flagellum is assembled.\\
\hline
\caption{Proposed mechanisms for flagellar length control. }
\label{flag_len_cont_table}
\end{longtable*}

\section{Modeling regeneration in a multiflagellated cell: shared resources}
\label{sec-biflagRegen}

Regeneration of severed flagella observed first \cite{chen50} almost seventy years ago, motivated many subsequent investigations of the consequences of amputation of flagella of multiflagellated unicellular eukaryotic organisms using several different experimental techniques with increasing sophistication  \cite{lefebvre86}. Pioneering experiments \cite{rosenbaum69,coyne70,dentler77} used either paralyzed strains or applied compression through a coverslip to hold the cells under study for direct viewing (see, for example, \cite{rosenbaum69}). However, both paralysis and compression are likely to affect intracellular processes thereby  causing significant deviation from regeneration under normal physiological conditions. In order to avoid possibilities of such adverse effects, ingenious experimental methods have been developed over the last decade  \cite{ludington12}. 

For simplicity, we consider the scenario where one of the two flagella is selectively severed to a length  $f L^{ss} $  ($0 \leq f < 1$) while the other flagellum remains intact, the special case of this situation corresponding to $f=0$ is usually referred to as ``long-zero case''. One of the most striking observations was the resorption of the unsevered flagellum in the ``long-zero case''. More precisely, after amputation of one of the flagella, the unsevered flagellum was found to resorb rapidly while the severed one began to elongate (see Fig.\ref{fig_flag_reg}(a)). When the resorbing unsevered flagellum and the regenerating amputated flagellum attained the same length, both elongated at the same rate till regaining their original (equal) steady-state lengths. In principle, a cell could sense the damage/amputation of a flagellum by two different pathways: (a) through the external fluid medium: driving fluid flow in the surrounding medium by flagellar beating depends crucially on the undamaged full-length normal flagellum and the loss of this function may be sensed by the cell, or (b) through internal sensing and communication.  
The pathway (a) was ruled out by the experimental demonstration that a paralyzed flagellum, which is disabled to perform its function of driving fluid flow, can still regenerate upon amputation \cite{ludington12}. Therefore, the pathway (b) seems to be used at least in the biflagellated {\it Chlamydomonas reinhardtii}. But, some of the challenging questions on this phenomenon that are still open are:
(i) how does the unsevered flagellum sense the amputation of its partner? (ii) how does it respond to this loss by initiating own initial resorption? (iii) how does it subsequently sense the equalization of its own length and that of its regenerating partner and begin  elongating instead of further shortening, and (iv) how thereafter the two maintain identical growth rate till regaining identical final steady-state lengths?

. 

The experimental observations mentioned above strongly indicate communication between the two flagella through the common shared pool of proteins at the base or in the cell body. But, the identity of the shared proteins remains far from established. Either some or all of the four different types of proteins, that we listed earlier as the key players in flagellar length control, could be the likely candidates for the shared pool; these are (i) the IFT trains, (ii) the motors proteins, (iii) the structural proteins like tubulin, (iv) the depolymerases.

\begin{figure}[htbp]
\begin{center}
{\bf Figure NOT displayed for copyright reasons}.
\end{center}
\caption{{\bf Regeneration of the selectively amputated flagellum: }(a) Regeneration of an amputated flagellum of a   biflagellate (Fig.2B of ref.\cite{ludington12}) and (b) regeneration of a pair of amputated flagella of a  quadriflagellate (Fig.3C of ref.\cite{ludington12}).}
\label{fig_flag_reg}
\end{figure}

\begin{figure}
\includegraphics[width=0.7\textwidth]{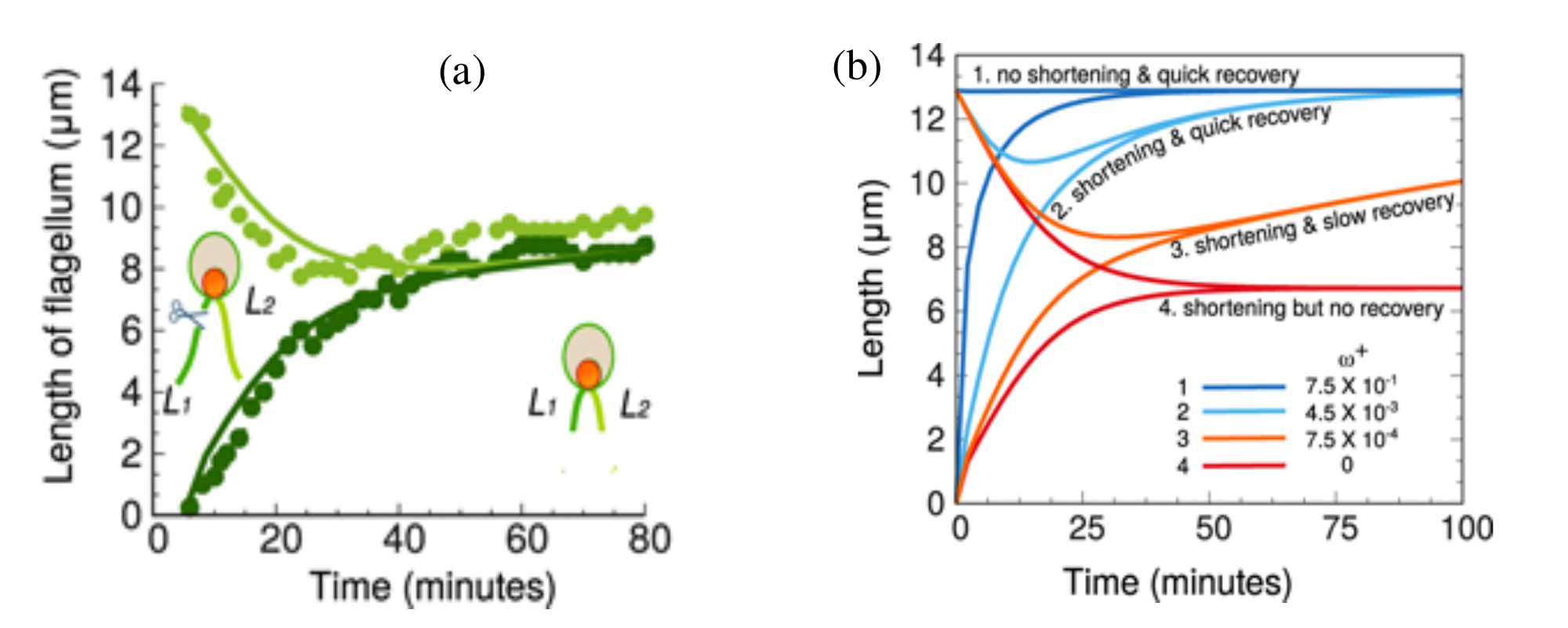}
\caption{{\bf Flagellar regeneration captured by theoretical models:} (a) Regeneration of amputated flagellum of a biflagellate captured by the model proposed by Patra et al (Fig.14B of ref.\cite{patra20a})  (b) Various possible scenarios during the regeneration of amputated flagellum predicted by Patra et al (Fig.12 of ref.\cite{patra20a})}
\label{fig_flag_reg_th}
\end{figure}

What is, perhaps, even more exotic phenomenon is that upon amputation of one flagellar pair of a qudriflagellate the the other pair shortened and then regrew after equalization of their length with that of the regenerating pair ((see Fig.\ref{fig_flag_reg}(b)). Additional questions to be addressed here is the following: how do the two members of the regenerating flagella maintain approximately equal length while the shortening unserved pair also do the same among themselves? The first quantitative study of the kinetics of regrowth of the shortened flagella was reported \cite{lewin53} soon after its first experimental discovery . However, we focus on improved theories developed over the last decade. 

Extending the phenomenological equation (\ref{rate_l2}) (more precisely, the Eq.(1) in the supplementary information of ref.\cite{ludington12}), to a biflagellate, Ludington et al.\cite{ludington12} write down the corresponding equations 
\begin{eqnarray}
\frac{d L_{1}(t)}{dt} &=& k_{1}~  C_{p}~ T(L_{1}) - k_{2}  \nonumber \\
\frac{d L_{2}(t)}{dt} &=& k_{1}~  C_{p}~ T(L_{2}) - k_{2} 
\label{eq-Luding12a}
\end{eqnarray}
In this description, the amount of material in the common pool is measured in the units of flagellar subunits. The rate of depletion of material in the common pool because of flagellar growth is $ k_{1}~  C_{p}~ [T(L_{1})+T(L_{2})]$. On the other hand, rate of gain of the material in the common pool arising from the shrinkage of the flagella is $2k_2$. In addition, it was assumed that, population dynamics of these materials in the common pool arising from the synthesis and degradation are such that, in the absence of flagellar growth and shrinking, the concentration of this material would tend to the steady value of $1$ at the rate $k_3$. Therefore, 
\begin{eqnarray}
\frac{dC_{p}}{dt} &=& 2k_2 - k_1 C_p [T(L_1) + T(L_2)] + k_3 (1-C_{p}).
\label{eq-Luding12b}
\end{eqnarray}
Note that the equations (\ref{eq-Luding12a}) and (\ref{eq-Luding12b}) are deterministic. In order to incorporate the  intrinsic stochasticity of IFT in their computer simulations, the transport rate $T(L)$ was obtained in each time step by multiplying the experimentally measured transport rate with a random number drawn from a uniform distribution  in the unit interval between 0 and 1. With this simulation Ludington et al.\cite{ludington12} could account for the experimental observation on regeneration of amputated flagellum (see Fig.\ref{fig_flag_reg}(a)).

An even more detailed microscopic model mentioned earlier \cite{patra20a} could also account for the same phenomenon (see Fig.\ref{fig_flag_reg_th}(a)). In fact, Patra et al. \cite{patra20a} showed that the nature of the kinetics of regeneration of the amputated flagellum depends on the kinetics of synthesis of the precursor proteins in the common pool and predicted a few different plausible scenarios (see Fig.\ref{fig_flag_reg_th}(b)).

In a multiflagellated cell, apart from sensing and maintaining the length of an individual flagellum, there must be certain mode of communication among the flagella for sensing the lengths also of the others. In most cases, the precursor pool is shared among the flagella and through the pool the flagella communicate among themselves.

\section{Growing and shortening of flagella in multiflagellated cells}
As can be seen in Fig.\ref{Ciliogen_cycle},  the monoflagellate can become transiently biflagellate or triflagellate whereas the biflagellates become quadriflagellate for a brief period during the ciliogenesis of the flagella prior to cell division. After cell division, the flagella grow to their respective steady state lengths. In Table.\ref{tab-trends_ciliogen} we have listed the observed trends of variation of the lengths of the flagella during ciliogenesis  and during regeneration after deflagellation in multiflagellated eukaryotic cells. 

Therefore, the general questions to address is:
how does a cell control and coordinate the growth and shrinkage of its multiple flagella simultaneously.  More specifically, the questions are: (i) what underlying mechanisms are responsible for the observed trends of variation of the different flagella with time? (ii) How does the cell know whether the current process is ciliogenesis or regeneration following deflagellation as the trends of growth are quite different in the two cases. In the Table.\ref{tab-trends_ciliogen}, we have listed some of the theoretical models which could successfully explain the experimentally observed trends.

\begin{longtable*} { |>{\centering\arraybackslash}m{0.2cm} | >{ \arraybackslash}m{2.5cm} |>{\arraybackslash}m{10cm} |>{\arraybackslash}m{4cm}| }
\hline 
 & {\bf ~~~~~~ Trend} & {\bf ~~~~~~~~~~~~~~~~~~~~~~~~~~~~Experiment} & {\bf ~~~~~~~~~~~~~Theory }\\
\hline
1 & \parbox[r]{1em}{\includegraphics[width=1in]{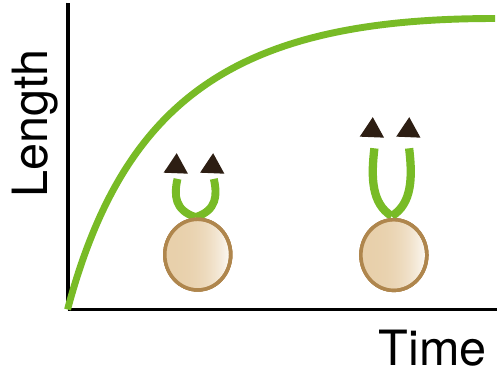}} & The pair of flagella of {\it Chlamydomonas reinhardtii} grow together during
\begin{itemize}
\item the ciliogenesis after cell division \cite{marshall01,rosenbaum69}.
\item the regeneration after deflagellation \cite{rosenbaum69}.
\end{itemize} &  Patra et al \cite{patra20a}, Marshall and Rosenbaum \cite{marshall01}, Hendel et al \cite{hendel18}, Ludington et al \cite{ludington15}\\
\hline 
\hline
2 & \parbox[r]{1em}{\includegraphics[width=1in]{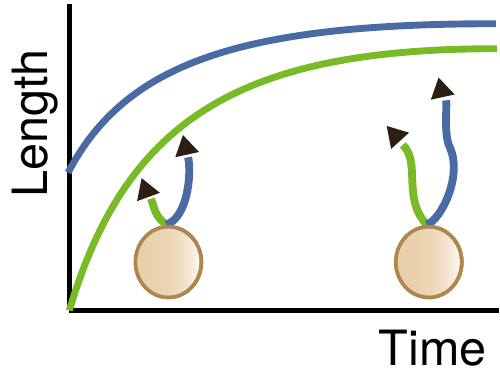}} & The daughter cells of {\it Nephroselmis stein} inherit one long and one short flagella which have not reached their full steady state length. Both  flagella keep elongating with the length dependent rates \cite{melkonian87}. & No theory yet.\\
\hline 
\hline
3 & \parbox[r]{1em}{\includegraphics[width=1in]{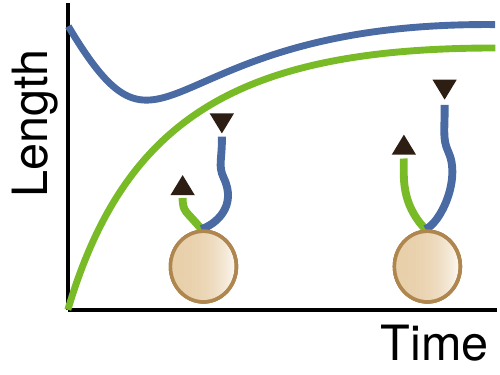}} & \begin{itemize}\item {\it Chlamydomonas}: During the regeneration of the flagellum of {\it Chlamydomonas} which is selectively amputated \cite{rosenbaum69,ludington12} the existing flagellum shrinks while the other grows. After they attain the same length they grow further together.
\item {\it Spermatozopsis}: During the ciliogenesis of a new shorter flagellum in the daughter {\it Spermatozopsis} cell which inherits one long and one short flagellum \cite{lechtreck97}.
\item {\it Leptomonas}: During the replication of flagellum in the monoflagellate {\it Leptomonas } cell which becomes transiently biflagellated prior to cell division \cite{he19,patra21}.
\end{itemize} & \begin{itemize} \item  {\it Chlamydomonas}: Patra et al \cite{patra20a}, Fai et al \cite{fai19} and Hendel et al \cite{hendel18}.
\item {\it Spermatozopsis}: No theory yet.
\item  {\it Leptomonas}: Patra and Chowdhury \cite{patra21}.
\end{itemize}\\
\hline 
\hline
4 & \parbox[r]{1em}{\includegraphics[width=1in]{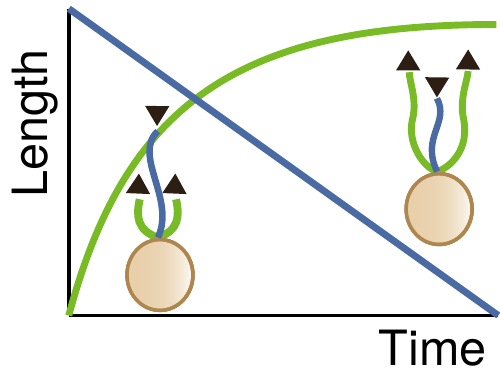}} &  Prior to the cell division, a pair of new flagella emerge
and elongate in {\it Peranema trichophorum} \cite{tamm67} and {\it Pseudopedinella elastica} \cite{heimann89} while the older flagellum undergo resorption.
 &   Patra and Chowdhury \cite{patra21}\\
\hline 
\hline
5 & \parbox[r]{1em}{\includegraphics[width=1in]{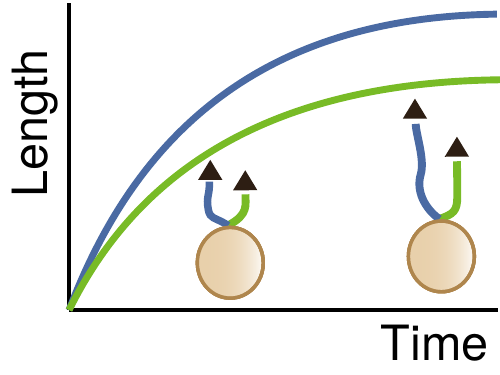}} & The two flagella of different length of {\it Nephroselmis stein} after deflagellation elongate with different rates \cite{melkonian87}.& No theory yet.\\
\hline
\hline
6 & \parbox[r]{1em}{\includegraphics[width=1in]{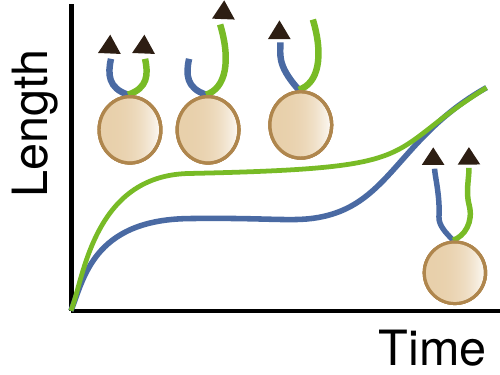}} & The flagella of {\it Volvox carteri}, grow together during the initial phase. Then one flagellum grows longer while the other one pauses. Later the growing one pauses while the pauses one resumes growing and catches up with the
longer paused flagellum. After the length equalization both the flagella start growing together \cite{coggin86}. & No theory yet but the emergence of the lag behaviour in the context of a single flagellum is explained by Patra et al \cite{patra20a} .
\\
\hline
\hline
7 & \parbox[r]{1em}{\includegraphics[width=1in]{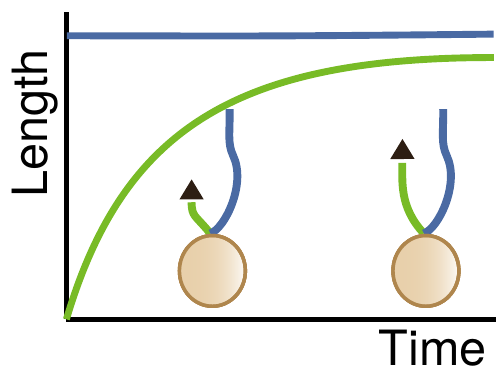}} & During the pre division when the flagellar replication occurs in {\it Pedinomonas tuberculata} , the older flagellum maintains its length during the elongation of the new flagellum \cite{heimann89}. & {\it Pedinomonas tuberculata}: Patra and Chowdhury \cite{patra21}
\\
\hline
\hline
8 & \parbox[r]{1em}{\includegraphics[width=1in]{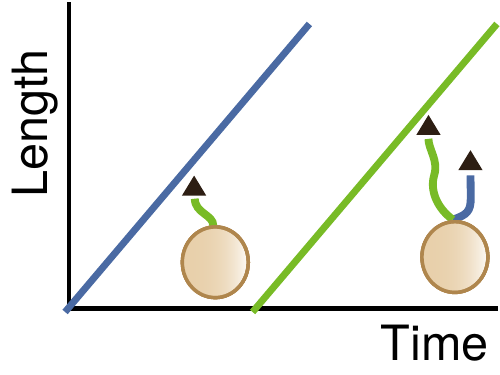}} & During the regeneration after deflagellation in {\it Spermatozopsis}, both the longer and shorter flagella elongate with the same rate but the shorter ones start
regenerating after a lag phase \cite{schoppmeier03}. & No theory yet.
\\
\hline
\hline
9 & \parbox[r]{1em}{\includegraphics[width=1in]{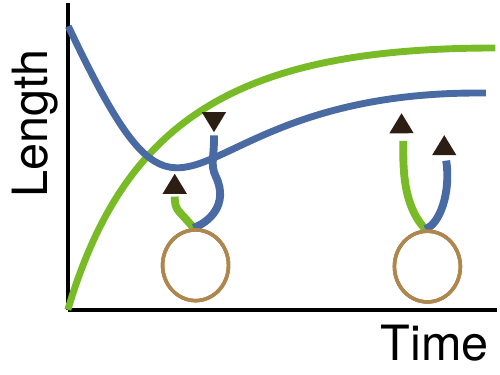}} & The daughter cells of {\it Epiphyxis pulchra} inherit one new flagellum undergoing ciliogenesis and an old flagellum undergoing retraction. But the old flagellum undergoes partial retraction and regrow to attain a shorter steady state length while the new flagellum becomes the longer one \cite{wetherbee88}. & No theory yet.
\\
\hline 
\caption{Growing and shortening trends of the flagella in the monoflagellated and multiflagellated eukaryotic cells}
\label{tab-trends_ciliogen}
\end{longtable*}

\section{Beyond mean values: implications of length fluctuations and correlations}

\subsection{Length fluctuations of a flagellum in the steady-state: Level-crossing statistics}

Because of the intrinsic stochasticity of  incorporation of fresh subunits  at the flagellar tip and that of the ongoing turnover of the building blocks from the tip  as well as the randomness in the synthesis and degradation of the precursor proteins in the pool at the base of the protrusion, the length of the flagellum fluctuates. As a result, the tip performs, effectively, a confined random walk even after the mean length achieves a constant value in the steady state. Fluctuations drive the instantaneous length away from mean value whereas the active length sensing mechanisms restores the length back to mean level and this gives rise to a confined random walk. The motion of the tip can be mapped onto that of a hypothetical Brownian particle subjected to linear restoring drift and hence, best described by the Ornstein-Uhlenbeck process
\begin{equation}
d L=-\frac{1}{\gamma}Ldt+\sqrt{2D}dW(t)
\end{equation}
where $\gamma$ is the corresponding relaxation time and $D$ the diffusion constant. The macroscopic parameters give a coarse grained description of the flagellar length fluctuations. They can be related to microscopic elements like the traffic properties of IFT, the kinetics of assembly and disassembly, using more detailed models. 

As discussed in section \ref{sec-length_fluctuations}, the statistics of various level crossing quantities can be estimated for flagellar length fluctuations \cite{patra20b}. 
Measuring such quantities can help identifying the mechanisms of flagellar length control. At present there are few distinct models, based on alternative scenarios described in section \ref{sec-flag-len-cont},  all of which can account for the experimentally observed time-dependence of the mean flagellar length. In case some distinct models predict the same mean steady-state length but different level crossing statistics in the steady state, it could be possible to rule out at least one (or more) of these alternatives, thereby narrowing the set of the possible models, by comparing the theoretically predicted level crossing statistics with the corresponding experimental data. In other words, the level-crossing statistics will impose additional stringent tests for the validity of the models of flagellar length control \cite{patra20b,patra20a,ludington15}.

\begin{figure*}[htbp]
\begin{center}
{\bf Figure NOT displayed for copyright reasons}.
\end{center}
\caption{{\bf Level crossing statistics of flagellar length fluctuations:} (Adapted from Fig.3 of ref.\cite{patra20b}) (a) Mean escape time as a function of the width of the zone. (b) Estimates of mean upcrossing and downcrossing times. (c) Number of peaks per unit time as a function of sojourn time beyond three thresholds. (d) Mean range scanned and (e) the maximum and the minimum length attained by the flagellum whose length fluctuates about the mean value in the steady state.  }   
\label{flagellum_level_cross_stat}
\end{figure*}

\begin{figure*}[htbp]
\begin{center}
{\bf Figure NOT displayed for copyright reasons}. 
\end{center}
\caption{ {\bf Correlation in flagellar length fluctuations:} (a) Nature of  correlation  in flagellar length fluctuations (a1) anticorrelated  (a2) correlated (a3) uncorrelated. (b)  Scatterplot shows fluctuations in the lengths of both flagella in a {\it Chlamydomonas} cell during a 10-min time-step.$\Delta$L1 and $\Delta$L2 indicate the change in length of the two flagella in a cell during a single 10-min time interval, plotted one against the other. (Fig. 2C of ref.\cite{bauer20}). Length fluctuations in {\it Chlamydomonas} are correlated.(c) Correlations in length fluctuation between the regenerating flagellum which is selectively amputated  and the  intact flagelllum  in  {\it Chlamydomonas} cell. Initially the pool gets exhausted and the length fluctuations are anticorrelated indicating symbiotic dependence but later they become uncorrelated indicating the  pool getting replenished with the synthesis of fresh precursors. (Fig. 15(b) of ref.\cite{patra20a}) (d) Correlations in length fluctuation between mother and daughter flagelllum during its replication in  {\it Leptomonas} cell which is a monoflagellate cell but become biflagellate prior to the cell division. Initially the length fluctuations are anticorrelated indicating symbiotic dependence but later they become uncorrelated indicating complete separation of pool and completion of cell division. (Fig. 8B of ref.\cite{patra21}) }    
\label{fig_flag_corr}
\end{figure*}

\subsection{Correlated length fluctuations in multiflagellates: ciliogenesis and regeneration}
Measuring the correlations in length fluctuations of two flagella of the same cell (see Fig.\ref{fig_flag_corr}(a)), we can probe the cooperativity among multiple flagella of the cell and infer rules of sharing precursors and other essential proteins during different stages. Here we mention some of the recent developments that shed light on both the theoretical and experimental fronts. 

(i) {\it Correlated length fluctuations:} The experimental data published by Bauer et al indicated that the correlations are positive, indicating that some common factor controls length \cite{bauer20} (see Fig.\ref{fig_flag_corr}(b)).  One possibility is that due to the translational bursting, the amount of precursors available in the basal pool changes in bursts. Following such a bursting event there is increased supply of precursors in the pool that allows both the flagella to grow simultaneously importing precursors which could be the reason for positive correlations of their length fluctuations. Growth of the flagella leads to depletion of precursors in the common basal pool and, hence, inability to import precursors to replace building blocks that were removed due to turnover at the tips. Therefore, under such circumstances, both the flagella could shorten simultaneously resulting in a positive correlation in the length fluctuations. Length fluctuations of the pair of daughter flagella undergoing ciliogenesis prior to  cell division in  monoflagellates are also positively correlated \cite{patra21}.

(ii) {\it Anticorrelated length fluctuations:} 

Under several different circumstances, there may be huge mismatch between the demand and supply of the structure-building components of the two flagella of the same cell. Under such circumstances one flagellum can grow at the expense of another by exchanging subunits  through their common basal pool, resulting in anticorrelation between their length fluctuations. Examples of such circumstances include regeneration of a selectively amputated flagellum \cite{patra20b} (see Fig.\ref{fig_flag_corr}(c)) or the replication of a daughter flagellum \cite{patra21} (see Fig.\ref{fig_flag_corr}(d).

(iii) {\it Uncorrelated length fluctuations:} Prior to cell division, when monoflagellate cells are in the multiflagellated phase, the new daughter flagella emerge from a common pool shared also with the existing mother flagellum. Depending on the pair of flagella, initially the correlation in their length fluctuations can either be positive or negative. But slowly, the correlations die or the fluctuations become uncorrelated, indicating a separation of the pools \cite{patra21} (see Fig.\ref{fig_flag_corr}(d)). 

In the future, measurement of  correlations of length fluctuations could reveal how anisokont octoflagellated cells like {\it Giardia} \cite{mcinally19} assemble and maintain their multiple flagella of unequal length  and what the rules for molecular communication among the flagella are.

\section{Open questions for multiflagellated and ciliated eukaryotes}
In section \ref{sec-biflagRegen} we have reviewed the current status of understanding of flagellar length control in multiflagellated eukaryotes. However, all the theoretical work so far has focussed almost exclusively on isokont biflagellates. In this section we list the challenging open questions on flagellar length control in anisokont biflagellates as well as those in cells with larger number of flagella and in ciliated cells.

\subsection{Maintaining flagella of different length in anisokont biflagellates}

How do anisokont biflagellated cells maintain flagella of different lengths simultaneously during the interphase? 
Although answer to this question remains elusive, we list here some possible mechanisms that such a cell might use to achieve this feat:\\
(i) {\bf Separate pools for different flagella:} Having separate pools for each of the flagella, which have different amounts of precursors and IFT trains, would result in different flux of loaded IFT trains in each of the flagella which, in turn, could maintain different flagella at different lengths in steady state. \\
(ii) {\bf Variation in the transition zone:}  If the transition zones of the flagella vary from one flagellum to the other causing variation in the rates of entry of the loaded trains, then flagella of different steady-state lengths could also be assembled.\\
(iii) {\bf Post-translational modification of axonemal proteins:}  If the axonemal proteins at the flagellar tip undergo certain post-translational modifications that alter the polymerization and/or depolymerization kinetics of the of the tip, flagella of different lengths in steady-state could be maintained.\\
(iv) {\bf Different length sensors:}  Having rulers with different length sensing kinetics for different flagella could also help in maintaining flagella of different lengths. For example, timers with different flipping rate (as used in model based on time-of-flight), or diffusive rulers with different diffusion constants (as used in diffusive ruler model) are some of the viable options.
\\
(v) {\bf Different amount of tip stabilizing or destabilzing proteins:}  If different amount of MT stabilizing proteins (like capping molecules) or MT destabilizing proteins (like depolymerase) are present at the tip of different flagella, then flagella of different length could be observed in the interphase. 
 
For detailed discussion, readers are referred to the recent review article by Bertiaux and Bastin \cite{bertiaux20}.

\begin{figure}
\includegraphics[scale=0.60]{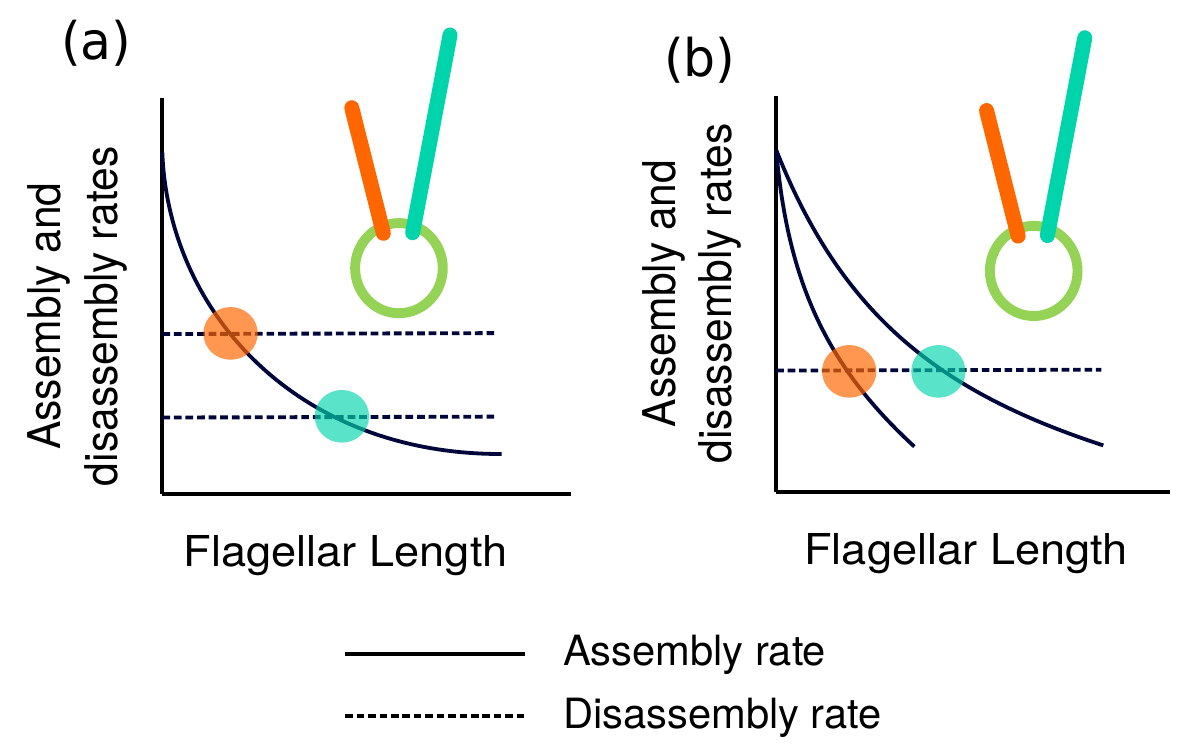}
\caption{{ Balance point for anisokonts}   }
\label{bpm}
\end{figure}

\begin{figure}
\includegraphics[scale=0.7]{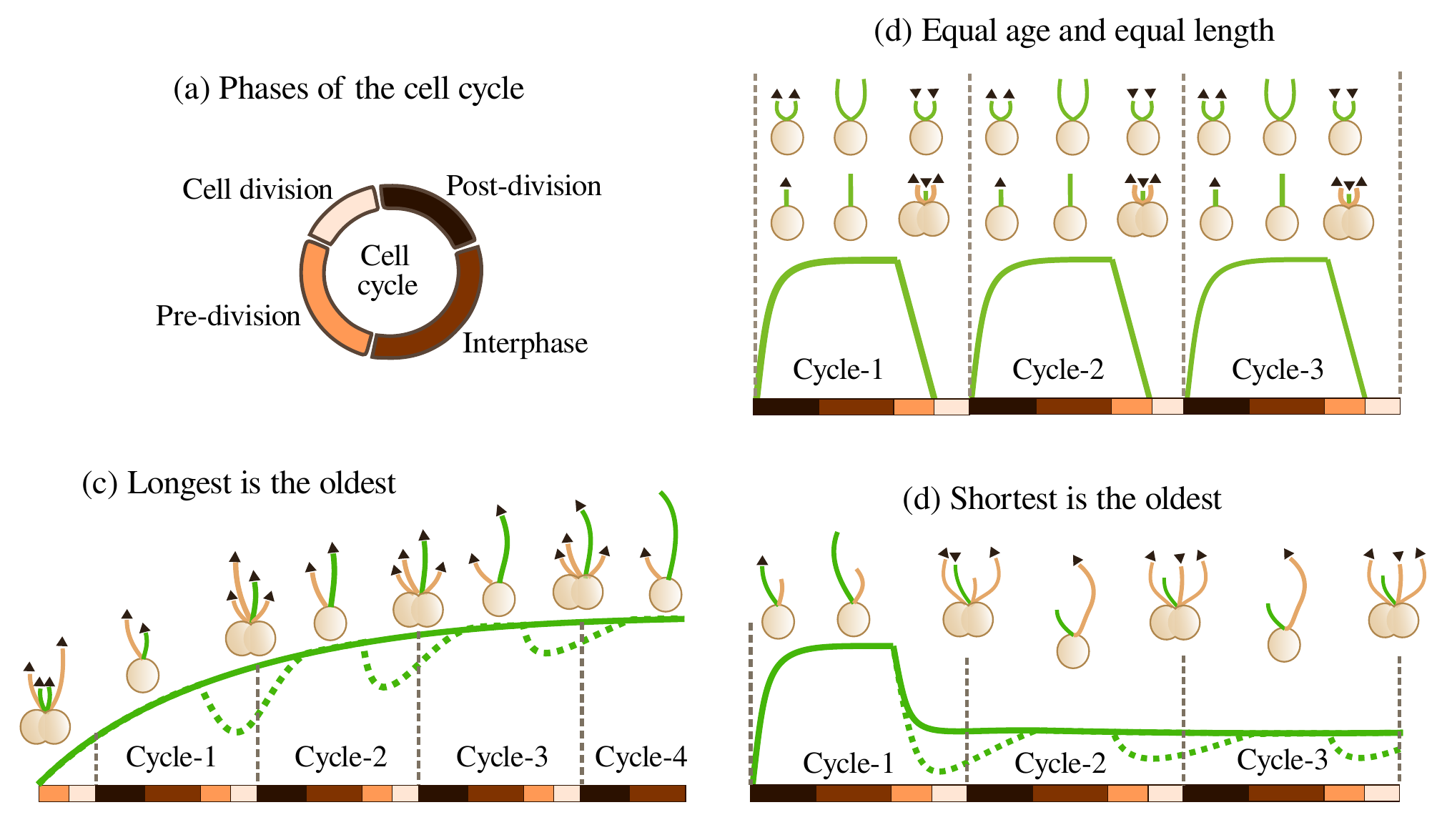}
\caption{{\bf Age-length dependency:} (a) Color code marking the different phases of a cell cycle. (b-d) Temporal evolution of a selected flagellum (depicted in green color) of a multi-flagellated cell over multiple cell cycles. (b) All flagella during the interface are of the same length and age and are normally a single cell cycle old. (c) The longest flagellum is the oldest which gets inherited by one of the daughter cells during the cell division and matures over multiple cell cycle. (d) The shortest flagellum is the oldest which gets inherited by one of the daughter cells during the cell division and keeps to attain the final length. (c-d) Both the longest and the shortest flagellum which have attained their final length may partially shorten during the cell division as indicated by the green dotted lines in (c) and (d). }
\label{fig-flagella-age}
\end{figure}

\subsection{Age length dependency in multiflagellates}      
As depicted in Fig.\ref{fig-flagella-age}(a-d), we can classify the flagella age-length relations into the following three categories : \\
(i) {\bf Flagella are of the same length and age :} As seen in {\it Chlamydomonas} \cite{rosenbaum69}, both the flagella elongate together in the newborn daughter cells to attain equal steady state length and before the next round of cell division, these two flagella undergo resorption together. Hence, both the flagella are of same length and age (Fig.\ref{fig-flagella-age}(b)). 
Moreover, unless they have undergone resorption (or amputation), followed by regeneration, both the flagella have the same age as the cell.\\
(ii) {\bf Longest is the oldest :} The flagellum which emerges prior to the cell division of the mother cell of the first generation, is the shortest one in the second generation daughter cell which inherits it and then becomes the longer one in the third generation of daughter cells. It could keep elongating till the completion of 2-3 subsequent cell cycles and finally attains a steady state matured length. Such thing is observed in case of {\it Nephroselmis olivacea} \cite{melkonian87} (solid green curve in Fig.\ref{fig-flagella-age}(c)). In case there is shortage of precursors during the cell division, the flagellum could undergo partial shortening while the new flagella assemble as in seen in case of {\it Spematozopsis} (dashed green curve in Fig.\ref{fig-flagella-age}(c)) \cite{lechtreck97}. (See the temporal evolution of the selected flagellum shown in green color in Fig.\ref{fig-flagella-age}(c).) \\
(iii) {\bf Shortest is the oldest :} Unlike the previous case, the flagellum which emerges prior to the cell division of the mother cell of the first generation, elongates to become the longest one in the second generation daughter cell which inherits it but thereafter it starts to shorten and may attain its final steady state length after 2-3 rounds of cell cycle. The shortening flagellum ultimately becomes the shortest and the oldest one as reported in {\it Giardia} \cite{dawson10} (solid green curve in Fig.\ref{fig-flagella-age}(d)). The  partial  shortening during the pre-division period may be caused by the exhaustion of the precursors because of the assembly of new flagella, as seen in {\it Epiphyxis pulchra } \cite{wetherbee88} (dashed green curve in Fig.\ref{fig-flagella-age}(d)). See the temporal evolution of the selected flagellum shown in green color in Fig.\ref{fig-flagella-age}(d). \\


\begin{figure}[htbp]
\begin{center}
{\bf Figure NOT displayed for copyright reasons}.
\end{center}
\caption{{\bf Ciliated cells :} (a) Spatial arrangement of cilia in (a1) unicellular microorganisms (website of Guillermina Ramirez) (a2) Olfactory epithelium of a rat (Fig.3(a) of ref.\cite{challis15}). (b) Mechanisms for controlling cilia number (b1) in a dividing {\it Tetrahymena} cell (Fig.5(a) of ref.\cite{soares19}) (b2) and in a proliferating multiciliated cell (Fig.1(a) of ref.\cite{nanjundappa19}). }
\label{fig-cilium_number}
\end{figure}

\subsection{Control of length and spatial positioning of cilia}

As shown in Fig.\ref{fig-cilium_number}(a1), in unicellular organisms, the cilia are arranged in specific patterns. Ciliary length varies across the tissue or organ and sometime have temporal dependence also (see Fig.\ref{fig-cilium_number}(a2)). These spatio-temporal variation are indirect indication that in a healthy cell or body, building cilia of correct length is very important. If they are too short, they may not be able to collect enough signalling molecules from the surroundings which the cell needs for activating further cascades. If they are too long, it will be a loss of cell's energy as well as collecting too much data by longer cilia can harm or tire the cell. Incorrect length may also hamper the spatiotemporal distribution of various fluids which the motile cilia help to circulate.  In Table.\ref{tab-cilia-length}, we have listed the diverse kind of cilium present on the mammalian cells, their length, functional role etc.

Different cells take different routes in order to ensure that the ciliated cells are equipped with the correct number of cilia.  We have listed some of these:\\
{\bf (i) Replication followed by distribution:} If a ciliated cell bears a certain number of protrusions during its interphase, then during cell divisions, the cell must manage how its protrusions are distributed among the new daughter cells. The mother cell doubles the number of protrusions prior to cell division and the two daughter cells receive equal number of protrusions post cell division. Thus, the age of the protrusions beared by the cell may vary. The cilia replication prior to cell division in {\it Tetrahymena} is shown in Fig.\ref{fig-cilium_number}(b1). The process is quite complicated in the case of ciliated {\it Paramecium} cell in which the cilia distribution on the surface is not symmetric \cite{pearson09} and length of cilia in different regions of the cell surface also vary. For discussion of cilia replication and distribution in unicellular microorganisms, readers are referred to the review article by Soares et al \cite{soares19}. \\
{\bf (ii) Amplification followed by distribution:} Cells with hundreds of protrusions like postmitotic multiciliated cells, the multiple protrusions are built by specialised structures called deuterosomes which quickly nucleate multiple centrioles from the cytoplasm (see Fig.\ref{fig-cilium_number}(b2)). Later these centrioles are released and get docked at the cell surface where they initiate ciliogenesis \cite{dehring13, nanjundappa19}.

\begin{longtable*} { |>{\centering\arraybackslash}m{1.0cm} |>{ \arraybackslash}m{3cm} |>{\arraybackslash}m{2 cm}  |>{\arraybackslash}m{5cm} |>{\arraybackslash}m{6cm}|}
\hline 
{\bf Sl.no} & {\bf  Name of cilia} & {\bf Organ} & {\bf Length ($\mu m$)} & {\bf Function} \\
\hline
1 & Olfactory cilia \cite{challis15} & Nose & 3 (Posterior septum)  &  Sensing odorants\\
& & & 10 (Mid region)
 &  Clearance of mutants\\
& & & 15 (Anterior septum) & \\
& & & 
30 (Dorsal recess) & \\
\hline
2 & Kinocilium & Ears & & Sound transduction \\
\hline
3 & Ocular cilia \cite{may17} & Eyes & 2-7 (Corneal endothelial cells) & Sensing the \\ & & & 5-6 (Trabecular Meshwork) & composition of intraocular fluid \\
& & & & changes in vascular pressure \\
& & & & light signals \\ 
\hline
4 & Airway Cilia  & Respiratory  & 6 (Trachea) & Air circulation \\
& & tract & 4.7 (Third generation bronchi) & Mucociliary clearance \\
& & & 3.9 (Fifth generation bronchi) & \\
& & & 3.7 (Seventh generation bronchi) & \\
\hline
5 & Chondrocyte cilia \cite{yuan15} & Cartilage  & 1.1 (Superficial layer) & Cartilage organization\\ 
& & and bones & 1.4 (Deep layer) & Skeletal patterning \\
\hline
6 & Cholangiocyte & Liver & 7.3 (Large bile duct) & Osmosensation \\
& primary cilia \cite{huang06, masyuk08} & & 3.5 (Small bile duct) & Chemosensation\\
\hline
7& & Kidney & & Sensing urine flow \\
\hline
8 & & Heart & & Orchestrating of cardiac left-right symmetry\\
\hline
{\bf Sl.no} &  & {\bf Organism} & {\bf Length ($\mu m$)} & {\bf Function} \\
\hline
9 & & Tetrahymena \cite{rajagopalan09} & 4.2 (Wildtype) & Swimming \\
& & & 3.4 (KO-DYHC2 mutant) & Phagocytosis\\
\hline
10 & & Paramecium \cite{valentine12} & 11.7 &  Swimming \\
& & & & Phagocytosis \\
\hline
\caption{Diverse kind of cilia associated with mammalian cells and uniccelular microorganisms. }
\label{tab-cilia-length}
\end{longtable*}

\part{Other non-ciliary protrusions}

The focus of the part I of this article was to review various generic aspects of length control, namely sensing and regulation of length, genesis and regeneration of long cell protrusions as well as communication between different protrusions of a cell through their bases and cell body. One particular example of such a protrusion is eukaryotic flagella (cilia), an appendage that has beed used extensively in experimental studies of various aspects of length control. 

As a case study with a specific example, in part II of this article we have critically reviewed in detail all the theoretical models developed for understanding the experimentally observed phenomena and quantitative results for eukaryotic flagella. The scope of part III is, however, very limited. Here we compare and contrast the mechanisms of control of the length of a few non-ciliary protrusions of eukaryotic cells and some common protrusions of prokaryotic cells with the mechanisms discussed in parts I and II. We hope that these comparisons 
the mechanisms of length control in different protrusions will promote exchange of ideas across different sub-disciplines thereby enriching science as a whole.


\section{Protrusions in eukaryotes}
\subsection{Microtubule based protrusions: axon length sensing by a neuron}

Most of the mechanisms of length sensing that work satisfactorily for protrusions of the order of tens of microns may be too noisy to give reliable estimates of length in case of protrusions like axons that can be as long as meters. Another additional complexity of length sensing arises when axon of a cell hits its target early in the stage of development; in that case the axon has to stretch as the distance between its two ends grow with the ongoing  developmental growth of the organism.  Hence, as summarized in Fig.\ref{fig-axon}(a) and in the review article \cite{albus13}, the axon may use different mechanisms during different stages of growth. Here we discuss the detailed mechanisms based on motor driven oscillations for the length sensing of the axons. 

The molecular motors plying between the axon terminal and the cell body are in charge of  controlling certain chemical oscillations in the cell body \cite{albus13} as shown in Fig.\ref{fig-axon}(b). The frequency of these oscillation ($\omega$) is a function of the time of flight of the motors moving with velocity ($v$) from the axon tip to the cell body and the time of flight ($T_\text{tof}=L_\text{axon}/v$) is proportional to the axon length ($L_{axon}$). The frequency of these oscillations is an indicator of the axon length. The key physical principle underlying this mechanism is the well known fact that negative feedback loops with a {\it time delay} can result in oscillating signals \cite{paszek10}. This concept was extended in the context of axonal length control \cite{rishal12} by explicitly demonstrating that if the time delay required for such oscillation is provided by motor-dependent axonal transport then the length-dependent frequency of temporal oscillation can also encode spatial information.

In the model developed by Folz et al.\cite{folz19} motors are assumed to transport molecular signals $I$ and $O$ along microtubules at velocity $v$, from the soma to the growth cone (input) and from the growth cone to the soma (output), respectively (see Fig.\ref{fig-axon}(b1)). Let $C_{I}(t)$ denote the concentration of an incoming signaling molecules $I$ at the growth cone at time $t$; it triggers the transport of outgoing signal molecules $O$ from the soma to the growth cone. Let $C_{O}(t)$  denote that of the concentration of the output signal molecules at the soma at time $t$ and it suppresses further transport of incoming signal molecules $I$. Signal $I$ at the growth cone is assumed also to stimulate  actin polymerization at the leading edge of the growth cone as well as to inhibit the response $R$ where $R$ itself induces actin network contraction (see Fig.\ref{fig-axon}(b1) for details).

\begin{figure}[htbp] 
\begin{center}
{\bf Figure NOT displayed for copyright reasons}.
\end{center}
\caption{{\bf Axon:} (a) Different mechanisms of length sensing and length control in axon during its different stages of development. (Fig.3 of ref.\cite{albus13}). (b) Model of axon length sensing. (b1) Schematics of the stochastic model. (Fig.1 of ref.\cite{folz19}). (b2) Dependence of chemical oscillations on axonal length .  }
\label{fig-axon}
\end{figure}

The axon length is assumed to be regulated by two mutually opposite processes \cite{folz19}. The extension axon at its leading edge is driven by actin polymerization whereas its shrinkage is caused by the actin network contraction induced by contractile force generated by ATP-powered motors in the axon. Recalling that actin polymerization is stimulated by $I$ and actin network contraction is induced by $R$, in the simplest formulation, one assumes that  $v_g C_{I}$ (with $v_{g} > 0$) and  $v_s C_{R}$ (with $v_{s} > 0$) are the velocities of growth and shrinkage, respectively, of the axon. Note that $\tau=L/v$ is the duration of motorized travel of signal along an axon of length $L$ with a constant velocity $v$. Therefore, in this model the dynamical evolution of the axonal length is governed by the equation \cite{folz19}
\begin{eqnarray}
\frac{dL}{dt} = v_g C_{I}(t) - v_s C_{R}(t)
\label{eq-FolzLength}
\end{eqnarray} 
and the time-dependence of $C_{I}(t)$, $C_{R}(t)$ and $C_{O}(t)$ are given by the equations
\begin{eqnarray}
\frac{dC_{R}(t)}{dt} &=& J [1-f_{\kappa}(C_{I}(t))] - \gamma C_{R}(t) \nonumber \\
\frac{dC_{I}(t)}{dt} &=& J_{max} - J f_{\kappa}(C_{O}(t-\tau)) - \gamma C_{I}(t) \nonumber \\
\frac{dC_{O}(t)}{dt} &=& J f_{\kappa}(C_{I}(t-\tau)) - \gamma C_{O}(t) 
\label{eq-FolzSignals}
\end{eqnarray}
where $J_{max}$ is the maximum incoming flux of $I$, and $J$ is the strength of the coupling between the pairs $(I,R)$ and $(I,O)$ while $\gamma$ is decay rate of all the three signals (assumed to be equal only for convenience).
The Hill function 
\begin{equation}
f_{\kappa}(c) = \frac{c^n}{\kappa^n+c^n}, 
\end{equation}
with sufficiently high value of the index $n$, is intended to mimic a switch.  The delay $\tau$ entering the equations are characteristic features of delay-differential equations. Numerical solution of the coupled four equations (\ref{eq-FolzLength})-(\ref{eq-FolzSignals}), for the given initial condition $L(t=0)=0$, shows that the qualitative trend of growth of the axonal length is very similar to that we have seen earlier for flagellar length. However, in the long time limit, the axonal length keeps oscillating with an amplitude that is less than $1\%$ of the average length. 

The retrograde signal $C_{O}$ oscillates with a length-dependent frequency (see Fig.\ref{fig-axon}(b2)) whose approximate analytical expression is \cite{folz19} 
\begin{equation}
\omega \sim \left( \sqrt{\frac{L_\text{axon}}{v} \left(\frac{L_\text{axon}}{3v}+\frac{1}{\gamma} \right) } \right)^{-1}~.
\label{eq-axon}
\end{equation}
The Eq.(\ref{eq-axon}) implies decreasing frequency of oscillation with increasing length of the axon. 
This expression shows explicitly how the length of the axon is `encoded' in the frequency of oscillation of the retrograde signal just as the length of a flagellum is encoded in the expression $t_{\text{ToF}}= (L(t)/v_a) +(L(t)/v_r)+\tau$ for the time of flight. 

It is interesting to note that frequency-encoded signals are exploited to estimate distance in radar and sonar technologies where those signals are referred to as `chirp' to distinguish them from `bang' (pulse) signals \cite{bloch73}. Moreover, there is an analogy between the model of Folz et al.\cite{folz19} and Laughlin's proposal \cite{laughlin15}  on how clocks can function as rulers. In the latter, which exploits the language of lasers, length regulation is argued to be possible by resonant chemical waves.  Terms corresponding to growth and shrinkage of an axon are analogs of the gain and loss terms of a laser. In both the systems, as the system `tunes' its length it reaches steady state where two competing terms balance each other. This happens concomitant with not only the selection of a specific frequency but also phase locking of different oscillating signals ($C_{I}(t)$ and $C_{O}(t)$ in Folz et al.'s model), the latter being analogous to a cavity resonance in Laughlin's Laser-based scenario \cite{laughlin15}. Two challenging open questions need to be addressed: (a) the identify of the signal molecules, and (b) how the neuron decodes the length encoded in the frequency. \\

\subsection{Actin based protrusions}

As stated earlier in this article, the primary focus of our detailed review are the microtubule-based tubular protrusions. Our discussions on actin-based protrusions, however, are not intended to be equally comprehensive overview. Instead, the main aim of this subsection is to highlight some of the special features of the common actin-based protrusions that make their length control mechanisms more challenging than those of the microtubule-based protrusions. {\it Sheet-like} actin-based protrusions, like lamellipodia  \cite{krause14}, that have branched networks of actin at their leading edge, will not be discussed in this review. Two concrete examples of actin-based {\it finger-like} cell protrusions that we consider, namely stereocilia and microvilli, consist of essentially tubular outgrowths from the cell membrane.  The main motivation for choosing these two types of protrusions from among the varieties of actin-based protrusions is that multiple copies of each of these two types of protrusions cluster together to form higher-order structures. How a cell maintains the observed relative lengths of the protrusions in a given cluster is a fundamental question in this context.

\begin{figure}
\includegraphics[width=0.65\textwidth]{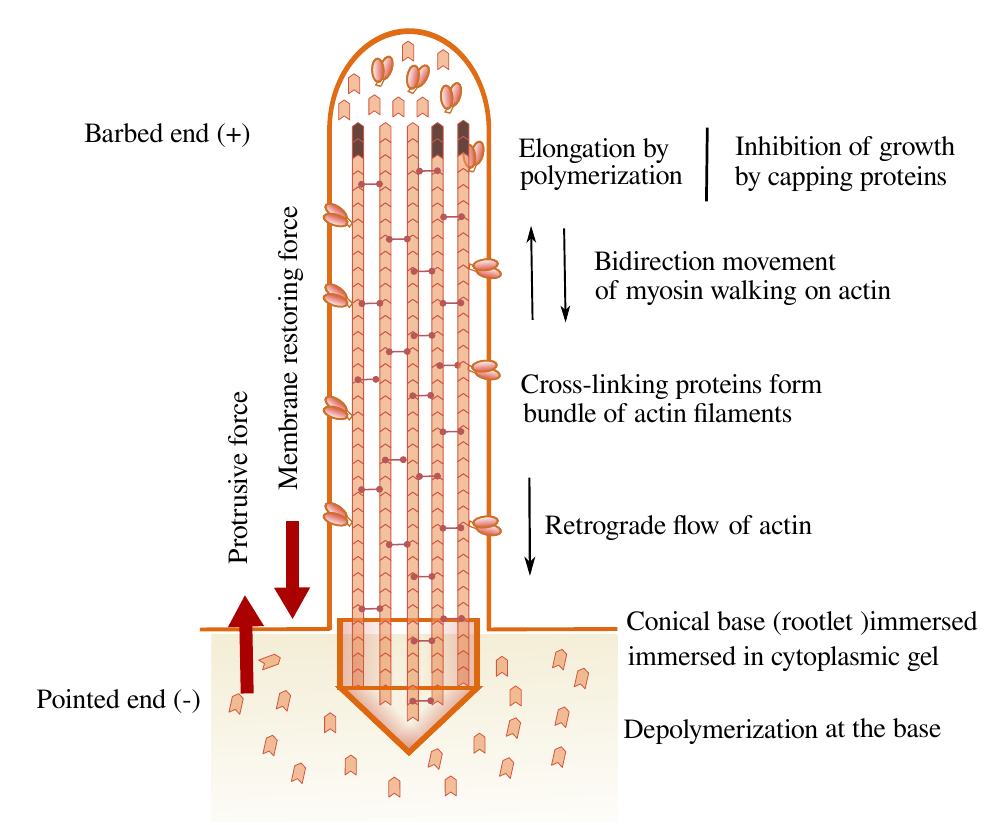}
\caption{{ Actin based protrusion: Sites of polymerization and depolymerization, the modes and direction of transport of precursors (actin monomers) for the elongation and maintenance of the protrusions and direction of various forces emerging from the interaction of various elements in the protrusion. }}
\label{fig-actin-based}
\end{figure}

 In spite of their structural and functional diversity, stereocilia and microvili share several common features in their length control mechanisms. 
The barbed ends of the actin filaments (analog of the plus ends of microtubules) are at the tip of the finger-like protrusions while their pointed ends (analogue of the minus ends of microtubules) are at the base (Fig.\ref{fig-actin-based}). In both types of protrusions, the tip is crowded with various proteins that are believed to have important regulatory roles. 
The linear actin filaments are assembled into bundles of different dimensions in the different protrusion types to provide the appropriate mechanical properties required for their respective dynamics and biological functions.  Although both types of protrusions stabilize the bundles by crosslinking the actin filaments, the specific cross-linking molecules used by different types of protrusions and the frequency of crosslinking can be different. The packing density of the filaments in different types of protrusions are also different. 

These actin-based protrusions are also dynamic like the microtubule-based protrusions. But, there are crucial differences:\\ 
(i) Both the incorporation and removal of subunits, that make the microtubule-based protrusions dynamic, take place almost exclusively at the plus end of the microtubules that are located at the  distal tip of the protrusion.  In contrast, two different types of incorporation/removal of actin subunits are possible in actin-based protrusions. One possibility is that, at least in the steady state, actin monomer incorporation and removal occurs only at ,or very near, the distal tip of the protrusion. In the other alternative mechanism, the actin monomers are continuously added at the barbed (+) end of the filaments at the tip, driven baseward like a treadmill, and removed from the pointed (-) end of the filament at the base. In the latter mechanism a steady (constant) average length of a filament can be maintained by balancing the rates of addition at the tip and removal at the base; the numerical values of these rates, however, can vary widely from one type of protrusion to another. Which of these two alternative ideal mechanisms describes the actual dynamics of a given protrusion needs thorough investigation. If the mechanism turns out to be of treadmill type, the length of the finger-like protrusions can be controlled by regulation of at least one of the three major processes \cite{lin05}: (a) the actin filament polymerization at the tip, (b) the baseward flux that characterizes treadmilling, and (c) filament disassembly at the base. Which of these dominates in a particular type of cell or cell protrusion, and under what circumstances, need further systematic studies. 

Actin polymerization at the protrusion tip can be regulated chemically or physically \cite{lin05}. In the chemical mode, binding or unbinding of capping proteins can downregulate or upregulate actin polymerization. In the physical mode increase or decrease of the force exerted by the membrane at the tip can alter the rate of filament elongation. Finally, myosin motors can either indirectly influence chemical regulation by transporting regulator molecules to the protrusion tip or directly alter the polymerization rate by physically knocking out bound capping proteins or steric constraints arising from membranes, etc. \\

((ii) A cell can have multiple actin-based protrusions of a given type. Exotic architectures formed by the cluster of cell protrusions in stereocilia and microvili, whose morphogenesis remain poorly understood, are also discussed briefly to emphasize the formidable intellectual challenges posed by their spatio-temporal organization. Just like flagella and other microtubule-based protrusions, the actin-based protrusions of a cell may  be correlated through shared common pool of material resources. However, more importantly, additional correlations between the actin-based protrusions arise from the physical links that connect the neighboring protrusions laterally by linking proteins, as we'll see in the specific case of stereocilia. 

The model developed by Orly et al.\cite{orly14} does not distinguish between the detailed structures of the different finger-like protrusions but captures only what are believed to be their  essential common features. In this model the equation for the dynamics of an actin-based finger-like protrusion is obtained from a force balance condition where the {\it protrusive} forces are balanced by the the {\it restoring} forces. The dominant protrusive force is assumed to be generated  by polymerization of actin. The elastic energy of deformation of the membrane at the growing tip is a dominant contributor to the restoring force. The myosin motors that walk on the actin filaments towards the tip of the protrusion also generate restoring force.   In Fig.\ref{fig-actin-based}, we have schematically shown the sites of polymerization and depolymerization, the modes and direction of transport and force. The steady-state solution of these equations describe the geometric shape of the protrusion in terms of its length as well as the radius of the hemispherical tip. The traffic of myosin motors in such actin-based protrusions is an important phenomenon in its own right because of the interesting density patterns displayed in this one-dimensional system far from equilibrium \cite{pinkoviezky14,pinkoviezky17,graf17}.

The friction between the membrane and action-based cytoskeleton is taken into account in the formulation of the dynamical equations of the model developed by Orly et al. \cite{orly14}. In contrast, in none of the models of microtubule-based protrusions known to us the friction between the membrane and the microtubule-based cytoskeleton has been incorporated in the dynamical equations. Moreover, in the models of microtubule-based protrusions kinesin and dynein motors are treated almost exclusively only as transporters and any possible contributions of these motors towards protrusive force generation have been ignored.

\subsubsection{Stereocilia}

\begin{figure}
\includegraphics[width=0.60\textwidth]{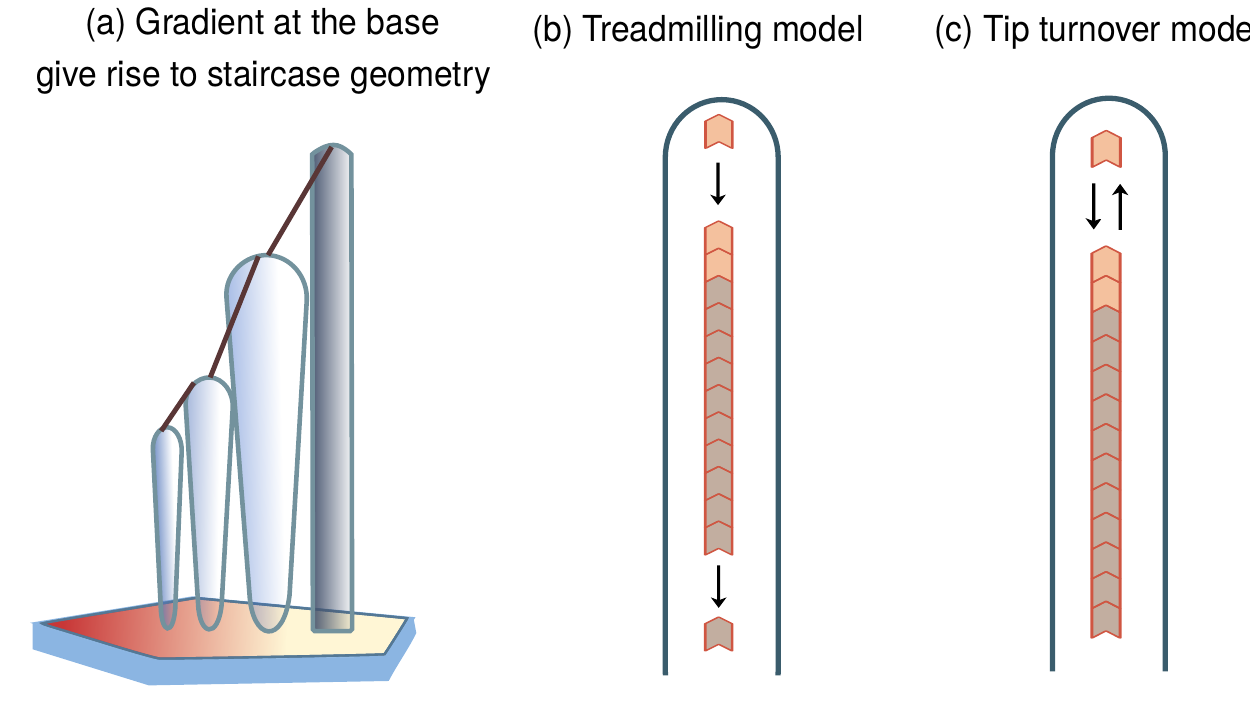}
\caption{{\bf Stereocilia}: (a) Presence of a gradient at a base can lead to the staircase geometry of the stereocilia. (b-c) Two alternative models of actin turnover in stereocilia; (b) treadmilling model and (c) tip turnover model. (See text for the details) }
\label{fig-stero2models}
\end{figure}

Stereocilia are actin based protrusions emerging from the hair cells in the inner ear of mammals. They are the {\it acoustic sensors} as they excite the auditory neurons by transforming the sound vibrations into electrical signals i.e, by mechano-electrical transduction \cite{hudspeth00}. The name of these protrusions is a misnomer in the sense that although part of its name carries the word `cilia', its internal cytoskeletal structure is very similar to the class of actin-based microvilli rather than microtubule-based cilia. How a hair cell selects the number of these protrusions and their positions, which characterize the pattern of the bundle of stereocilia, are challenging open questions \cite{jacobo14,ebrahim16}.

The lengths of different stereocilia in a single cell vary  over a wide range: typically from $1$ to $100\mu$m \cite{manor08}.
The most remarkable feature of stereocilia is the `staircase' geometry in which the stereocilia in the first row are the tallest, those in the second row are somewhat shorter while those in the third row are even shorter. However, in a given row the lengths of all the stereocilia is approximately equal. Another special feature, distinct from filopodia and microvili, is that the tips of the shorter stereocilia are connected to the sides of the adjacent taller row by tip link proteins. These tip links are coupled to the mechanosensitive ion channels at the tips of the shorter stereocilia. The longest stereocilia are attached to a microtubule-based protrusion known as kinocilium \cite{wang21}. Mechanical tension generated in the tip links by deflection of the bundle of stereocilia results in the opening of the ion channels thereby triggering mechano-electrical transduction.

The intrinsic factors that play key role in controlling the length of steriocilia are: (a) molecules that regulate the polymerization and depolymerization of actin filaments at the plus and minus ends; (b) molecules that crosslink the actin filaments in the core of a stereocilium; and (c)  myosin motors and other molecules that bind with/unbind from mysion during myosin's operational cycles. 
The extrinsic factors that also influence the steriocilia length are (i) the crosslinkers of the adjacent steriocilia, (ii) the molecules that transmit the plasma membrane tension to the stereocilia tips; and (iii) kinocilial links that connect the kinocilia with the adjacent stereocilia.

In the context of length control of stereocilia, the biggest puzzle is how the cell controls the lengths of these  protrusions at different levels of the staircase in a position-dependent way  \cite{tarchini13,tarchini16}? 
 Orly et al \cite{orly15} extended their original model \cite{orly14} for the length control of actin-based finger-like single protrusion by coupling multiple protrusions to an apical surface that is assumed to be spatially inhomogeneous. This inhomogeneity can arise, for example, from a spatial gradient of the viscosity $\gamma_c$ of the cytoplasm. Orly et al. \cite{orly15} claimed this scenario to be superior to some other alternative possibilities in explaining  the staircase like arrangement of the stereocilia. For example, the model with a linear gradient in $\gamma_c$ develops into the the characteristic staircase pattern of stereocilia  in the steady state \cite{orly15} (see Fig.\ref{fig-stero2models}(a)). However, to our knowledge, the key assumption of this model, namely the spatial inhomogeneity of the apical surface required for the emergence of the staircase pattern of the stereocilia is yet to receive full experimental justification.

Two alternative explanations of the staircase pattern  seem plausible, at least in the steady-state after completion of the morphogenesis (see Fig.\ref{fig-stero2models}):\\
(i) {\bf Treadmilling model}:  This model, based on early experimental work \cite{rzadzinska04}, assumed  \cite{manor08} that  treadmilling takes place in longer stereocilia at a proportionately higher rate. A direct implication of this assumption is that, in any given bundle of stereocilia, the actin monomers would take the same duration of time to treadmill through the length of the steriocilia irrespective of their individual lengths. Consequently, the entire bundle of stereocilia would renew synchronously thereby enabling a dynamic stability of the staircase structure (see Fig.\ref{fig-stero2models}(b)). Obviously, this model also assumes a spatial inhomogeneity, namely that of the treadmilling rate, in the steady state to maintain the staircase geometry. This explanation, if true, raises the next obvious question: how does a hair cell sense the lengths of the stereocilia and what physical or chemical mechanism does it adopt to ensure that the velocities of their treadmilling are proportional to their respective lengths? \\ 

(ii) {\bf Tip turnover model}: More recent experiments \cite{narayanan15,mcgrath17}, however, suggest that, except for a very narrow strip of the actin bundle at the tip the remainder shaft remains practically static. Actin monomers assemble and disassemble only at the tip; the constant length of each stereocilium is maintained because of the balancing of the rates of assembly and disassembly only at the tip (see Fig.\ref{fig-stero2models}(c)). In other words, treadmilling is practically nonexistent and actin turnover takes place only at the tip and none at the base. If this result withstands further experimental scrutiny, explanation of the emergence of the staircase pattern through morphogenesis of the stereocilia will have to be found.

During morphogenesis the rate of actin assembly at the tip must be higher than that of disassembly \cite{mcgrath17}. However, the kinetics of binding-unbinding of the actin cross-linkers \cite{prost07,lenz10} and the cargo transport as well as force generation by myosin motors \cite{ebrahim16} may also play important roles in the emergence of the shapes of the stereocilia. The final design of a fully formed bundle of stereocilia is necessary for mechano-transduction.  What would be more interesting if it is firmly established that during the  formation of that bundle mechano-transduction actively shapes their design including the lengths of the stereocilia \cite{caberlotto11,krey20}.

\subsubsection{Microvilli, microridge, brush border }

Another class of actin based protrusions are microvilli. They cover the apical surface of various epithelial cells \cite{sauvanet15}, enterocytes which form the intestinal surface \cite{crawley14}, trophoblasts of placenta and the surface of oocytes \cite{courjaret16} and lymphocytes \cite{orbach20}. Microvilli protruding from the cell surface cause manyfold increase of the effective surface area of the cell without changing the cell volume. They are involved in absorption and secretion, serve as reservoirs of membrane and actin monomers. When a cell needs additional membrane to expand its size \cite{figard16}, it resorbs microvilli (see Fig.\ref{fig-microvilli}(a)). Actin monomers released after resorption are utilised by the leading edge of the migrating cells \cite{ubelmann13} and supply the actin needed from epithelial to mesenchymal transition. Microvilli form ridges on the surface of epithelial cell (see Fig.\ref{fig-microvilli}(b)).


\begin{figure}[htbp]
\begin{center}
{\bf Figure NOT displayed for copyright reasons}. 
\end{center}
\caption{{\bf Microvilli:} (a) Microvilli serve as reservoirs of membrane and actin monomers. (b) Short microvilli forming ridges on the surface of epithelial cells. (c)  Clustering of microvilli in the intestinal brush border. (Fig.4 of ref.\cite{crawley14}). }
\label{fig-microvilli}
\end{figure}

A model for the dynamics of a single microvili  \cite{gov06} was formulated in terms of force balance. For short microvilli, the equation of motion for the length $h$ (referred to as `height' in ref.\cite{gov06}) of a single microvili was argued to have the form 
\begin{equation}
\frac{\partial h}{\partial t}=-\omega h+An
\end{equation}
where $\omega$ characterizes the strength of the restoring force exerted by the membrane and $A$ relates the membrane protein density $n$ to the forward velocity $v=An$ induced by actin polymerization. 
This mechanism can be interpreted as a variant of the {\it balance point model} where the length-dependent first term competes against the length-independent second term. At the balance point, which corresponds to the steady-state, the length of a single microvili is given by $v/\omega$ as one would have anticipated intuitively. 

Normally microvili do not appear in isolation; on many cells the arrangement of microvilli resemble long ridges (see Fig.\ref{fig-microvilli}(b)). Interaction between the neighboring microvili that tends to suppress length difference between neighboring microvili can account for the observed narrow length distribution of these protrusions along a ridge \cite{gov06}. However, convincing evidence in support of the physico-chemical origin of such interaction is still lacking. Other interesting ordered arrays of microvilli include (i)  {\it brush border} \cite{crawley14}, (ii) whorls  \cite{depasquale18} for which, to our knowledge, no mechanism of length control have been reported till now in the literature.

\section{Bacterial appendages}

\subsection{Bacterial flagella}

Bacterial flagella are long protrusions that emerge from the bacterial cell surface. Flagella serve the dual purpose of swimming and swarming; it is the most extensively studied  among all the prokaryotic motility structures \cite{jarrell08}. It consists of roughly three substructures: (a) a basal body complex, (b) the hook, and (c) the filament. 
The basal body is anchored in the bacterial membrane. Its main components are the rotary motor and an export machinery. This rotary motor that is powered by an ion-motive force. A {\it rod}, which is a constituent of the basal body, extends from the rotor of the rotary motor and is encircled by a series of coaxial rings. One end of the hook gets connected with this rod while the other end joins the whip-like filament. Thus, the flagellar hook transmits the rotational movement generated at the basal body to the filament.  It is the rotation of the flagellar filament that propels the bacterial cell in the aqueous medium.  It is worth pointing out that unusual flagella can deviate from the general spatial organization mentioned above. To our knowledge, the most unusual are the periplasmic flagella of spirochaetes. The filament of these flagella do not project out of the cell surface; instead they remain confined between the cytoplasmic and outer membranes of the cell. We'll not consider length control mechanisms of such uncommon flagella. 

The biogenesis of bacterial flagella proceeds in the direction base-to-tip, i.e., in the sequence first basal body, then hook and, finally, the filament \cite{chevance08}. 
The sequential order of this morphogenesis is maintained by the cell with a combination of two control systems:\\
(1)  The first implements a sequential gene expression, i.e., sequential {\it production} of structure building materials. In the early stage of flagellar growth the proteins synthesized are mainly those that are constituents of the rod and the hook while in the late stages structural proteins of the filament are the dominant products of this regulated gene expression system.\\
(2) The second control system ensures the correct sequential {\it delivery} of the required proteins at the growing tip of the flagellum. For this purpose the flagellar export machinery must have a switching mechanism whereby its substrate specificity switches from rod-/hook-type to filament-type \cite{ferris06}. Such a switch would guarantee that proteins required for the growth of the rod and the hook are exported in the early stages and the export of the filament proteins begin only after the synthesis of the hook is completed.\\ 
The two control systems for production and delivery must be coupled to each other for the correct sequential growth of the flagellum. Unravelling the molecular identity of the components of these two control systems and the kinetics of their operation are challenging tasks and are essential for understanding the mechanisms of length control of bacterial flagella. In this subsection we review the growth of both the hook and the filament. Another bacterial cell protrusion called injectisome \cite{erhardt10}, that has many close similarities with the bacterial flagellum, will be discussed in the next subsection.  

\subsubsection{Bacterial flagellar hook }

It is generally believed that the hook of each bacterial species needs to be of a particular length. Hooks much shorter than an optimal length cannot generate sufficient bend angle whereas those much longer than the optimal may not transmit the torque efficiently \cite{waters07}. As an example of typical lengths of flagellar hooks, consider 
wild type {\it S. enterica} whose flagellar hook in straightened conformation has an average length $55.0 \pm 5.9$ nm \cite{waters07}. The standard deviation, about $10 \%$ of the mean, is quite large. It strongly suggests that  the control of flagellar hook length is not tight and  indicates the possibility of a stochastic mechanism. Interestingly, the diameter (~20 nm) of the hook is of the same order as its length (~55 nm).

The first experimental evidences reported about half a century ago \cite{silverman72}, and all the others reported since then, have established that FliK protein plays a key role in the flagellar length control. However, the actual mechanism through which FliK achieves this feat remained mysterious for the following reasons \cite{aizawa12}: (i) FliK itself is not incorporated into the flagellar structure, (ii) FliK is soluble and secreted, (iii)  FliK seems to perform two distinct functions; it somehow controls the hook length and also triggers substrate specificity switching. 

In the first half of the first decade of this century the debate on the mechanism of length control of a growing flagellar hook heated up with the claim of two competing models \cite{makishima01,journet03}  reported by two different research groups based on their interpretations of the respective experimental observations. Let us begin by summarizing those two groups of models:\\
{\bf (ia) Finite measuring cup model \cite{makishima01}: } Among the coaxial rings encircling the basal rod, the ring at the cytosolic end called the C-ring. Prior to the initiation of the assembly the subunits accumulate in the cavity of the C ring and the hook growth stops when this pool of subunits in the cup gets exhausted. So, in this model, it is the size of the C ring  which governs the length of the protrusion . Studying the lengths of large varieties of targeted mutants, it has been concluded \cite{aizawa12} that although the C ring is necessary for hook formation it does not determine the hook length thereby invalidating the original version of the measuring cup model. In fact, it may be approriate to treat the C-ring as a `docking station', rather than as a cup, for the hook subunits before they enter the channel  \cite{hughes12b}.\\

{\bf (ib) Waiting room model \cite{aizawa12}: } In an attempt to rectify the shortcomings of the measuring cup model without scrapping it altogether, an improved model, called {\it waiting room model}, was proposed.  This model is essentially the measuring cup model with an additional step of the process that is assumed to involve a `ruler' molecule. In this model it is assumed that the flagellar proteins accumulate within and around the C ring before secretion.  Initially, the accumulated subunits at the base block the contact of the ruler molecule with the base. Once the subunits get exhausted because of the growth of the hook, the ruler molecule easily interacts with the base and change its conformation in such a way that further export of the subunits is inhibited (see \cite{hughes12b} for critical comments). Note that in this model the ruler does not directly determine the length of the fully grown hook. 
Because of the observed dramatic effects of the FliK protein on the hook length \cite{silverman72} it is considered to be a possible candidate for the role of the `ruler' \cite{aizawa12} although the validity of this model itself still remains controversial \cite{hughes12,hughes12b}. \\ \\

{\bf (iia) Static molecular tape model \cite{journet03}}: In this model the hypothetical ruler molecule is assumed to directly control the length of the hook by performing the role of a `measuring tape'. In this scenario, one end of the tape is assumed to be attached to the growing tip of the hook whereas the other end loosely hangs inside the cytoplasm. As the hook  elongates, the loosely hanging tip of the tape comes closer to the base. On interaction with the base, it induces substrate switching thereby inhibiting further export of subunits of the hook. FliK is a possible candidate also for the tape in this model. However, this model is too simple to explain length control of bacterial flagellar hook.  Several arguments against the validity of this model have been listed (see, for example, ref.\cite{aizawa12}).\\ 
{\bf (iib) Molecular clock model \cite{moriya06}}: It was proposed in this model that hook length is dependent on the rate $R$ of hook polymerization. The start of the hook polymerization switches on a molecular clock (or, is it more appropriate to call it a `timer'?) that decides the hook length $L$ by the simple equation $L = T/R$ where $T$ is the time on the clock when the substrate switching takes place thereby stopping further elongation of the hook. Possible candidate for the molecular clock was also speculated. However, some investigators regard it as just a variant of the molecular tape (ruler) model  where the timer is essentially a ruler \cite{aizawa12,aizawa12b}. \\ 
{\bf (iic) Infrequent  dynamic ruler model \cite{erhardt11}}: This model proposes that a molecular ruler is secreted into the protrusion randomly from time to time and this ruler protein is later released outside the cell. As the protrusion elongates , the time of passage through the protrusion would increase and the end which is inside the cell would linger around the base for longer period of time. This, in turn, would increase the probability of its interaction with the base and, upon interaction, it would induce substrate switching thereby halting the elongation of the hook \cite{erhardt11,wee15}. (see \cite{aizawa12b} for critical comments). \\

\begin{figure}[htbp] 
\begin{center}
{\bf Figure NOT displayed for copyright reasons}. 
\end{center}
\caption{{\bf Temporally evolving length of bacteria flagella:} (a) Assembly of bacterial flagellum. Flagellin subunits synthesized in the cell are unfolded and injected into the growing flagella which is of tubular shape. The flagellin subunits diffuse inside the narrow channel and on reaching the tip, the flagellin subunits get folded and polymerized into the existing chain of flagellin monomers forming the flagella. 
(b) Frequent pauses made by the growing flagella in multiflagellated bacteria (Adapted from Fig.4 of ref.\cite{zhao18}).   }
\label{fig-bflagella}
\end{figure}

\subsubsection{Bacterial flagellar filament}
 
The flagellar filament is approximately 10 to 15 $\mu$m long but its diameter is only about a couple of nanometers. 
Flagellin are the subunits which assemble to form the flagellar filament. At the base of the flagellum, an export machinery pushes the flagellin in (fully or partially) unfolded form into the narrow channel of bacterial flagellar filament \cite{lee15} (see \ref{fig-bflagella}(a)). Inside the channel, the unfolded flagellin moves solely by diffusing. The narrowness of the channel does not allow overtaking of one flagellin by another \cite{chen17,renault17}.  Flagellin transport is modelled as a single file diffusion \cite{stern13} along the narrow channel. On reaching the tip, the flagellin folds, gets added to the tip of the existing flagellar filament and this polymerization of flagellin elongates the flagellum by one unit at time \cite{chen17,renault17}. In their recent review, Zhuang and Lo \cite{zhuang20b} have presented  a timeline of important  experiments and models which summarize how the experimental and theoretical research of flagellar filament construction and loss have evolved over the past five decades.

A mathematical model for this process was proposed by Keener \cite{keener06} (see \cite{renault17} for a simplified outline). The flux $J(x,t)$ of the monomers is defined as the number of unfolded monomers passing through $x$ per unit time at time $t$. Accordingly, the dynamical equation governing the flagellar length is given by 
\begin{equation}
\frac{dL}{dt} = \beta J(L,t)
\label{eq-bactFlagEq}
\end{equation}
where $\beta$ is the increment of flagellar length caused by incorporation of a single monomer. Since the process is diffusive inside the channel, $J(L,t)$ on the right hand side of (\ref{eq-bactFlagEq}) couples the filament growth with the diffusion of the monomers within the channel. Since the rate of flagellar growth is small compared to the average velocity of the monomers, the time-dependent density profile of the monomers is well approximated by the corresponding stead-state profile. In other words, because of this approximation, justified by the time scale separation, $J(L,t)$ is approximated by $J_{ss}(L)$ which is obtained by solving the time-indent diffusion equation under the boundary conditions that realistically capture the conditions at the two boundaries of the channel. This model also has not gone unchallenged; for the arguments and evidences reported against this model as well as for the counter-arguments and counter-evidences see refs.\cite{turner12,evans13,renault17,chen17,hughes17,schmitt11}.

Different species of bacterial cells can have different numbers of flagella that are distributed spatially on the cell for convenience of their biological function \cite{aizawa98}. How a bacterial cell controls the number and position of its flagella is an interesting area of research but this topic is beyond the scope of this review \cite{schuhmacher15}.  
The flagella of multiflagellated cell {\it Escherichia coli} make  frequent pauses during the flagellar development, as captured by the plot given in Fig.\ref{fig-bflagella}(b); these pauses are suspected to be caused by the exhaustion of flagellin monomers \cite{zhao18}. Whether the flagella of a multi-flagellated bacterium communicate through the cell body resulting in any correlation between their length changes is not known.  \\ \\
{\bf Loss and regeneration of bacterial flagella:} Three different modes of loss of flagella are known.\\
{\bf (i) Loss by breaking: } Bacterial flagella gets shortened partially by breaking at arbitrary positions when subjected to continuous shear while swimming.  But the cell can successfully overcome this loss by regenerating the flagellum  with the continuous supply of the flagellin at the tip. The rate of elongation of the regenerating flagellum is a function of the flagellar length. \\
{\bf (ii) Loss by amputation:} In contrast, when the flagella is amputated by using lasers in a controlled manner, the  flagella is unable to regenerate. Lasers induce damage to the flagella in such a way that no new flagellar monomers could be added to the new tip \cite{paradis17}. \\
{\bf (iii) Loss by ejection:} The bacterial cells eject their protrusion when there is dearth of nutrition and regenerate them back when sufficient nutrition is available \cite{ferreira19,zhu20,zhuang20}.

\subsection{Injectisome}

The type III injectisome of Gram-negative bacteria is closely related to the bacterial flagellum \cite{erhardt10}. The resemblance is not only structural but also compositional: about twenty proteins are essentially common constituents of the basal bodies of both. However, in place of the hook and the filament of the flagellum the injectisome has a straight needle. But, the structure of the export machinery, through which most extracellular components of the flagellum and injectisome are exported, are also closely related to each other. Just like the flagellum, the injectisome is also assembled sequentially from the base to the tip of the needle. Here we are interested in the length control of the needle complex. Because of the similarity, the injectisome and flagellar hook are often studied side by side.

At least two alternative models have been proposed in the literature for the control of the length of the needle of injectisome; these are summarized below.

{\bf (i) Ruler-based model}: According to the ruler mechanism, a specific ruler protein, which is secreted into the needle periodically, interacts with the base whereas the other end of the protein connects with the growing tip of the needle. This needle and the ruler protein  interaction serves as an indicator of the needle length to the base. When the ruler protein gets fully extended, the base begins the secretion of tip proteins which leads to the needle maturation. This is same as the {\it infrequent dynamic ruler mechanism} discussed in the context of flagellar hook. Analyzing their experimental data, Cornelis and co-workers claimed that in {\it Yersinia}, YscP can serve as the ruler protein \cite{cornelis10,nariya20}. 


{\bf (ii) Substrate switching model}: As observed in {\it Salmonella} and {\it Shigella}, an inner rod inside the base complex spans between the inner and the outer membrane. In this scenario,  simultaneous assembly of the inner rod and the outer needle takes place till the termination of the assembly of the inner rod. This model assumes that termination of the assembly of the inner rod triggers substrate switching so that export of protein subunits of the outer rod from the base is replaced by the secretion of tip proteins. One consequence of this mechanism of needle length control is that in {\it Salmonella} overexpression of the inner rod protein, PrgJ resulted in the formation of shorter needles, and deletion of PrgJ lead to the creation of very long needles \cite{marlovits06}. 

\begin{figure}[htbp] 
\begin{center}
{\bf Figure NOT displayed for copyright reasons}. 
\end{center}
\caption{{\bf Mechanisms of length control of injectisome:}  (a) (b) Ruler based mechanism (Fig.1B of ref.\cite{nariya20}). (b) Substrate switching mechanism (Fig.1 of ref.\cite{nariya16}).}
\label{fig-injectisome}
\end{figure}

\subsection{Pili and fimbriae}

Non-flagellar appendages on bacterial surfaces were first observed in electron-microscopic studies of outer membranes of Gram-negative bacteria \cite{telford06}. One of the two leading research groups during the first decade of research on these protrusions called these `pili' while the other group referred to these appendages as `fimbriae'. We'll use the two terms interchangeably. Subsequently the pili were found also on the surface of Gram-positive bacteria. One distinction between the pili of Gram-positive and Gram-negative bacteria is that the monomeric subunit are bonded covalently in the former but non-covalently in the latter \cite{ramirez20}.

Pili or fimbriae are involved in wide range of functions including motility as well as adhesion to other cells and environmental surfaces. Pilus of correct wildtype length is essential for various functions of the bacteria \cite{chang19}; this raises the question of the mechanisms of length control for pili. In stark contrast to the most of the other protrusions discussed in this review, which elongate by adding subunits at their growing distal tips, pili elongate and retract by adding or removing monomers at their base \cite{hospenthal17}. 
Pili of a class of bacteria attain a steady length whereas those of certain other class are dynamically unstable and keep switching  between phases of elongation and retraction with brief pauses in between \cite{koch21}.
The  relevant question in the latter case is: how does the cell control the switching of the pilus between the growing and shrinking phases and how do the observed statistics of the tip trajectory arise from a (possibly stochastic) kinetic model of the control system.

\subsubsection{Pilus length control in Gram-positive bacteria}

Pilus biogenesis in Gram-positive bacteria is believed to be a biphasic process which takes place in two distinct but coupled phases \cite{ramirez20}. In the first phase, the polymerization of the pilus shaft and this is followed by the second phase in which the polymerised pilus is anchored on the cell wall. At least three alternative scenarios, possibly valid in three different species of bacteria, have been discussed in the literature.

{\bf (i) Length control by the base pilin:} In {\it Corynebacterium diphtheriae}  the deletion of the base pilin SpaA leads to a pilus of increased length whereas its overexpression leads to pilus of shorter length than the wild type pili. These observations indicate that the base pilin SpaA acts as a switch for the termination of these pilus polymerization \cite{mandlik08}. However, what mechanism triggers the incorporation of the SpaA pilin at the base, thereby stopping further elongation of the pilus remains to be established.

{\bf (ii) Length control by the shaft pilin:} In another class of pili in {\it Corynebacterium diphtheriae},  overexpression of the shaft pilin SpaH usually results in longer pilus filament. Thus, for these pili  the level of the shaft pilin SpaH is a major determinant of the length of the pilus.\cite{swierczynski06}. 

{\bf (iii) Length control by an enzyme:} These special classes of pili do not have a base pilin. In such species of bacteria, a special enzyme, that has two structural components, is believed to regulate the length of the pilus. If one of these two structural components of the enzyme is genetically modified, a pilus shorter than the wild type is assembled whereas if the other structural component is altered genetically, a pilus longer than the wild type is assembled. These experimental observations strengthen the claim for a crucial role of this special enzyme, called housekeeping sortase, in pilis biogenesis in a special group of species of Gram-positive bacteria \cite{chang19}.

\begin{figure}[htbp] 
\begin{center}
{\bf Figure NOT displayed for copyright reasons}. 
\end{center}
\caption{{\bf Temporally evolving length of pili:} (a) Trajectory depicting the extension and retraction
dynamics of pilus length for seven individual pili (roman numerals) which emerge from the same pole of the 
same cell without surface contact (Fig 2(c) of ref.\cite{koch21}). (b) Schematic of the model proposed
in ref.\cite{koch21} for explaining dynamic features pilus extension and retraction 
(Fig 3(a) of ref.\cite{koch21}).}
\label{fig-pili}
\end{figure}

\subsubsection{Pilus with dynamic length in Gram-negative bacteria} 

As stated earlier, the pilins in Gram-negative bacteria are linked by non-covalent protein-protein interactions although their modular arrangement is similar to to that of Gram-positive bacteria, namely the three modules: tip, shaft and base. One of the distinct features of the pili of Gram-negative bacteria is their dynamic nature. The pilus length can be modulated, for example, by\\  
{\bf Addition and removal of pilin subunits by extension and retraction motors:} The length T4 pilus of {\it Pseudomonas aeruginosa} exhibits random cycles of extension and retraction where the periods of elongation and shortening as well as that of the pause in between are random variables \cite{koch21}. The statistical quantities that characterize this process quantitatively are: distribution of pilus length, and distribution of  timescales like the extension time, retraction time and dwell time \cite{koch21}.  Important roles of extension and retractions motors in this process have been suspected for a long time. However, a more concrete stochastic model for this phenomenon has been reported only very recently \cite{koch21}. This model is based on the assumption \cite{koch21} that (a) when the binding site at the base of the pilus is free each of the two types of motors has a finite probability to attach, and (b) not more than one motor can bind that site simultaneously. In other words, the extension and retraction motors compete for binding and both the binding and unbinding are stochastic events.
In this minimal model, which is essentially a 3-state Markovian model, the basal body of the motor switches between three states: (1) unbound, (2) bound to the extension motor and (3) bound to the retraction motor. The six key parameters of the model are: (1) the speeds of (a) growth and (b) retraction,  (3) the rates of (c) binding ($k_\text{ext,on}$) and (d) unbinding ($k_\text{ext,off}$) of the extension motor,  and the rates of (e) binding ($k_\text{ret,on}$) and (f) unbinding ($k_\text{ret,on}$) of the retraction motor . The results of the model are consistent with the experimental data (see ref\cite{koch21}). However, this pioneering work marks only the beginning of quantitative investigations  on the mechanisms of length control in Gram-negative bacteria.


\section{Summary, conclusion and outlook}


Tubular cell protrusions are ubiquitous and perform their respective biological functions. Interestingly, for essentially the same biological function different species of cells may deploy protrusions made of different components and having structures which might have descended from different molecular ancestors. In this article we have compiled a comprehensive list of cell protrusions that differ from each other in terms of molecular components and structures although we have not considered their evolutionary origin. For the convenience of the readers, this list has been presented in a tabular form (see Table-I) where the key physical properties of the protrusions have been mentioned. In spite of the diversity of components, structures and dynamics, the cell protrusions of different types share a common set of structural modules. Attention of the readers have been drawn to the universal features of these modular structures by comparing the architectures of different protrusions in Table-II.

The main focus of this article is the mechanisms of controlling the length of protrusions that are known to attain a particular characteristic length for performing their respective specific functions. Unlike physics, there are no universal principles that govern a specific phenomenon in all biological systems. So, not surprisingly, length of different types of protrusions of a cell may be controlled by different mechanisms. Besides, different species of cells may adopt different mechanisms for controlling the length of the same type of protrusion. Nevertheless, based on our survey, we have presented a comprehensive list of all varieties of mechanisms of length control of cell protrusions observed or speculated in the literature. For each such mechanism, the experimental evidences for and against its applicability to some specific examples have also been pointed out in the list.  For the convenience of readers who might not be interested in details but wish to get a bird's eye view of the whole field, a summary of these mechanisms is  presented in table-III. Tables have been used extensively in this review for comparative assessment of the systems, phenomena and models.

Experimental studies of living systems have always been extremely challenging. Collecting highly accurate quantitative data on many different aspects of a cell protrusion for a specific cell type would help in establishing the underlying mechanism. Unfortunately, the limited data available at present are often not adequate for this purpose. Consequently, for almost every type of cell protrusion more than one competing mechanism seem to be consistent with the experimental observations. In case of eukaryotic flagella, which we have reviewed in most detail, we have presented all the competing models in comparable detail without any bias or prejudice. Although the authors are theorists, the experimentally established facts on the core issues are summarized so that the reader gets a balanced picture of the current status of this field of research. At the risk of hazarding a guess, we present a list of areas that need attention of both experimentalists and theorists:\\

${\bullet}$ Although mechanisms based on a hypothetical ruler or timer seem to have successfully described at least the qualitative trends in the experimental data for some protrusions, the identity of the molecular ruler or the timer still remain elusive. Identification of the timers or/and rulers need urgent attention. \\

${\bullet}$ For ruler-based mechanisms the identification of the ruler explains how its length decides the length of the protrusion. But, it leads to the obvious next question: what is the mechanism that decides the length of the ruler itself?\\

${\bullet}$ For timer-based mechanisms, even after proper identification of the timer, one challenging open question would have to be addressed for a complete understanding. A timer gives feedback to the cell as to the instantaneous length of the corresponding protrusion in the time or frequency domain. What is the hardware and software of the cell that decodes this information on the length that is encoded in a time or frequency? \\

${\bullet}$ As we have seen in this review, important information on the length control mechanism can be extracted from statistical analysis of protrusion length fluctuations. Time may be ripe now for experimental measurement  of the fluctuations and noise in the context of length of protrusions. \\ 

${\bullet}$ Gene expression is known to be bursty \cite{kaern05,raser05,raj08}. What effects, if any, do such noisy production of structural proteins have on the elongation kinetics and, hence, on the length control mechanisms \cite{zhao18}?

${\bullet}$ Most of the protrusions, like the eukaryotic flagella, are membrane-bound appendages. The contribution of the membrane elasticity in the generation of force against the elongation of protrusion has been incorporated in the models of actin-based protrusions, but neglected in the studies of microtubule based protrusions. The role of the ciliary membranes in the transport of IFT trains along the narrow passage in between the axoneme and the ciliary membrane remains a mystery and, therefore, a possible topic for future studies.\\

${\bullet}$ Anterograde as well as retrograde transport of the protrusion-building materials as molecular cargo is obviously a process essential not only for the growth and shrinkage of the protrusion but also for maintenance of its steady length because of the high turnover rate. The molecular motors driving the active transport have been studied extensively while the cargo carriers like IFT trains have received far less attention. The fusion and fission of IFT trains during their journey along an eukaryotic flagellum was discovered a few years ago, but further progress has been slow since then. \\

${\bullet}$ Surprisingly, the effects of the congestion of IFT trains, and similar other motorized cargo carriers, on the traffic flow along protrusions have been modelled only theoretically. The effects of traffic congestion, and possible jamming, on the regulation of  the rates of assembly and disassembly of cell protrusions should be explored experimentally {\it in-vivo}. \\

${\bullet}$ Many actin-based protrusions form clusters whose unique spatial organization is essential for the biological function of those protrusions. How the necessary length distribution of those clusters arise and are maintained in the steady state by the collective dynamics need both experimental investigation and theoretical modelling. 

Finally, no doubt, size matters \cite{bonnerbook}.  However, it is worth pointing out that the search for size control mechanisms is not limited to only to individual living organisms. A single colony of social insects, like ants and bees,  is known to self-organize to an optimum size that is ``convenient'' for collective foraging and decision making \cite{dornhaus12}. How such a super-organism \cite{holldoblersuperorg} can sense and control its size is a question similar to the questions we have addressed in this article at the sub-cellular level. It is not an exaggeration to say that exploration of the mechanisms of growth and sustaining size pervades almost all branches of natural and social sciences starting from cells to cities and civilizations \cite{smilbook}; however, a comparative study of those mechanisms operating from micro-scale to mega-scale is far beyond the scope of this review and left as a multi-disciplinary future endeavour.

\section*{{\bf Acknowledegements: }} 

DC thanks SERB (India) for supporting this research through a J.C. Bose National Fellowship.


\section*{{\bf Author contributions: }} 

SP, DC and FJ selected the topics to be included, planned the organization of the contents as well as the depth of analysis and breadth of the perspective of this review.  A large part of this review is adapted from SP's thesis \cite{patrathesis}, written under the supervision of DC and submitted to IIT Kanpur for the Ph.D. degree. SP designed all the original graphical contents and the tables. SP, DC and FJ wrote the text and all the authors approved the final manuscript.


\widetext


\end{document}